\documentclass[a4paper,11pt]{article}
\usepackage{fullpage}
\usepackage[pdfencoding=auto]{hyperref}
\usepackage{graphicx}
\usepackage[english]{babel}
\usepackage{amsmath}
\usepackage{amssymb}
\usepackage{dsfont}
\usepackage{bm}
\usepackage{tikz}
\usetikzlibrary{decorations.pathmorphing, decorations.markings}

\newcommand{\normal}[1]{{\hspace{-0.1ex}:\hspace{-0.6ex} #1 \hspace{-0.6ex}:\hspace{0.2ex}}}
\newcommand{\s}{\hspace{0.2ex}}
\newcommand{\barred}[1]{\hspace{0.2ex}\overline{\hspace{-0.2ex} #1}}
\newcommand{\circcolon}{\raisebox{4pt}{\tiny{$\circ$}}\hspace{-3.8pt}\raisebox{-0pt}{\tiny{$\circ$}}}
\newcommand{\circnormal}[1]{\circcolon #1 \circcolon}

\newcommand{\rrangle}{\rangle\hspace{-2pt}\rangle}

\numberwithin{equation}{section}

\title{\textbf{Angular Quantization in CFT}}
\author{Nicholas Agia and Daniel L. Jafferis}
\date{\vspace{1ex}\emph{Center for the Fundamental Laws of Nature, Harvard University, Cambridge, MA, USA}}

\begin{document}

\maketitle

\begin{abstract}
The most common quantization scheme in which to study a conformal field theory is radial quantization, wherein a Hilbert space of states is defined on a sphere, whose Hamiltonian when mapped to the plane is the dilatation generator and which boasts a state/operator correspondence. In this paper, we consider an alternative quantization scheme for 2d CFTs in which the plane is foliated by constant-angle slices, as opposed to concentric circles, whose Hamiltonian is the rotation generator. In this angular quantization, there is no state/operator correspondence but instead an ``asymptotics/operator correspondence''. A central feature is that the quantization slice ends on two operators, and a regulator must be chosen by excising holes around each operator and imposing suitable boundary conditions such that the holes shrink to the desired local operators. This angular quantization may be viewed as constructing CFTs in Minkowski space, or separately as studying non-conformal (but local) boundary conditions in CFTs. We provide explicit Fock space constructions for various free 2d CFTs. In addition to the motivation of applying angular quantization in string theory situations where traditional radial quantization is insufficient, we comment on its relation to modular Hamiltonian approaches to entanglement entropy.
\end{abstract}

\tableofcontents

\section{Introduction and Motivation}

A conformal field theory is defined, in part, by its spectrum of local operators together with their correlators on a conformal equivalence class of manifolds. However, to make contact with fundamental principles of QFT such as unitarity and locality, one must construct a Hilbert space of the CFT by foliating spacetime in spatial slices, allowing one to discuss the states of the CFT and an associated Hamiltonian. The most common choice of quantization scheme in the literature for a $d$-dimensional CFT is radial quantization, where the spatial slices are the spheres of the Lorentzian cylinder $\mathds{R}\times S^{d-1}$, which upon analytic continuation to Euclidean signature becomes the plane $\mathds{R}^d$ foliated in concentric spheres via the usual exponential map. A great deal is known about CFT Hilbert spaces in radial quantization, such as possessing a unique ground state with a gapped spectrum, primarily due to the famous state/operator correspondence providing an isomorphism between the local operators and the Hilbert space of states on the sphere. 

Of course, one is free to choose any foliation to define a quantization scheme, and neither the local operators nor their correlators can depend on this choice, as the latter are intrinsic to the CFT. When two foliations differ by a continuous deformation, the Hilbert spaces so constructed are isometrically isomorphic, and in particular each state in one Hilbert space has a unique and convergent expansion in a basis of the other. This isometry is clear from the point of view of a path integral, where the expansion coefficients are given by the path integral over the bulk region between the two slices with fixed boundary wavefunctionals.

However, the story is far more subtle for Euclidean foliations which are not deformable into each other. The simplest example is that of Euclidean $\mathds{R}^d$ foliated in concentric spheres $S^{d-1}$ versus Euclidean $\mathds{R}^d$ foliated in Cartesian planes $\mathds{R}^{d-1}$. The former upon analytically continuing the radial direction becomes the CFT on the Lorentzian cylinder $\mathds{R}\times S^{d-1}$, while the latter upon analytically continuing the Cartesian direction becomes the CFT on Minkowski space $\mathds{R}^{d-1,1}$. Far less is known about the structure of general CFTs on Minkowski space, whose Hilbert space is not isomorphic to the one in radial quantization and indeed is not even gapped. Of course, that the Lorentzian cylinder Hilbert space is gapped is not surprising because the cylinder has some finite radius $R$ and the physical energy of a state is $E = \frac{\Delta}{R}$, where $\Delta$ is the scaling dimension of the associated operator under the state/operator correspondence. One way to obtain the CFT on Minkowski space is to take the $R\rightarrow\infty$ limit of the theory on the Lorentzian cylinder, upon which the energy spectrum generically becomes a gapless continuum, and it becomes possible for a moduli space of vacua to arise\footnote{It is commonly believed that a moduli space of vacua can only exist in free or supersymmetric theories.}. This limit is necessarily singular, and the relation between the radially-quantized Hilbert space and the flat Minkowski space Hilbert space is far from obvious.

In this paper we shall describe in detail yet another quantization scheme in the specific context of two-dimensional CFTs wherein the Euclidean plane is foliated by constant-angle rays extending from the origin to infinity, which we generically call ``angular quantization''. In the sphere conformal frame, one considers an operator $\mathcal{O}_i$ at the origin and an operator $\mathcal{O}_j$ at its antipodal point (one or both of which may be the identity), and the goal is to construct the angularly-quantized Hilbert space $\mathcal{H}_{ij}$ associated to the spatial slice ending on those two operators and its associated Hamiltonian $H_{\text{R}}$, which we shall call the ``Rindler Hamiltonian''. This procedure requires regularization near the endpoints where the two operators sit, so we shall excise holes of radius $\varepsilon$ and place suitable boundary conditions which shrink to the appropriate operators, constructing the desired Hilbert space as the $\varepsilon\rightarrow 0$ limit of the regulated one. This construction is performed very explicitly for the free field examples given below, though its generalization to arbitrary CFTs is straightforward, albeit abstract. 

The concept of angular quantization is hardly new. It is the analytic continuation of the Rindler quantization performed by a uniformly accelerating observer in Minkowski space, which is why we call $H_{\text{R}}$ the Rindler Hamiltonian despite being in Euclidean signature. Furthermore, the description of the modular Hamiltonian in the context of entanglement entropy \cite{Casini09, Cardy16, Michel16} is essentially a special case of angular quantization. This latter construction defines a modular Hamiltonian $K$ from a reduced density matrix $\rho$ via $\rho = e^{-2\pi K}$, and the von Neumann entropy of the reduced density matrix is computed from the $n^{\text{th}}$ R\'{e}nyi entropy $\mathrm{Tr}[e^{-2\pi n K}]$ analytically continued in $n$. The connection to the entanglement entropy literature is commented on at the end. It should be emphasized that the angular quantization constructed here is equivalent to considering the CFT on Minkowski space, which is carefully constructed from the Hilbert space on a very wide strip in the limit that the width of the strip becomes infinite. As such, it can be a useful stepping stone to considering Minkowski space CFTs more generally.

Finally, there is another important application of the angular quantization formalism which is actually our main motivation. In conventional string theory, asymptotic scattering states are described via BRST-invariant worldsheet vertex operators in radial quantization \cite{Polchinski981}. This suffices to describe, for example, the perturbative S-matrix of excitations above the Minkowski vacuum, but is not sufficient to capture all the on-shell states in more interesting scenarios involving black hole backgrounds \cite{Jafferis21}; even string theory in Rindler space cannot be described by vertex operators in radial quantization alone \cite{Witten18}. This can be seen in a lack of unitarity of the standard scattering amplitudes, indicating the existence of additional physical string states. In contrast, quantum field theory in such backgrounds is perfectly unitarity when all modes, on both sides of any horizon, are included. The distinction is because strings are extended objects that can be bisected by horizons. 

Spacetime states in string theory should be defined via a Keldysh field contour for the complexified time boson $X^0$, and the aforementioned cases where radial quantization may be insufficient arise when the Keldysh contour is chosen such that $X^0$ is compact in the Euclidean section, whence on-shell Euclidean time-winding operators may arise. One place such operators are prominent is in the work \cite{Jafferis21}, which presented a Lorentzian description of the FZZ duality between the $\mathrm{SL}(2,\mathds{R})/\mathrm{U}(1)$ cigar black hole \cite{Witten91} and sine-Liouville theory \cite{Kazakov00}, as well as  three-dimensional uplifts of this duality. There, the dual sine-Liouville description of the cigar black hole involves a condensate of Euclidean time-winding operators. 

The need for angular quantization for such worldsheet CFTs with BRST-invariant vertex operators that wind Euclidean time was already described in \cite{Jafferis21}, so here we shall only briefly summarize the logic based on the Witten $i\varepsilon$ prescription. In order for a string theory to make sense, there must be a well-defined prescription for computing observables from correlators, which must involve a treatment of potential divergences in moduli space integrals. In \cite{Witten13}, Witten showed that the way to regulate a region in Euclidean moduli space where one scattering state vertex operator approaches another that results in causal propagation in spacetime is to cut out a disk of radius $\varepsilon$ around the coincident point and glue in a Lorentzian cylinder. With this prescription, the worldsheet time must be analytically continued from Euclidean to Lorentzian signature so that as one vertex operator approaches another, it continues along an infinite Lorentzian tube rather than reaching the singular point.

For ordinary scattering states, Witten's $i\varepsilon$ contour prescription thus dictates using radial quantization, or any continuous deformation thereof, since the codimension-one gluing locus between the Lorentzian cylinder section and the remaining Euclidean worldsheet is topologically a circle, and the gluing slice is the fixed locus of a local $\mathds{Z}_2$ time-reflection symmetry. However, this $i\varepsilon$ prescription fails for time-winding vertex operators because the modified contour with the Lorentzian tube does not avoid the multi-valued cut in the OPE due to the winding of $X^0$. In order for the moduli space integrand to be finite and single-valued, the vicinity of the time-winding vertex operator must be replaced with a Lorentzian Rindler wedge, on which the OPE is indeed single-valued and finite. Therefore, the correct Witten $i\varepsilon$ prescription for time-winding vertex operators dictates using angular quantization, as the gluing locus between the Euclidean worldsheet and the Lorentzian Rindler wedge is topologically a semi-infinite ray extending away from the would-be singularity. We may always locally choose the gauge such that the gluing locus in this Witten contour is at constant angle $\theta$ so that the constant-time slices are indeed the constant-$\theta$ slices of angular quantization. Hence describing string states corresponding to time-winding vertex operators in angular quantization is a formal requirement of the Witten $i\varepsilon$ prescription in order that the string theory make sense in the first place\footnote{One might try to gain intuition on the need for angular quantization near Euclidean time-winding operators using logic from effective string theory by fixing static gauge in order to define the spacetime string states. The gauge choice fixing the radial direction away from such an operator to be $X^0$ is inconsistent with the shift in $X^0$ around a Euclidean time-winding operator. However angular gauge, fixing $X^0$ in terms of the angular coordinate, is consistent and leads to a real condition in the combined Lorentzian continuation. Such static gauges are incompatible with having a flat worldsheet metric unless the equation of motion for $X^0$ is satisfied, so one would either have to work with a dynamical metric mode which complicates the BRST quantization or to confine oneself to classical string theory.}. 

As described in \cite{Jafferis21}, the spacetime string states associated to Euclidean time-winding operators are not scattering states in the usual sense, and the ones appearing in that work (and also in \cite{Maldacena05}) are anchored at infinity but extend to finite locations in the bulk; we shall present a detailed analysis of the Hilbert space of the resulting ``long strings'' in \cite{Agia22}. The present paper can be viewed as a self-contained reference for the CFT techniques and results which we shall subsequently use in the string theory context.

The structure of this paper is as follows. In \hyperref[general construction]{Section \ref{general construction}}, we provide a brief general outline of the construction of angular quantization on the plane/sphere and its main characteristics. In \hyperref[free boson]{Section \ref{free boson}}, we study the simplest example of a noncompact free boson in detail, including different non-conformal boundary conditions that shrink to exponential vertex operators which are either ``Neumann-like'' or ``Dirichlet-like''. In \hyperref[free CFTs]{Section \ref{free CFTs}}, we present the main results for other free CFTs for which the angular quantization admits an explicit Fock space description, in particular for the linear dilaton theory, the compact boson and the $bc$ ghost system. From these results, one may apply the formalism to BRST quantization in limits of NL$\sigma$M worldsheet string theory containing free or linear dilaton target space directions. Lastly, in \hyperref[EE]{Section \ref{EE}} we mention the connection to entanglement entropy as a special case of the formalism; while the work in this paper does not provide any new results in that application, it does allow one to have a more systematic operator method of computing certain entanglement entropies which could potentially be useful even in non-conformal theories. More importantly, the rationale behind obtaining local operators by shrinking boundary conditions allows one to study more carefully the admissibility of boundary conditions necessary to define entanglement entropy in continuum quantum field theory. For the entirety of this paper we work in Euclidean signature. A collection of some operator algebra calculations is contained in two appendices.

\section{General Construction on the Sphere}\label{general construction}

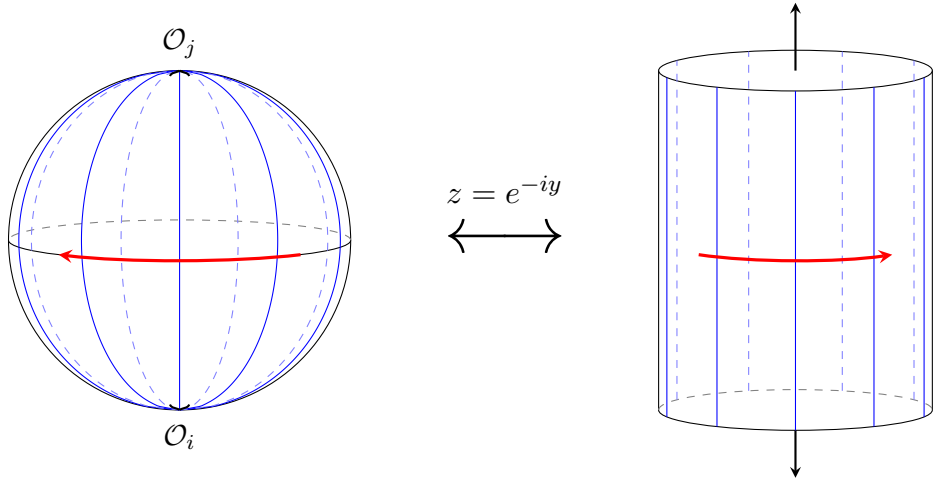
\begin{figure}
\centering
\begin{tikzpicture}
\begin{scope}[scale=0.9]
\begin{scope}[xshift=-0.75cm]
\draw (4,0) circle (2.5);
\draw (6.5,0) arc (0:-180:2.5 and 0.3);
\draw[dashed,black!50] (6.5,0) arc (0:180:2.5 and 0.3);
\foreach \x in {-2,...,2}
{\draw[blue,thin,domain=-2.5:2.5,variable=\t] plot[smooth,samples=100] ({(sqrt((2.5^2)-(\t)^2))*cos(90+35*\x)+4},{\t});}
\foreach \x in {3,...,6}
{\draw[blue!50,thin,dashed,domain=-2.5:2.5,variable=\t] plot[smooth,samples=100] ({(sqrt((2.5^2)-(\t)^2))*cos(90+40*\x)+4},{\t});}
\draw[-stealth,red,very thick] ({4+2.5*cos(45)},{0-0.3*sin(45)}) arc (-45:-135:2.5 and 0.3);
\draw[thick] (4,2.5) to[out=-35,in=110] ({4+0.14},{2.5-0.08});
\draw[thick] (4,2.5) to[out=215,in=70] ({4-0.14},{2.5-0.08});
\draw[thick] (4,-2.5) to[out=35,in=-110] ({4+0.14},{-2.5+0.08});
\draw[thick] (4,-2.5) to[out=-215,in=-70] ({4-0.14},{-2.5+0.08});
\node[scale=1] (ketop) at (4,-2.9) {$\mathcal{O}_i$};
\node[scale=1] (braop) at (4,2.9) {$\mathcal{O}_j$};
\end{scope}
\node[scale=2.25] (arrow) at (8,0) {$\longleftrightarrow$};
\node[scale=1.1] (map) at (8,0.75) {$z = e^{-iy}$};
\begin{scope}[xshift=0.75cm]
\draw (9.5,2.5) arc (-180:180:2 and 0.3);
\draw (9.5,2.5) -- (9.5,-2.5);
\draw (9.5,-2.5) arc (-180:0:2 and 0.3);
\draw (13.5,-2.5) -- (13.5,2.5);
\draw[-stealth, thick] (11.5,2.5) -- (11.5,3.5);
\draw[-stealth, thick] (11.5,-2.8) -- (11.5,-3.5);
\draw[dashed,black!50] (9.5,-2.5) arc (180:0:2 and 0.3);
\foreach \x in {-2,...,2}
{\draw[blue,thin] ({11.5 + 2*sin(35*\x)},{-2.5 - 0.3*cos(35*\x)}) -- ({11.5 + 2*sin(35*\x)},{2.5 - 0.3*cos(35*\x)});}
\foreach \x in {3,...,6}
{\draw[blue!50,thin,dashed] ({11.5 + 2*sin(40*\x)},{-2.5 - 0.3*cos(40*\x)}) -- ({11.5 + 2*sin(40*\x)},{2.5 - 0.3*cos(40*\x)});}
\draw[stealth-,red,very thick] ({11.5+2*cos(45)},{0-0.3*sin(45)}) arc (-45:-135:2 and 0.3);
\end{scope}
\end{scope}
\end{tikzpicture}
\caption{The angular quantization slicing on the sphere and on the infinite cylinder, related by the exponential map. The red arrows denote Euclidean time evolution via the Rindler Hamiltonian in the appropriate conformal frame.}
\label{angular slicing}
\end{figure}

Angular quantization is most naturally discussed on the genus zero Riemann sphere; for correlators at higher genera, one may use a standard pair-of-pants decomposition to express everything in terms of genus zero quantities. So consider a sphere correlator arranged via a global $\mathrm{PSL}(2,\mathds{C})$ transformation to have an operator $\mathcal{O}_i$ inserted at the origin $z=0$ and an operator $\mathcal{O}_j$ inserted at $z = \infty$, the latter of which is understood as being at the origin of the inverted patch with transition function $z' = 1/z$. Either operator could be chosen to be the identity so that the procedure works for all $n$-point correlators including the partition function. We would like to define a Hilbert space $\mathcal{H}_{ij}$ of states defined on the line between the two operators which is time-evolved with a Rindler Hamiltonian $H_{\text{R}}$ in angles around the origin, resulting in all sphere correlators being computed as ``thermal traces'' over this angularly-quantized Hilbert space. We shall often freely pass between the sphere and cylinder conformal frames, related  by $z = e^{-iy}$, where we write the cylinder complex coordinate as $y = y_1 + iy_2$ with $y_1 \sim y_1 + 2\pi$ the circle direction; see Figure \ref{angular slicing}.

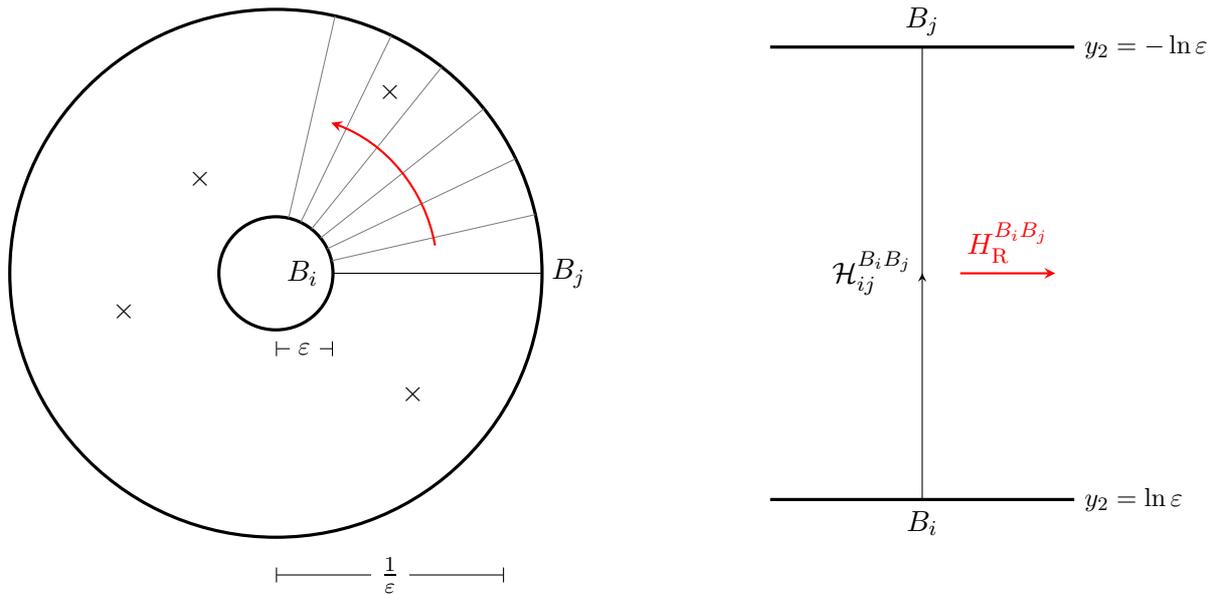
\begin{figure}
\centering
\begin{tikzpicture}
\begin{scope}[xshift=-4cm]
\draw[very thick] (0,0) circle (0.75) node[scale=1,xshift=0.35cm]{$B_i$};
\draw[very thick] (0,0) circle (3.5) node[scale=1,xshift=3.85cm]{$B_j$};
\draw[|-|] (0,-1) -- node[midway,fill=white,scale=0.9]{$\varepsilon$} (0.75,-1);
\draw[|-|] (0,-4) -- node[midway,fill=white]{$\frac{1}{\varepsilon}$} (3,-4);
\node[scale=1] (op1) at (1.5,2.4) {$\times$};
\node[scale=1] (op2) at (-1,1.25) {$\times$};
\node[scale=1] (op3) at (1.8,-1.6) {$\times$};
\node[scale=1] (op4) at (-2,-0.5) {$\times$};
\draw (0.75,0) -- (3.5,0);
\foreach \x in {1,...,6}{
\draw[gray] ({0.75*cos((\x)*360/28)},{0.75*sin((\x)*360/28)}) -- ({3.5*cos((\x)*360/28)},{3.5*sin((\x)*360/28)});
}
\draw[red,thick,-stealth] ({2.125*cos(10)},{2.125*sin(10)}) arc (10:70:2.125);
\end{scope}
\begin{scope}[xshift=4.5cm]
\draw[very thick] (-2,-3) -- node[midway,below,scale=1]{$B_i$} (2,-3) node[right,scale=0.9,xshift=0cm]{$y_2 = \ln\varepsilon$};
\draw[very thick] (-2,3) -- node[midway,above,scale=1]{$B_j$} (2,3) node[right,scale=0.9,xshift=0cm]{$y_2 = -\ln\varepsilon$};
\draw[postaction={decorate,decoration={markings,mark=at position 0.5 with {\arrow{stealth}}}}] (0,-3) -- node[midway,left,scale=1]{$\mathcal{H}_{ij}^{B_i B_j}$} (0,3);
\draw[red,thick,-stealth] (0.5,0) -- node[midway,above,scale=1]{$H_{\text{R}}^{B_i B_j}$} (1.75,0);
\end{scope}
\end{tikzpicture}
\caption{Left: The finite-regulator set-up for sphere correlators obtained by excising holes of radius $\varepsilon$ around the endpoint operators and imposing suitable boundary conditions $B$ such that the operators are reproduced in the shrinking limit $\varepsilon\rightarrow 0$; at finite regulator it is manifest that correlators are computed as traces over some Hilbert space after evolving $2\pi$ in angular time. Right: The appropriate Hilbert space $\mathcal{H}^{B_i B_j}_{ij}$ and its Hamiltonian $H_{\text{R}}^{B_i B_j}$ as defined on the very wide strip with the local boundary condition $B_i$ on the bottom and the local boundary condition $B_j$ on the top. The Hilbert space $\mathcal{H}_{ij}$ of angular quantization is obtained in the $\varepsilon\rightarrow 0$ limit of $\mathcal{H}^{B_i B_j}_{ij}$.}
\label{finite regulator}
\end{figure}

The immediate problem is that the picture in Figure \ref{angular slicing} does not  define a trace over a Hilbert space because of the fixed points of the angular evolution at $z=0$ and $z = \infty$, and so a regularization prescription must be given. A natural regulator is provided by excising a hole of some radius $\varepsilon$ around an endpoint operator $\mathcal{O}$ and imposing a boundary condition $B$ such that the boundary ``shrinks'' back to the operator $\mathcal{O}$ in the limit $\varepsilon\rightarrow 0$; we are of course free to consider independently-sized holes cut around the operator at the origin and the one at infinity, but it is easiest to take the symmetric regulator where the excised hole around infinity has the same radius $\varepsilon$ in the $z' = 1/z$ patch. That is, in lieu of the operators $\mathcal{O}_i(0)$ and $\mathcal{O}_j(\infty)$, we shall place a boundary condition $B_i$ at $|z| = \varepsilon$ and a boundary condition $B_j$ at $|z| = \frac{1}{\varepsilon}$, as shown on the left of Figure \ref{finite regulator}, performing all calculations at finite $\varepsilon$ and then taking the ``shrinking limit'' $\varepsilon\rightarrow 0$ at the end. Then, if we choose $B_i$ and $B_j$ to be \emph{local} boundary conditions, they allow us to define the regulated Hilbert space $\mathcal{H}^{B_i B_j}_{ij}$ of a very wide strip between the two boundary conditions, and the Rindler Hamiltonian $H^{B_i B_j}_{\text{R}}$ is just the generator of time translations on this strip, as shown on the right of Figure \ref{finite regulator}. The angular quantization Hilbert space $\mathcal{H}_{ij}$ with associated Rindler Hamiltonian $H_{\text{R}}$ is then extracted from the limit $\varepsilon\rightarrow 0$ by constructing the equivalence classes of normalizable, finite-energy states\footnote{The Minkowski space Hamiltonian will generally differ from $\lim_{\varepsilon\rightarrow 0}H^{B_i B_j}_{\text{R}}$ by an infinite constant, so one really means that the states have finite energy above some vacuum. However, the identification of the physical states in the shrinking limit is still subtle due to their gapless nature; we shall be more explicit about this limit in \cite{Agia22}.}. To avoid some notational clutter, we shall usually leave the dependence on $\varepsilon$ implicit, as everything in this paper is at strictly finite regulator.
Now, the procedure to compute correlators is clear from Figure \ref{finite regulator}; one computes the trace over the evolution operator $e^{-2\pi H_{\text{R}}^{B_i B_j}}$ on the two-holed sphere in the Hilbert space $\mathcal{H}_{ij}^{B_i B_j}$ and takes the limit $\varepsilon\rightarrow 0$ at the end. That is,
\begin{equation}
\lim_{\varepsilon\rightarrow 0}\text{Tr}_{\mathcal{H}^{B_i B_j}_{ij}}\hspace{-3pt}\left[\mathcal{O}\cdots\mathcal{O}e^{-2\pi H_{\text{R}}^{B_i B_j}}\right]_{S^2} \propto \left\langle\mathcal{O}_i(0)\mathcal{O}\cdots\mathcal{O}\mathcal{O}_j(\infty)\right\rangle_{S^2},
\end{equation}
where $\mathcal{O}\cdots\mathcal{O}$ represents any number of bulk operator insertions, and we explicitly use an $S^2$ subscript to remind us of the conformal frame. Of course, the correlator on the right side of this equation vanishes simply because $\mathcal{O}_i$ and $\mathcal{O}_j$ are infinitely far separated in this conformal frame. Once we write more careful expressions below to extract the finite quantities, it will be important to determine the (finite) normalization, which depends only on $B_i$ and $B_j$.

\subsection{Shrinking of Boundary Conditions}

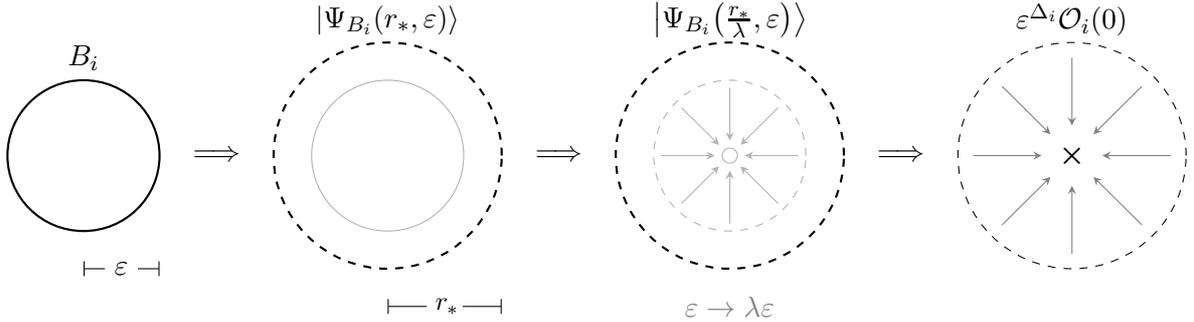
\begin{figure}
\centering
\begin{tikzpicture}
\draw[thick] (0,0) circle (1) node[scale=1,yshift=1.3cm]{$B_i$};
\draw[|-|] (0,-1.5) -- node[midway,fill=white,scale=1]{$\varepsilon$} (1,-1.5);
\node[scale=1] (arrow1) at (1.75,0) {$\Longrightarrow$};
\draw[black!30] (4,0) circle (1);
\draw[dashed,thick] (4,0) circle (1.5) node[scale=1,yshift=1.8cm]{$|\Psi_{B_i}(r_*,\varepsilon)\rangle$};
\draw[|-|] (4,-2) -- node[midway,fill=white,scale=0.9]{$r_*\hspace{-2pt}$} (5.5,-2);
\node[scale=1] (arrow2) at (6.25,0) {$\Longrightarrow$};
\draw[dashed,thick] (8.5,0) circle (1.5) node[scale=1,yshift=1.8cm]{$\big|\Psi_{B_i}\big(\tfrac{r_*}{\lambda},\varepsilon\big)\big\rangle$};
\draw[dashed,black!30] (8.5,0) circle (1);
\foreach \x in {0,...,7}{
\draw[black!30,-stealth] ({8.5+0.9*cos((\x)*45)},{0.9*sin((\x)*45)}) -- ({8.5+0.2*cos((\x)*45)},{0.2*sin((\x)*45)});
}
\draw[black!30] (8.5,0) circle (0.1);
\node[scale=1,gray] (shrink) at (8.5,-2) {$\varepsilon\rightarrow\lambda\varepsilon$};
\node[scale=1] (arrow3) at (10.75,0) {$\Longrightarrow$};
\draw[dashed,thin] (13,0) circle (1.5) node[scale=1,yshift=1.8cm]{$\varepsilon^{\Delta_i}\mathcal{O}_i(0)$};
\foreach \x in {0,...,7}{
\draw[gray,-stealth] ({13+1.3*cos((\x)*45)},{1.3*sin((\x)*45)}) -- ({13+0.4*cos((\x)*45)},{0.4*sin((\x)*45)});
}
\node[scale=1] (operator) at (13,0) {$\bm{\times}$};
\end{tikzpicture}
\caption{The procedure of determining which operator is obtained upon shrinking a given boundary condition. One path integrates from $|z| = \varepsilon$ to some fixed larger radius $r_*$, takes the limit $\varepsilon\rightarrow 0$ in the resulting state (for example by taking $\varepsilon\rightarrow\lambda\varepsilon$ followed by $\lambda\rightarrow 0$) and then back-evolves this state to the origin using the dilatation generator.}
\label{boundary condition shrinking}
\end{figure}

The first step in this construction is to determine the set of admissible boundary conditions $B_i$ that will shrink to a given operator $\mathcal{O}_i$. The shrinking condition may be characterized as follows --- the boundary condition $B_i$ placed at $|z|=\varepsilon$ shrinks to the operator $\mathcal{O}_i$ if and only if
\begin{equation}\label{shrinking condition}
\lim_{\varepsilon\rightarrow 0}\left\langle\mathcal{O}\cdots\mathcal{O}\right\rangle_{B_i(\epsilon)} = \mathcal{N}_{B_i}\left\langle\mathcal{O}_i(0)\mathcal{O}\cdots\mathcal{O}\right\rangle
\end{equation}
for all bulk operator insertions $\mathcal{O}\cdots\mathcal{O}$ outside the hole at all steps, where the correlator on the left is that in the presence of the hole of radius $\varepsilon$ with boundary condition $B_i$ and that on the right is in the original theory; $\mathcal{N}_{B_i}$ is a normalization constant which depends on the boundary condition $B_i$ chosen. Abstractly, the necessary and sufficient condition which $B_i$ must satisfy is depicted in Figure \ref{boundary condition shrinking}, which follows a similar logic as for the state/operator correspondence in radial quantization.
For simplicity, suppose one constructs an abstract path integral description of the CFT. Inside any correlator with the boundary condition $B_i$ at radius $\varepsilon$, one may perform the partial path integral over the annular region $\varepsilon \leqslant |z| \leqslant r_*$ for any $r_*$ smaller than the radius of the nearest bulk operator insertion. The result of this path integral produces a state $|\Psi_{B_i}(r_*,\varepsilon)\rangle$ in the wavefunctional basis (in radial quantization), in the sense that it is a linear functional which maps the rest of the configuration to the number obtained by the path integral over all $|z| \geqslant r_*$ with the initial condition set by $\Psi_{B_i}$. Let $|\Psi_{\mathcal{O}}(r_*)\rangle$ likewise denote the state in the wavefunctional basis produced by the path integral in the original theory over the disk $|z| < r_*$ with the operator $\mathcal{O}(0)$ inserted at the origin. We may always expand the state $|\Psi_{B_i}\rangle$ in the wavefunctional basis $\{|\Psi_{\mathcal{O}}\rangle\}$, i.e.~
\begin{equation}
|\Psi_{B_i}(r_*,\varepsilon)\rangle = \sum_{\mathcal{O}} c^i_{\mathcal{O}}(r_*,\varepsilon)|\Psi_{\mathcal{O}}(r_*)\rangle,
\end{equation}
where the sum is over all independent local operators in the theory, which we may take to have definite scaling dimensions (both primaries and descendants being included explicitly). We would now like to find the state $|\Psi_{B_i}(r_*)\rangle$ obtained when shrinking the excised hole away, i.e.~when taking $\varepsilon\rightarrow 0$, and so we need to know how the coefficients $c^i_{\mathcal{O}}(r_*,\varepsilon)$ depend on $\varepsilon$. To do so, consider shrinking $\varepsilon\rightarrow\lambda\varepsilon$ for a finite constant $0 < \lambda < 1$. Since the state obtained at $|z| = r_*$ with the transformed boundary condition at $|z| = \lambda\varepsilon$ is equivalent to the state obtained at $|z| = \frac{r_*}{\lambda}$ with the boundary condition at $|z| = \varepsilon$ unchanged\footnote{\label{footnote conformal transformation} It is implicitly understood that the conformal transformation of the boundary state also involves transforming the boundary condition itself so that the resulting overlaps with the bras that represent the rest of the correlator are identical before and after the transformation. If this transformation of the boundary condition is nontrivial, one says that it breaks conformal symmetry because the set of conformal transformations preserving the boundary condition does not leave the correlators invariant. Note also that we slightly abuse notation by directly comparing boundary states in Hilbert spaces $\mathcal{H}_{S^1}$ of different radii because we have an isometric isomorphism between them provided by the dilatation generator.}, the fact that $|\Psi_{B_i}\rangle$ is a state in the original CFT and so evolves with the plane dilatation generator $r^{-D}$ diagonalized by $\{|\Psi_{\mathcal{O}}\rangle\}$ tells us that
\begin{equation}
|\Psi_{B_i}(r_*,\lambda \varepsilon)\rangle = \big|\Psi_{B_i}\big(\tfrac{r_*}{\lambda},\varepsilon\big)\big\rangle = \left(\frac{r_*/\lambda}{r_*}\right)^{-D}|\Psi_{B_i}(r_*,\varepsilon)\rangle = \sum_{\mathcal{O}}c^i_{\mathcal{O}}(r_*,\varepsilon)\lambda^{\Delta_{\mathcal{O}}}|\Psi_{\mathcal{O}}(r_*)\rangle,
\end{equation}
where $\Delta_{\mathcal{O}} = h_{\mathcal{O}} + \widetilde{h}_{\mathcal{O}}$ is the conformal dimension of $\mathcal{O}$, with $h$ and $\widetilde{h}$ the conformal weights. Hence we see that $c^i_{\mathcal{O}}(r_*,\varepsilon) = b^i_{\mathcal{O}}(r_*)\varepsilon^{\Delta_{\mathcal{O}}}$, where the coefficients $b^i_{\mathcal{O}}(r_*)$ are independent of $\varepsilon$. Since $\lambda^{\Delta_{\mathcal{O}}}|\Psi_{\mathcal{O}}(r_*)\rangle = |\Psi_{\mathcal{O}}(\frac{r_*}{\lambda})\rangle$, the above equation also says that $b^i_{\mathcal{O}}(r_*)$ is in fact independent of $r_*$, a totally obvious fact from quantum mechanics that time-evolving a Schr\"{o}dinger-picture state amounts to time-evolving each basis state in an expansion starting with fixed coefficients. Thus, the wavefunctional state defined by Figure \ref{boundary condition shrinking} is
\begin{equation}\label{wavefunctional boundary state}
|\Psi_{B_i}(r_*,\varepsilon)\rangle = \sum_{\mathcal{O}}b^i_{\mathcal{O}}\varepsilon^{\Delta_{\mathcal{O}}}|\Psi_{\mathcal{O}}(r_*)\rangle,
\end{equation}
where $b^i_{\mathcal{O}}$ are pure constants. Then, in the limit $\varepsilon\rightarrow 0$, the leading contribution in $|\Psi_{B_i}(r_*,\varepsilon)\rangle$ is just from the state with the lowest scaling dimension, and therefore, up to normalization, the boundary condition $B_i$ shrinks to the lowest-dimension operator $\mathcal{O}_{\text{min}}$ that appears in \eqref{wavefunctional boundary state}; this last step is identical to the state/operator correspondence in radial quantization, wherein one simply back-evolves $|\Psi_{\mathcal{O}_{\text{min}}}(r_*)\rangle$ to the origin via the dilatation generator; since $|\Psi_{\mathcal{O}_{\text{min}}}(r_*)\rangle$ was defined as the state obtained by path integrating over $\mathcal{O}_{\text{min}}(0)$, note that \eqref{wavefunctional boundary state} says that the coefficient when shrinking the boundary condition to the origin of the plane decays as $\varepsilon^{\Delta_{\mathcal{O}_{\text{min}}}}$. Of course, if there are multiple independent states of lowest scaling dimension in the above expansion, then the boundary condition would shrink to the linear combination of those operators that appears in the expansion. It should be noted that, in the order of limits described above, it is important that $\varepsilon$ be sent to zero at fixed and finite $r_*$ first.

Generically, boundary conditions are specified by the limiting behavior of bulk correlators as an operator approaches the boundary. Since the state $|\Psi_{B_i}(r_*,\varepsilon)\rangle$ above is the linear functional on configurations of bulk operators which by definition gives the respective correlator with the boundary condition $B_i$ imposed at $|z| = \varepsilon$, the abstract boundary condition $B_i$ is uniquely encoded in the set of coefficients $b^i_{\mathcal{O}}$ appearing in \eqref{wavefunctional boundary state}. Thus, the abstract procedure for enumerating all boundary conditions which shrink to a multiple of an operator $\mathcal{O}_i$, which we assume to be the only operator of dimension $\Delta_i$ for simplicity, is very simple --- set $\mathcal{O}_{\text{min}} = \mathcal{O}_i$ (so that $b^i_{\mathcal{O}} = 0$ for all $\Delta_{\mathcal{O}} < \Delta_{i}$ and $b^i_{\mathcal{O}_i}$ is some finite number) and set $b^i_{\mathcal{O}}$ for all $\Delta_{\mathcal{O}} > \Delta_i$ to be anything one wants; by the above discussion, the boundary condition so constructed shrinks to the operator $b^i_{\mathcal{O}_i}\varepsilon^{\Delta_i}\mathcal{O}_i$. Thus there are always infinitely many boundary conditions which shrink to a given operator, since the shrinking procedure kills all contributions except that from $b^i_{\mathcal{O}_i}$ itself. ``Most'' boundary conditions in a unitary CFT will simply shrink to the identity operator, as $b_{\mathds{1}}$ would have to be tuned to zero for this not to occur. In order to have an angular quantization Hilbert space as defined in Figure \ref{finite regulator}, we should also demand an appropriate notion of locality; we shall not attempt to provide a systematic discussion of the condition of locality here, as the explicit boundary conditions described in the following sections are manifestly local. It should also be emphasized that these finite-regulator boundary conditions are generically non-conformal. Indeed, a condition that a 2d conformally-invariant boundary condition must satisfy is that the integral of $v^{\alpha}T_{\alpha\beta}n^{\beta}$ around the boundary vanish (when inserted into correlators), where $T_{\alpha\beta}$ is the stress-energy tensor and $v^{\alpha}$ is any vector field whose diffeomorphism flow preserves the boundary with outward normal vector $n^{\beta}$. Acting with this integral on \eqref{wavefunctional boundary state} will not generically annihilate it because there is no relation between the coefficients $b^i_{\mathcal{O}}$ for a primary and its descendants. This shrinking procedure may therefore be regarded as a generalization of Cardy's construction of conformally-invariant boundary states out of Ishibashi states \cite{Cardy89, Ishibashi88}. In particular, the Ishibashi states are the very special states of the form \eqref{wavefunctional boundary state} where the $b^i_{\mathcal{O}}$ are only nonvanishing for the conformal family of a single scalar primary with the descendant coefficients determined by the primary coefficient so that the boundary integral of $v^{\alpha}T_{\alpha\beta}n^{\beta}$ does annihilate it, and the conformally-invariant boundary states are the linear combinations of Ishibashi states which further satisfy the Cardy condition of ``modular invariance'' between the open and closed string channels. For the free CFT examples discussed below, we shall not need to make contact with this abstract prescription since boundary conditions can be defined on the fields themselves; the connection to and generalization of Cardy states could be useful for performing angular quantization calculations in other CFTs, for example in Liouville CFT, by setting up and solving bootstrap-like equations in the shrinking limit.

Since there are infinitely many ways to choose the coefficients $b^i_{\mathcal{O}}$ so that the boundary condition shrinks to the desired operator $\mathcal{O}_i$, \eqref{wavefunctional boundary state} is not necessarily the most useful characterization of the shrinking condition. Consider the case of a primary operator $\mathcal{O}_i$. Taking an orthonormal basis of two-point functions, it suffices to demonstrate the original shrinking condition \eqref{shrinking condition} for just a single bulk primary operator insertion; specifically, one need only show that the limit of $\langle\mathcal{O}(z,\barred{z})\rangle_{B_i(\epsilon)}$ is proportional to the nonvanishing two-point function when $\mathcal{O} = \barred{\mathcal{O}}^i$ is the conjugate primary of $\mathcal{O}_i$ and is zero when $\mathcal{O}$ is any other primary. Then \eqref{shrinking condition} will be satisfied for all insertions of two bulk operators by an application of the OPE, and so on. In the free theory examples considered in the following sections, checking these one-point functions is a simple way to prove the boundary conditions there indeed shrink to the primaries claimed.

\subsection{Matching Conditions and Asymptotics/Operator Correspondence}

Having discussed the shrinking of a single boundary condition to a local operator, the set-up of angular quantization as depicted in Figure \ref{finite regulator} follows from imposing boundary condition $B_i$ at $|z| = \varepsilon$ which shrinks to $\mathcal{O}_i(0)$ and simultaneously imposing boundary condition $B_j$ at $|z| = \frac{1}{\varepsilon}$ which shrinks to $\mathcal{O}_j(\infty)$. The usual two- and three-point functions of conformal primaries on the sphere are
\begin{align}
\left\langle\mathcal{O}_i(z_i,\barred{z}_i)\mathcal{O}_j(z_j,\barred{z}_j)\right\rangle_{S^2} & = \frac{\delta_{ij}}{z_{ij}^{2h}\barred{z}_{ij}^{2\widetilde{h}}}
\\ \left\langle\mathcal{O}_i(z_i,\barred{z}_i)\mathcal{O}_k(z_k,\barred{z}_k)\mathcal{O}_j(z_j,\barred{z}_j)\right\rangle_{S^2} & = \frac{C_{ikj}}{z_{ik}^{h_i+h_k-h_j}z_{kj}^{h_k+h_j-h_i}z_{ji}^{h_j + h_i - h_k}\barred{z}_{ik}^{\widetilde{h}_i+\widetilde{h}_k-\widetilde{h}_j}\barred{z}_{kj}^{\widetilde{h}_k+\widetilde{h}_j-\widetilde{h}_i}\barred{z}_{ji}^{\widetilde{h}_j+\widetilde{h}_i-\widetilde{h}_k}},
\end{align}
where $C_{ijk}$ are the three-point coefficients. 

The thermal trace on the two-holed sphere over $\mathcal{H}^{B_i B_j}_{ij}$ as depicted on the left of Figure \ref{finite regulator} (with no bulk insertions) should reproduce in the shrinking limit the two-point function above with $z_i = 0$ and $z_j = \infty$; the thermal trace with $\mathcal{O}_k(z_k,\barred{z}_k)$ inserted is likewise supposed to reproduce in the shrinking limit the three-point function above with $z_i = 0$ and $z_j = \infty$. However, we must be careful to keep track of the various powers of $\varepsilon$ which cause various quantities to decay or diverge. For example, both correlators vanish since $z_{ji} = \infty$, which is purely an artifact of working in a conformal frame with two operators at infinite coordinate separation; finite quantities are obtained by simply stripping off this zero. 

Furthermore, as elucidated above, there are powers of $\varepsilon$ incurred by shrinking the boundary conditions themselves. In particular,  $B_i$ shrinks to (a finite constant) times $\varepsilon^{\Delta_i}\mathcal{O}_i(0)$; for aesthetic reasons explained below, we shall separate out a spin factor of $i^{-s_i}$ to write the shrinking limit as $b_{\mathcal{O}_i}i^{-s_i}\varepsilon^{\Delta_i}\mathcal{O}_i(0)$, where $b_{\mathcal{O}_i}$ is a finite constant. For the other endpoint, consider placing the operator $\mathcal{O}_j$ at large but finite position $z_j = \frac{1}{\varepsilon}$ and transition to the inverted patch $z_j' = 1/z_j$ in which $\mathcal{O}_j'(\varepsilon) = (-1)^{s_j}\varepsilon^{-2\Delta_j}\mathcal{O}_j(\frac{1}{\varepsilon})$, where $s_j = h_j - \widetilde{h}_j$ is the operator's spin. Then, shrinking $B_j'$ at $|z'| = \varepsilon$ is identical to shrinking $B_j$ at $|z| = \varepsilon$, and hence the original $B_j$ at $|z| = \frac{1}{\varepsilon}$ shrinks to $b_{\mathcal{O}_j}i^{s_j}\varepsilon^{-\Delta_j}\mathcal{O}_j(\infty)$. That is, angular quantization traces on the sphere shrink according to
\begin{multline}
\mathrm{Tr}_{\mathcal{H}_{ij}^{B_i B_j}}\left[\mathcal{O}_1(z_1,\barred{z}_1)\cdots\mathcal{O}_n(z_n,\barred{z}_n)e^{-2\pi H_{\text{R}}^{B_i B_j}}\right]_{S^2} 
\\ \stackrel{\varepsilon\rightarrow 0}{\longrightarrow} b_{\mathcal{O}_i}b_{\mathcal{O}_j}i^{s_j - s_i}\varepsilon^{\Delta_i - \Delta_j}\big\langle \mathcal{O}_i(0)\mathcal{O}_1(z_1,\barred{z}_1)\cdots\mathcal{O}_n(z_n,\barred{z}_n)\mathcal{O}_j(\infty) \big\rangle_{S^2}.
\end{multline}

Finally, it is $\mathcal{O}_j'(0)$ which has a finite two-point function with $\mathcal{O}_i(0)$ and a finite three-point function with $\mathcal{O}_i(0)\mathcal{O}_k(z,\barred{z})$, which are extracted via further multiplication by $\varepsilon^{-2\Delta_j}$ to strip off the infinite-separation zero. Altogether, therefore, the matching conditions which must be satisfied in angular quantization are
\begin{align}
\label{2-point matching} \lim_{\varepsilon\rightarrow 0}\left(i^{s_i - s_j}\varepsilon^{-\Delta_i-\Delta_j}\mathrm{Tr}_{\mathcal{H}_{ij}^{B_i B_j}}\left[e^{-2\pi H_{\text{R}}^{B_i B_j}}\right]_{S^2}\right) & = \mathcal{N}_{ij}^{B_i B_j}\delta_{ij}
\\ \label{3-point matching} \lim_{\varepsilon\rightarrow 0}\left(i^{s_i - s_j}\varepsilon^{-\Delta_i-\Delta_j}\mathrm{Tr}_{\mathcal{H}_{ij}^{B_i B_j}}\left[\mathcal{O}_k(z,\barred{z})e^{-2\pi H_{\text{R}}^{B_i B_j}}\right]_{S^2}\right) & = \frac{\mathcal{N}_{ij}^{B_i B_j}C_{ikj}}{z^{h_i + h_k - h_j}\barred{z}^{\widetilde{h}_i+\widetilde{h}_k - \widetilde{h}_j}},
\end{align}
where $\mathcal{N}_{ij}^{B_i B_j}$ is a \emph{finite} normalization constant that depends on which boundary conditions we use to shrink to the two operators (but not on $\varepsilon$) which factorizes between $B_i$ and $B_j$ because the excision regulators are local and independent.

It may seem peculiar that we are defining these matching conditions from the limit of sphere correlators involving $\mathcal{O}_j(z_j,\barred{z}_j)$ evaluated at $z_j = \frac{1}{\varepsilon}$ even though the sphere correlators have no relation to the artificial excision regulator $\varepsilon$. Doing so simply defines the normalization and phase convention of the operators in angular quantization. The statement \eqref{2-point matching} taken by itself is very weak because it cannot disentangle the Hamiltonian $H_{\text{R}}^{B_i B_j}$ from the Hilbert space $\mathcal{H}_{ij}^{B_i B_j}$. That is, the ``matching condition'' \eqref{2-point matching} is really just a partial definition of the numbers $\mathcal{N}_{ij}^{B_i B_j}$, which we demand to be finite without loss of generality. With this understanding, the \emph{bona fide} matching condition \eqref{3-point matching} becomes a highly nontrivial constraint on the asymptotic spectrum of the Rindler Hamiltonian and the matrix elements of all primaries in the states of angular quantization. Note also that \eqref{2-point matching} says that the limit of the thermal trace over $\mathcal{H}_{ij}^{B_i B_j}$ must vanish if the endpoint operators do not lie in conjugate Verma modules, and this vanishing cannot come from an overall positive power of $\varepsilon$ without contradicting \eqref{3-point matching}, meaning that the Rindler Hamiltonian $H_{\text{R}}^{B_i B_j}$ is not generally Hermitian. It may seem odd that a non-Hermitian Hamiltonian describes a unitary theory, but we shall see in our examples below that its non-Hermiticity is of a rather mild form, essentially that the only non-Hermitian part\footnote{For quantum field theory in asymptotically-flat or -AdS space, specifications of the asymptotic behavior of correlators is part of the definition of the theory, as it cannot be changed by any finite-energy excitations in the bulk. The Rindler Hamiltonian here is only non-Hermitian under a na\"{i}ve application of Hermitian conjugation descended from radial quantization, which can change the asymptotic conditions. One can modify the definition of Hermitian conjugation so that it does not mix asymptotic conditions, under which the Rindler Hamiltonian is perfectly Hermitian.} of $H_{\text{R}}^{B_i B_j}$ is what provides the orthogonality $\delta_{ij}$ needed in \eqref{2-point matching}. Despite our notation and language, it is an assumption for the moment that there exist choices of boundary conditions $B_i$ such that $\mathcal{H}_{ij}^{B_i B_j}$ is an honest Hilbert space; one is always free to define a vector space $\mathcal{H}_{ij}^{B_i B_j}$ and scalar operator $H_{\text{R}}^{B_i B_j}$ based on reproducing the finite-regulator thermal partition function and one-point functions, but there is no guarantee \emph{a priori} that there exists a Lorentzian continuation thereof such that $\mathcal{H}_{ij}^{B_i B_j}$ is a Hilbert space of normalizable states with a local operator algebra obeying microcausality.

It suffices to consider only primary endpoint operators. That is, the map $\mathcal{O}_i\times\mathcal{O}_j\rightarrow\mathcal{H}_{ij}$ is actually well-defined on the full Verma module product. Indeed, if one of the endpoint operators is a descendant, we rewrite it as the appropriate contour integrals surrounding a primary and then excise the primary with a boundary condition contained inside the stress-energy contour integrals. Thus, the angular quantization Hilbert space for endpoint descendants is the same as the Hilbert space for their primaries, and the entire effect of the descendants is described by stress-energy insertions in the thermal traces which may be reabsorbed into a new Rindler evolution operator. 

Finally, we emphasize that there is no state/operator correspondence in angular quantization. In radial quantization, the state/operator correspondence provides an isomorphism between the Hilbert space of states on $S^1$ and local operators. The construction of angular quantization that we have described in this section gives an entirely different map from all pairs of local operators to all angularly-quantized Hilbert spaces, $\mathcal{O}_i\times\mathcal{O}_j\rightarrow\mathcal{H}_{ij}$. Notably, this correspondence requires two operators instead of one, and they are associated with an entire Hilbert space and not just a single state within a Hilbert space. We shall call the map $\mathcal{O}_i\times\mathcal{O}_j\rightarrow\mathcal{H}_{ij}$ the ``asymptotics/operator correspondence'' instead. From the abstract definition provided in this section, the ``asymptotics'' part of this correspondence is given by the Hilbert space with suitable boundary conditions on a very wide strip in the limit that the width of the strip tends to infinity. After the shrinking limit, the states in the angularly-quantized Hilbert space $\mathcal{H}_{ij}$ live on an infinite line and obey an asymptotic fall-off condition; the latter is most easily seen in the wavefunctional basis, where ``shrinking'' the finite-regulator boundary conditions has the effect of determining the asymptotic form of the wavefunctionals $\Psi[\mathcal{O}(y_1,y_2)]$, where $\mathcal{O}$ stands for all local operators of the theory, as $y_2 \rightarrow \pm \infty$. While the map $\mathcal{O}_i\times\mathcal{O}_j\rightarrow\mathcal{H}_{ij}$ described above requires a pair of endpoint operators, the asymptotic fall-off condition in the wavefunctional basis holds for the endpoint operators individually. In the free CFT examples described in detail below, the asymptotic fall-off conditions associated to given endpoint operators will be even more explicitly written in terms of the free field description. This asymptotics/operator correspondence is central to angular quantization applications to the $\mathrm{NL}\sigma\mathrm{M}$ limit of string theory, as we shall describe in \cite{Agia22}.

\subsection{The Weyl Anomaly}

It is important to note that the thermal trace depicted on the left of Figure \ref{finite regulator} computes correlators in the sphere conformal frame after shrinking $\varepsilon\rightarrow 0$ whereas the actual definition of the angular quantization Hilbert space on the right of Figure \ref{finite regulator} is on the very wide strip which becomes the cylinder conformal frame after identifying $y_1 \sim y_1 + 2\pi$. As such, we must be careful of the frame dependence of the Hamiltonian. Despite the fact that the boundary conditions involved are non-conformal, the only contribution to the Weyl transformation of the angular quantization traces here is from the usual bulk Weyl anomaly. 

First recall that, if a 2d CFT on a Riemann surface $\Sigma$ with metric $g_{\alpha\beta}$ and boundary embedding $f:\partial\Sigma\hookrightarrow\Sigma$ has partition function $Z[g]$, then the partition function under the local Weyl transformation $g_{\alpha\beta} \mapsto g'_{\alpha\beta} = e^{-2\omega}g_{\alpha\beta}$ is given by\footnote{Recall that this result comes from insertions of the stress-energy tensor in correlators being written as $T_{\alpha\beta} = -\frac{4\pi}{\sqrt{g}}\frac{\delta W}{\delta g^{\alpha\beta}}$, where $Z[g] = e^{-W[g]}$, whose trace is $T^{\alpha}_{\phantom{\alpha}\alpha} = -\frac{2\pi}{\sqrt{g}}\frac{\delta W}{\delta\omega}$. The usual Weyl anomaly action for a 2d CFT then follows from the curved-space one-point function $\langle T^{\alpha}_{\phantom{\alpha}\alpha}\rangle_g = -\frac{c}{12}R - \frac{c}{6}\delta(\Sigma-\partial\Sigma)K$, where $\delta(\Sigma-\partial\Sigma)$ is the covariant delta function localizing to the boundary, together with the Weyl transformation of the scalar curvature being $R' = e^{2\omega}(R + 2\nabla^2\omega)$; i.e.~$S_{\text{W}}$ is the unique diffeomorphism-invariant action whose Weyl variation is exactly that dictated by the trace of the stress-energy tensor.}
\begin{align}
Z[g'] & = e^{-S_{\text{W}}}Z[g]
\\ \label{Weyl action} S_{\text{W}} & = \frac{c}{24\pi}\int_{\Sigma} d^2\sigma\sqrt{g}\left(g^{\alpha\beta}\partial_{\alpha}\omega\partial_{\beta}\omega + \omega R\right) + \frac{c}{12\pi}\int_{\partial\Sigma}d\ell\sqrt{f^* g}\s\s\omega K,
\end{align}
where $R$ is the scalar curvature associated to $g_{\alpha\beta}$ and $K$ the (trace of the) extrinsic curvature of the boundary whose invariant line element is $d\ell\sqrt{f^* g}$; the Euler characteristic of $\Sigma$ of genus $g$ with $b$ boundaries is
\begin{equation}
\chi = \frac{1}{4\pi}\int_{\Sigma}d^2\sigma\sqrt{g}\s\s R + \frac{1}{2\pi}\int_{\partial\Sigma}d\ell\sqrt{f^* g}\s\s K = 2 - 2g - b.
\end{equation}
We shall always discuss CFTs for which the holomorphic and antiholomorphic central charges agree, $c = \widetilde{c}$, so that there is no gravitational anomaly. On the very wide cylinder with complex coordinate $y = y_1 + iy_2$, where $\ln\varepsilon \leqslant y_2 \leqslant -\ln\varepsilon$ and $y_1 \sim y_1 + 2\pi$, we simply take the flat metric $g_{\alpha\beta} = \delta_{\alpha\beta}$ with scalar curvature $R = 0$ everywhere and boundary extrinsic curvature $K = 0$ at $y_2 = \mp \ln\varepsilon$ (where $d^2\sigma\sqrt{g} = dy_1 dy_2$ and $d\ell\sqrt{f^* g} = dy_1$). The two-holed sphere at finite regulator with complex coordinates $(z,\barred{z})$, where $\varepsilon \leqslant |z| \leqslant \frac{1}{\varepsilon}$, is related by the exponential map $z = e^{-iy}$. For the latter we wish to take the flat plane metric $ds^2 = dz \otimes_{\text{s}}d\barred{z} = e^{2y_2}dy\otimes_{\text{s}}d\barred{y}$ which has scalar curvature $R=0$ and boundary extrinsic curvature $K = \pm \frac{1}{|z|}$ at $|z| = \varepsilon^{\mp 1}$ (where now $\chi=0$ is due to this opposite orientation of the two boundaries). Hence the flat plane metric $g'_{\alpha\beta} = \delta_{\alpha\beta}$ is obtained from the flat cylinder metric via the Weyl transformation $\omega = -y_2$, for which the Weyl anomaly action above is
\begin{align}
S_{\text{W}} = \frac{c}{24\pi}\int_{\ln\varepsilon}^{-\ln\varepsilon}dy_2\int_0^{2\pi}dy_1 = -\frac{c}{6}\ln\varepsilon.
\end{align}
So the Weyl anomaly says that the finite-regulator partition functions in the two conformal frames are related by
\begin{equation}\label{pure Weyl anomaly}
Z_{\text{flat plane}} = \varepsilon^{c/6}Z_{\text{flat cylinder}}.
\end{equation}

The fact that \eqref{pure Weyl anomaly} is the only factor in the Weyl transformation between the flat finite cylinder and the flat annulus requires some explanation. The thermal traces in angular quantization are just a different way of slicing up and computing correlators in any conformal frame. In particular, the thermal trace over the evolution operator is always equal to the partition function in that frame. Moreover, since the underlying correlators are those of a CFT, a conformal transformation acts on both the local operators \emph{and} the boundary conditions so that, up to the Weyl anomaly, the final number (or distribution) computed from a given configuration does not change; see footnote \ref{footnote conformal transformation} above. 

In path integral language, the boundary condition $B_i$ that shrinks to the operator $\mathcal{O}_i$ is imposed via a boundary action $S_{\mathcal{O}_i}[g,B_i]$; the dependence on $B_i$ means that the boundary action depends on a set of boundary condition parameters which should be thought of as background fields in the path integral. When performing a Weyl transformation, the change in the boundary action due to the metric variation is exactly compensated by the variation of the parameters in the boundary condition, so that schematically
\begin{equation}
\frac{\delta S_{\mathcal{O}_i}[g,B_i]}{\delta\omega} = 2g^{\alpha\beta}\frac{\delta S_{\mathcal{O}_i}[g,B_i]}{\delta g^{\alpha\beta}} + \frac{\delta S_{\mathcal{O}_i}}{\delta B_i}\frac{\partial B_i}{\partial \omega} = 0.
\end{equation}
Therefore, the Weyl transformation for the path integral whose action is $S_{\text{CFT}_0} + S_{\mathcal{O}_i}[g,B_i]$ is identical to that of the path integral whose action is $S_{\text{CFT}_0}$, i.e.~it is given by \eqref{pure Weyl anomaly}. 
Note that the stress-energy tensor is defined as the Noether current for translations so that the Hamiltonian is constructed from it in the usual way; that is, $T_{\alpha\beta}$ is computed from varying the metric with fixed boundary conditions, and $S_{\mathcal{O}_i}[g,B_i]$ does give a boundary stress-energy contribution which makes $T^{\alpha}_{\phantom{\alpha}\alpha}$ generically nonzero even in flat space with flat boundaries.

Including the other endpoint operator, the full path integral action is $S_{\mathrm{CFT_0}}[g] + S_{\mathcal{O}_i}[g,B_i] + S_{\mathcal{O}_j}[g,B_j]$, and the Weyl invariance of the boundary actions again results in the relation $Z[g'] = e^{-S_{\text{W}}}Z[g]$, where $S_{\text{W}}$ is the same Weyl anomaly action as before. Despite the fact that, barring the Weyl anomaly, the partition functions in any two conformal frames are equal, the boundary conditions $B_i$ and $B_j$ are not generically conformally invariant, which would have required equality of the partition functions when the boundary conditions are held fixed.

While angular quantization traces are equal to the appropriate partition functions on any surface, the flat cylinder frame plays a distinguished role. On the finite cylinder, the spatial and temporal directions are flat and perpendicular, so the equality $Z_{S^1\times\mathds{R}} = \mathrm{Tr}_{\mathcal{H}}[e^{-2\pi H_{\text{R}}}]_{S^1\times\mathds{R}}$ (together with its generalizations with operator insertions) describes the thermal trace at inverse temperature $\beta = 2\pi$. Due to this distinguished role, we define the normalization in angular quantization so that the shrinking limit on the cylinder produces the two endpoint operators with no powers of $\varepsilon$. Specifically,
\begin{multline}
\mathrm{Tr}_{\mathcal{H}_{ij}^{B_i B_j}}\left[\mathcal{O}_1(y_{1,1},y_{1,2})\cdots\mathcal{O}_n(y_{n,1},y_{n,2})e^{-2\pi H_{\text{R}}^{B_i B_j}}\right]_{S^1\times\mathds{R}} 
\\ \stackrel{\varepsilon\rightarrow 0}{\longrightarrow} b_{\mathcal{O}_i}b_{\mathcal{O}_j}\big\langle \mathcal{O}_i(0,\ln\varepsilon)\mathcal{O}_1(y_{1,1},y_{1,2})\cdots\mathcal{O}_n(y_{n,1},y_{n,2})\mathcal{O}_j(0,-\ln\varepsilon) \big\rangle_{S^1\times\mathds{R}};
\end{multline}
the $\varepsilon\rightarrow 0$ limit of the correlator of course contains powers of $\varepsilon$, but the point is that the relation between the trace and the correlator does not. Then, the shrinking of the boundary conditions in any other conformal frame is obtained from the above cylinder limit simply by applying conformal transformations to $\mathcal{O}_i(0,\ln\varepsilon)$ and $\mathcal{O}_j(0,-\ln\varepsilon)$. To reiterate, the partition functions are always the same (modulo the Weyl anomaly), but the relation between the thermal traces and the correlators changes between frames. Indeed, mapping the endpoint operators from the cylinder to the sphere is precisely why shrinking the boundary conditions on the plane acquires factors of $i^{-s}\varepsilon^{\pm\Delta}$, as we used to obtain the matching conditions \eqref{2-point matching} and \eqref{3-point matching} above.

In summary, the angular quantization thermal traces in the regulated (flat) sphere and in the regulated (flat) cylinder conformal frames are related by
\begin{equation}\label{Weyl transformation}
\text{Tr}_{\mathcal{H}_{ij}^{B_i B_j}}\left[\mathcal{O}\cdots\mathcal{O}e^{-2\pi H_{\text{R}}^{B_i B_j}}\right]_{S^2} = \varepsilon^{c/6}\text{Tr}_{\mathcal{H}_{ij}^{B_i B_j}}\left[\mathcal{O}'\cdots\mathcal{O}'e^{-2\pi H_{\text{R}}^{B_i B_j}}\right]_{S^1\times\mathds{R}},
\end{equation}
where any bulk operator insertions are also transformed in the usual way, and the Rindler Hamiltonian is the same on both sides. We are free to interpret the factor from this Weyl transformation as the shift in the Rindler Hamiltonian between the conformal frames, which is a local and covariant shift corresponding to a bulk cosmological constant. Unlike in radial quantization where the shift in energy is finite, the shift in energy in angular quantization tends to infinity; the difference, of course, is because the states in radial quantization live on a finite circle whereas angular quantization is really constructing the states in Minkowski space with potential sources at infinity\footnote{We should clarify that this shift is not a Casimir energy which, while still present, is proportional to $\frac{1}{\ln\varepsilon}$ and so vanishes in the shrinking limit. It is an example of a far more dramatic effect of an infinite energy shift between the Minkowski vacuum and the Rindler vacuum.}. This precise infinite shift will be important in the explicit calculations presented in the following sections. The ensuing derivations make it clear that the care required to take the shrinking limit properly manifests, in part, as an extreme sensitivity of angular quantization to factors of 2 and minus signs!

\section{Warm-Up: Noncompact Free Boson}\label{free boson}

We now illustrate the procedure outlined above for the simplest example, that of the noncompact free boson $X(z,\barred{z})$ with central charge $c = \widetilde{c} = 1$. Strictly speaking, the noncompact free boson is singular as a CFT, but nevertheless it is an instructive pedagogical exercise, and the pathology-free example of the compact free boson will be considered in the following section. We shall explicitly perform the angular quantization for endpoint exponential primary operators $\mathcal{O}_{k} \equiv \s\s\s\normal{e^{ikX}}$, which have conformal weights $h = \widetilde{h} = \frac{k^2}{4}$, with normalization
\begin{align}
\left\langle \mathcal{O}_{k_1}(z_1)\mathcal{O}_{k_2}(z_2)\right\rangle & = \frac{2\pi\delta(k_1+k_2)}{|z_{12}|^{k_1^2}}
\end{align}
and hence three-point coefficient
\begin{equation}
C_{k_1 k_2 k_3} = 2\pi\delta(k_1+k_2+k_3).
\end{equation}
Since the theory is free, we shall use canonical quantization to provide a Fock space construction of $\mathcal{H}_{k_1 k_2}$. In particular, we perform the excision regularization for two particularly simple classes of boundary conditions --- one which is ``Neumann-like'' and one which is ``Dirichlet-like''. As we shall show, pure Neumann and pure (indefinite) Dirichlet boundary conditions both shrink to the identity operator, but small modifications thereof will shrink to nontrivial exponential primaries. It is perhaps non-intuitive that Neumann and Dirichlet boundary conditions are equivalent in angular quantization, given how different they seem at finite regulator.

\subsection{Boundary Conditions for Exponential Primaries}

We must first find appropriate boundary conditions $B_{k_1}$ at $|z| = \varepsilon$ and $B_{k_2}$ at $|z| = \frac{1}{\varepsilon}$ which shrink to a multiple of $\mathcal{O}_{k_1}(0)$ and $\mathcal{O}_{k_2}(\infty)$, respectively; since this is a free theory, we may impose the boundary condition directly on $X(z,\barred{z})$ itself. We first present a heuristic derivation of the relevant boundary conditions; that the angular quantization constructed out of these boundary conditions correctly reproduces the matching equations \eqref{2-point matching} and \eqref{3-point matching} may be viewed as \emph{a posteriori} proof that they indeed shrink to the exponential primaries claimed. Nevertheless, it is useful pedagogically to derive rigorously that the boundary conditions shrink to the exponential primaries, as this property is logically independent from the construction of angular quantization, so we then provide this derivation for the Neumann-like boundary conditions.

\subsubsection{Heuristic Derivation from the OPE}

The easiest way to find such appropriate boundary conditions is via the OPE, which tells us for instance how $X(z,\barred{z})$ must behave in the vicinity of $\normal{e^{ikX(0)}}$. Our normalization convention is such that the free-field OPE at separated points on the plane is
\begin{equation}
X(z,\barred{z})X(0) = -\frac{1}{2}\ln|z|^2 + \normal{X(z,\barred{z})X(0)}.
\end{equation}
Since $X(z,\barred{z})$ is not a well-defined local operator in the CFT, we really should consider its derivatives $\partial X(z)$ and $\barred{\partial}X(\barred{z})$ instead. Their OPEs with an exponential primary are
\begin{align}
\partial X(z)\normal{e^{ikX(0)}} & = -\frac{ik}{2z}\normal{e^{ikX(0)}} + \normal{\partial X(z)e^{ikX(0)}}
\\ \barred{\partial}X(\barred{z})\normal{e^{ikX(0)}} & = -\frac{ik}{2\barred{z}}\normal{e^{ikX(0)}} + \normal{\barred{\partial} X(\barred{z})e^{ikX(0)}}.
\end{align}
The fact that the singular parts of these OPEs contain the same operator with which we started at the origin says that the operators $z\partial X(z)$ and $\barred{z}\barred{\partial}X(\barred{z})$ both behave like constants (in fact the same constant $-\frac{ik}{2}$) near the $\normal{e^{ikX(0)}}$ insertion; it is understood that $X(z,\barred{z})$ is analytically extended to take values in a curve in one complex dimension\footnote{Equivalently, one may stick to a real field and consider the analytic continuation to complex values of $k$. While the OPE of two exponential operators with non-real $k$ is no longer unitary, none of the analyticity properties relevant to the angular quantization procedure detailed here change for complex $k$.}. However, since we would like to perform canonical quantization, for which $X(z,\barred{z})$ and its conjugate momentum $\Pi(z,\barred{z})$ obey the usual canonical commutation relations at equal angle $\theta$, it is inconsistent to impose constancy of $z\partial X(z)$ and of $\barred{z}\barred{\partial}X(\barred{z})$ on the boundary separately since $\partial X$ and $\barred{\partial}X$ will not commute on the quantization slice. We could take one of these conditions or the other, or any linear combination of the two, but the most natural choices are to take their sum or to take their difference, which say that
\begin{align}
r\partial_r X(z,\barred{z})\normal{e^{ikX(0)}} & = -ik\normal{e^{ikX(0)}} + \normal{r\partial_r X(z,\barred{z})e^{ikX(0)}}
\\ \partial_{\theta} X(z,\barred{z})\normal{e^{ikX(0)}} & = \s\s\s\normal{\partial_{\theta} X(z,\barred{z})e^{ikX(0)}},
\end{align}
where $r\partial_r X = z\partial X + \barred{z}\barred{\partial}X$ and $\partial_{\theta}X = i(z\partial X - \barred{z}\barred{\partial}X)$ are the radial and angular derivatives on the plane. For $k = k_1$, we may then impose the constancy of $r\partial_r X$ or of $\partial_{\theta}X$ at $|z| = \varepsilon$; for $k = k_2$, the above OPEs hold in the inverted $z' = 1/z$ patch, so the boundary conditions at $|z| = \frac{1}{\varepsilon}$ receive a minus sign from the reversed orientation, i.e.~from $r'\partial_{r'}X'(z',\barred{z}') = -r\partial_r X(z,\barred{z})$ and $\partial_{\theta'}X'(z',\barred{z}') = -\partial_{\theta}X(z,\barred{z})$. Therefore, the two simplest boundary conditions on the plane that shrink to a multiple of $\normal{e^{ik_1 X(0)}}$ and $\normal{e^{ik_2 X(\infty)}}$ are
\begin{align}
\text{Neumann-like:} \qquad r\partial_r X\Big|_{|z| = \varepsilon} & = -ik_1 \phantom{\qquad \text{Neumann-like:}}
\\ r\partial_{r} X\Big|_{|z| = \frac{1}{\varepsilon}} & = +ik_2
\end{align}
and
\begin{align}
\text{Dirichlet-like:} \qquad \partial_{\theta}X\Big|_{|z| = \varepsilon} = \partial_{\theta}X\Big|_{|z| = \frac{1}{\varepsilon}} = 0. \phantom{\qquad \text{Dirichlet-like:}}
\end{align}
We call the former ``Neumann-like'' because it fixes the derivative normal to the boundary (along the spatial slice) and the latter ``Dirichlet-like'' because it fixes the derivative parallel to the boundary (along the time direction). Another easy way to see that the Neumann-like boundary condition shrinks to an exponential primary is by noting that the global translation current is $j_{\alpha} = i\partial_{\alpha}X$ under which $\normal{e^{ikX}}$ has charge $k$; since the translation charge of the operator at the origin can be computed as the constant-$|z|$ integral of $\frac{1}{2\pi}j_r$, the boundary condition $j_r = k/\varepsilon$ at $|z| = \varepsilon$ simply ensures that the boundary shrinks to an operator of charge $k$, and the lowest-dimension operator of charge $k$ is the primary $\normal{e^{ikX}}$. 

The Dirichlet-like boundary condition written above, on the other hand, seems insufficient to shrink to $\normal{e^{ikX}}$, as it does not seem to depend on $k$ at all. Indeed, all this boundary condition says by itself is that $X(z,\barred{z})$ is single-valued around the origin, so it only guarantees that the boundary condition shrink to a local operator (i.e.~not contain a bare field $X(z,\barred{z})$ that is neither differentiated nor exponentiated). Therefore, that the charge of the resulting operator be exactly $k$ must be fixed dynamically for this Dirichlet-like boundary condition, which will determine what the actual action for this regulator is. We shall show this dynamical fixing below, but roughly speaking it arises from the equation of motion for the boundary Lagrange multiplier which must be introduced to impose the boundary condition.

It may seem like we have ignored the simplest boundary condition consistent with the OPE, which would just set $X = -ik\ln\varepsilon$ at $|z| = \varepsilon$, fixing the actual value of the free boson itself more akin to what would usually be called a Dirichlet boundary condition in the literature. We have opted not to write this boundary condition because it is misleading for the compact boson discussed in the following section due to the need to split $X(z,\barred{z})$ into chiral and antichiral halves. The chiral splitting for the compact boson is pivotal to angular quantization and is in fact determined by the theory unlike in radial quantization where one usually says it amounts to a choice of branch cut which disappears at the end of the day for correlators of truly local operators. There is also a technical reason the boundary condition $X = -ik\ln\varepsilon$ is inappropriate for the noncompact boson --- due to the continuous spectrum, fixing the value of the field does not shrink to a well-defined primary. This latter fact holds even for the pure Dirichlet boundary condition $X = 0$, which does not quite shrink to the identity operator as may be verified by applying the framework below. This difficulty is of course due to the pathological continuous spectrum of the noncompact theory and hence does not persist in the compact theory. Nevertheless,  it is the Dirichlet-like boundary condition described above and below which most easily generalizes to one which shrinks to winding operators for the compact boson. We shall revisit this point later.

Since we wish to perform the canonical quantization on the very wide (but finite) strip, where $\ln\varepsilon\leqslant y_2 \leqslant -\ln\varepsilon$ and $y_1 \in \mathds{R}$ is not compactified, we merely need to transform the above boundary conditions via $z = e^{-iy}$. Since $X(z,\barred{z})$ is a scalar, the relevant boundary conditions on the strip are simply
\begin{align}\label{N noncompact boson}
\text{Neumann-like:} \qquad \partial_{y_2}X\Big|_{y_2 = +\ln\varepsilon} & = -ik_1 \phantom{\qquad \text{Neumann-like:}}
\\ \notag \partial_{y_2}X\Big|_{y_2 = -\ln\varepsilon} & = +ik_2
\end{align}
and
\begin{align}\label{D noncompact boson}
\text{Dirichlet-like:} \qquad \partial_{y_1}X\Big|_{y_2 = \pm\ln\varepsilon} & = 0. \phantom{\qquad \text{Dirichlet-like:}}
\end{align}

\subsubsection{Rigorous Derivation from the Path Integral}

Here we provide a rigorous derivation that the Neumann-like boundary condition $r\partial_r X = -ik$ at $|z|=\varepsilon$ shrinks to (a constant times) the operator $\normal{e^{ikX(0)}}$ as $\varepsilon\rightarrow 0$; this subsection is independent from the rest of the paper and included for completeness, so there is no loss of continuity for the reader wishing to jump ahead to the angular quantization Hilbert spaces and Rindler Hamiltonians in the following subsection. The simplest way to prove the desired shrinking is to derive the one-point function limit
\begin{equation}\label{exponential 1-point limit}
\lim_{\varepsilon\rightarrow 0}\left\langle\normal{e^{ik'X(z,\barred{z})}}\right\rangle_{\text{N}(k,\varepsilon)} = \frac{2\pi b\varepsilon^{\frac{k^2}{2}}\delta(k+k')}{|z|^{k^2}},
\end{equation}
where $\langle\cdots\rangle_{\text{N}(k,\varepsilon)}$ denotes the correlator of the free boson theory on the plane with the disk $|z| < \varepsilon$ cut out and the Neumann-like boundary condition $r\partial_r X = -ik$ imposed at $|z| = \varepsilon$, and $b$ is an $\varepsilon$-independent constant. The normalization constant $b$ is not relevant to prove that we obtain the correct operator in the shrinking limit, so we need not compute the one-loop determinant in the path integral; the exact normalization relevant to angular quantization will be obtained in the following subsection. 

The bare path integral on the excised plane $\Sigma(\varepsilon) = \mathds{C}\setminus D^2(\varepsilon)$ subject to the Neumann-like boundary condition at $|z| = \varepsilon$ is
\begin{equation}
Z[0] = \int\mathcal{D}X \ \exp\left[-\frac{1}{2\pi}\int_{\Sigma(\varepsilon)}\hspace{-5pt}d^2 z\s\s \partial X\barred{\partial}X + ik\int_0^{2\pi}\frac{d\theta}{2\pi}X(|z| = \varepsilon)\right].
\end{equation}
The boundary term is precisely how the boundary condition $r\partial_r X = -ik$ at $|z| = \varepsilon$ is imposed via the path integral, because the variation of the action is
\begin{equation}
\delta S = -\frac{1}{\pi}\int_{\Sigma(\varepsilon)}\hspace{-5pt}d^2 z\s\s \partial\barred{\partial}X \delta X + \int_0^{2\pi}\frac{d\theta}{2\pi}\left.\left(\sqrt{f^* g}\s n^{\alpha}\partial_{\alpha}X - ik\right)\delta X\right|_{|z|=\varepsilon},
\end{equation}
where the line element on the boundary is $d\theta\sqrt{f^* g} = \varepsilon d\theta$ and the outward unit normal vector is $n = n^{\alpha}\partial_{\alpha} = -\partial_r$, with the minus sign due to the chosen orientation of the boundary. The vanishing of the action variation on the boundary then indeed correctly imposes $r\partial_r X = -ik$ at $|z| = \varepsilon$. Before computing the propagator, we must first address the fact that the noncompact free boson has a non-normalizable zero-mode which causes the bare path integral to be ill-defined. The zero-mode here, of course, takes the form $X = x_0$ for constant $x_0$; without the boundary term, the bare path integral is infinity, while with the boundary term the bare path integral vanishes unless $k=0$ (when it is also infinity). When fermionic path integrals have zero-modes, the correct procedure is to compare correlators not with the bare path integral but with the path integral with the correct number of zero-modes inserted in order to render the result finite. Morally speaking, that should also be the correct prescription for non-normalizable bosonic zero-modes, namely we must separate out the zero-mode contribution from the rest of the path integral that has the zero-mode absorbed by a delta function. Then, if we schematically write $\mathcal{O} = \mathcal{O}_0 \mathcal{O}'$ for any operator as its zero-mode part times its nonzero-mode part (which is appropriate for the exponential operators, for instance), we compute correlators as
\begin{equation}
\left\langle\mathcal{O}_1\mathcal{O}_2\cdots\mathcal{O}_n\right\rangle_{\text{N}(k,\varepsilon)} = \left\langle \mathcal{O}_1\mathcal{O}_2\cdots\mathcal{O}_n\right\rangle_0\left\langle\mathcal{O}_1\mathcal{O}_2\cdots\mathcal{O}_n\right\rangle_{\text{N}(k,\varepsilon)}^{\prime},
\end{equation}
where
\begin{equation}
\left\langle\mathcal{O}_1\mathcal{O}_2\cdots\mathcal{O}_n\right\rangle_0 \equiv \int_{-\infty}^{\infty}dx_0 \left(\mathcal{O}_1\mathcal{O}_2\cdots\mathcal{O}_n\right)_0 e^{ikx_0}
\end{equation}
is the zero-mode contribution, and
\begin{equation}
\left\langle\mathcal{O}_1\mathcal{O}_2\cdots\mathcal{O}_n\right\rangle_{\text{N}(k,\varepsilon)}^{\prime} = \int \mathcal{D}X' \ \mathcal{O}_1'\mathcal{O}_2'\cdots\mathcal{O}_n'\exp\left[-\frac{1}{2\pi}\int_{\Sigma(\varepsilon)}\hspace{-5pt}d^2 z\s\s \partial X'\barred{\partial}X' + ik\int_0^{2\pi}\frac{d\theta}{2\pi}X'(|z| = \varepsilon)\right]
\end{equation}
is the part of the path integral orthogonal to the zero-mode. Note that the part of the path integral with the zero-mode removed has well-defined normalization, but the zero-mode part may be given arbitrary ($\varepsilon$-independent) normalization as usual for a continuous spectrum. For the exponential operators, we have $\mathcal{O}_i = \s\s\s\normal{e^{ik_i X}} = e^{ik_i x_0}\normal{e^{ik_i X}}\hspace{-0pt}^{\prime}$ so the zero-mode part of the path integral immediately gives $\langle \mathcal{O}_1\cdots\mathcal{O}_n\rangle_0 = 2\pi\delta(k+k_1+\dotsc+k_n)$ in the conventional continuous-momentum state normalization, and after the zero-mode has been removed the correlator of normal-ordered exponentials may be computed via the series expansion of the exponentials. In particular, the one-point function we wish to compute is given by
\begin{equation}\label{exponential 1-point function}
\left\langle\normal{e^{ik'X(z,\barred{z})}}\right\rangle_{\text{N}(k,\varepsilon)} = 2\pi b\delta(k+k')\sum_{n=0}^{\infty}\frac{(ik')^n}{n!}\left\langle\normal{X^n(z,\barred{z})}\right\rangle_{\text{N}(k,\varepsilon)}^{\prime},
\end{equation}
where $b$ is the normalization from the one-loop fluctuation determinant, which we do not need to compute here, and hence the remaining correlators $\langle \cdots\rangle_{\text{N}(k,\varepsilon)}^{\prime}$ are computed with the normalization $\langle \mathds{1}\rangle_{\text{N}(k,\varepsilon)}^{\prime} = 1$.

The last ingredients we need are the nonzero-mode one-point function $\langle X(z,\barred{z})\rangle_{\text{N}(k,\varepsilon)}^{\prime}$ and propagator $\langle X(z_1,\barred{z}_1)X(z_2,\barred{z}_2)\rangle_{\text{N}(k,\varepsilon)}^{\prime}$. These may be computed by evaluating the sourced partition function $Z'[J]$ in the usual way by shifting the variable of integration so as to complete the square and hence decouple $X$ from its source $J$. It is easier, however, to directly solve the free Schwinger-Dyson equation subject to the boundary condition, which for the two-point function read
\begin{align}
\left\langle \partial_1\barred{\partial}_1 X(z_1,\barred{z}_1)X(z_2,\barred{z}_2)\right\rangle_{\text{N}(k,\varepsilon)}^{\prime} & = -\pi\delta^2(z_{12})
\\ \left.\left\langle\left(z_1\partial_1 + \barred{z}_1\barred{\partial}_1\right)X(z_1,\barred{z}_1)X(z_2,\barred{z}_2)\right\rangle_{\text{N}(k,\varepsilon)}^{\prime}\right|_{|z_1| = \varepsilon} & = -ik\left\langle X(z_2,\barred{z}_2)\right\rangle_{\text{N}(k,\varepsilon)}^{\prime},
\end{align}
where $\partial_1$ and $\barred{\partial}_1$ denote the holomorphic and antiholomorphic derivatives with respect to $(z_1,\barred{z}_1)$, and which for the one-point function read
\begin{align}
\left\langle \partial\barred{\partial} X(z,\barred{z})\right\rangle_{\text{N}(k,\varepsilon)}^{\prime} & = 0
\\ \left.\left\langle\left(z\partial + \barred{z}\barred{\partial}\right)X(z,\barred{z})\right\rangle_{\text{N}(k,\varepsilon)}^{\prime}\right|_{|z| = \varepsilon} & = -ik,
\end{align}
where we have normalized $\langle \mathds{1}\rangle_{\text{N}(k,\varepsilon)}^{\prime} = 1$. The one-point function equations are uniquely solved by
\begin{equation}
\left\langle X(z,\barred{z})\right\rangle_{\text{N}(k,\varepsilon)}^{\prime} = -\frac{ik}{2}\ln\left|\frac{z}{\varepsilon}\right|^2 + \gamma,
\end{equation}
where $\gamma$ is an arbitrary constant independent of $\varepsilon$ which reflects the shift ambiguity in the free boson; since $\gamma$ is shifted by the zero-mode which has been removed and integrated over separately, we may simply set $\gamma = 0$. Note that the bulk equation of motion is indeed satisfied because $\partial\barred{\partial}(\ln|z|^2) = 2\pi\delta^2(z)$ which has no support on $\Sigma(\varepsilon)$ and so equals the zero distribution; it is also important to realize that the $\varepsilon$ dependence of the correlator is fixed by the invariance under the simultaneous $z \mapsto \lambda z$ and $\varepsilon \mapsto \lambda\varepsilon$ manifest in the action, which is why we know that the constant ambiguity $\gamma$ cannot in fact depend on $\varepsilon$. With this one-point function (with $\gamma = 0$), the two-point function Schwinger-Dyson equation with boundary condition is uniquely solved by
\begin{multline}\label{Neumann-like disk propagator}
\left\langle X(z_1,\barred{z}_1)X(z_2,\barred{z}_2)\right\rangle_{\Sigma(\varepsilon)}^{\prime} = -\frac{1}{2}\ln\left|\frac{z_{12}}{\varepsilon}\right|^2 - \frac{1}{4}\ln\left|\frac{\varepsilon}{z_1} - \frac{\barred{z}_2}{\varepsilon}\right|^2 - \frac{1}{4}\ln\left|\frac{z_1}{\varepsilon} - \frac{\varepsilon}{\barred{z}_2}\right|^2 + \frac{1}{4}\ln\left|\frac{z_1 z_2}{\varepsilon^2}\right|^2
\\ + \left(-\frac{ik}{2}\ln\left|\frac{z_1}{\varepsilon}\right|^2\right)\left(-\frac{ik}{2}\ln\left|\frac{z_2}{\varepsilon}\right|^2\right).
\end{multline}
It is worthwhile to note that taking separate holomorphic derivatives of the two-point function gives
\begin{equation}
\left\langle \partial X(z_1)\partial X(z_2)\right\rangle_{\text{N}(k,\varepsilon)}^{\prime} = -\frac{1}{2z_{12}^2} - \frac{k^2}{4z_1 z_2},
\end{equation}
and hence the nonzero-mode correlator of the stress-energy tensor $T(z) = -\normal{\partial X(z)\partial X(z)}$ is
\begin{equation}
\left\langle T(z)\right\rangle_{\text{N}(k,\varepsilon)}^{\prime} = \frac{k^2}{4z^2}.
\end{equation}
This result, together with its antiholomorphic counterpart, says that the boundary condition must shrink to a scalar (primary) operator with weights $h = \widetilde{h} = \frac{k^2}{4}$, for which the only candidates are $\normal{e^{\pm ikX}}$. It might seem that the one-point function of the stress-energy tensor is nonvanishing in the limit $\varepsilon\rightarrow 0$ for $k \neq 0$ since it does not even depend on $\varepsilon$, but we must remember that the full one-point function includes a $\delta(k)$ from the zero-mode and hence gives $\langle T(z)\rangle_{\text{N}(k,\varepsilon)} = 0$ as one would expect. It is very nice that, by separating out the zero-mode contribution, we can determine the weights of the operator to which the boundary condition shrinks before even computing the one-point functions of the exponential primaries.

Finally we may compute the one-point functions $\left\langle\normal{X^n(z,\barred{z})}\right\rangle_{\text{N}(k,\varepsilon)}^{\prime}$ needed in \eqref{exponential 1-point function}. To do so, we note that the one- and two-point functions computed above immediately tell us that the nonzero-mode sourced partition function, in which the integral over $iJ(z,\barred{z})X(z,\barred{z})$ is added to the exponential, is 
\begin{equation}
\frac{Z'[J]}{Z'[0]} = \exp\hspace{-2pt}\left[-\frac{1}{2}\int_{\Sigma(\varepsilon)}\hspace{-7pt}d^2 z\hspace{-2pt}\int_{\Sigma(\varepsilon)}\hspace{-7pt}d^2 z' J(z,\barred{z})K(z,\barred{z};z',\barred{z}')J(z',\barred{z}') + i\hspace{-2pt}\int_{\Sigma(\varepsilon)}\hspace{-7pt}d^2 z\left\langle X(z,\barred{z})\right\rangle_{\text{N}(k,\varepsilon)}^{\prime}\hspace{-2pt}J(z,\barred{z})\right],
\end{equation}
where the integral kernel $K(z_1,\barred{z}_1;z_2,\barred{z}_2)$ is the first line in \eqref{Neumann-like disk propagator}. In addition to taking $n$ functional derivatives with respect to $J$, one combinatorically subtracts off all divergences using Wick's Theorem before taking the multi-coincident limit to obtain the one-point function of $\normal{X^n(z,\barred{z})}$. Since the combinatorics of Wick's Theorem exponentiate \cite{Polchinski981}, the normal-ordered $n$-point function at distinct points is obtained by replacing the bulk kernel $K(z,\barred{z};z',\barred{z}')$ above with $K(z,\barred{z};z',\barred{z}') + \frac{1}{2}\ln|z-z'|^2$, and the subsequent multi-coincident limit allows one to express the desired one-point function as
\begin{equation}
\left\langle\normal{X^n(z,\barred{z})}\right\rangle_{\text{N}(k,\varepsilon)}^{\prime} = (-i)^n \frac{\delta^n}{\delta J(z,\barred{z})^n}\left.\exp\hspace{-2pt}\left[-\frac{1}{2}\hspace{-2pt}\left(\int_{\Sigma(\varepsilon)}\hspace{-7pt}d^2 z'\sqrt{A}J(z',\barred{z}')\right)^2 \hspace{-2pt}+\hspace{-2pt} \int_{\Sigma(\varepsilon)}\hspace{-7pt}d^2 z'\s\s B J(z',\barred{z}')\right]\right|_{J=0}\hspace{-2pt},
\end{equation}
where the two integral kernels are
\begin{align}
A(z',\barred{z}') & = \ln\varepsilon -\ln\left(1 - \frac{\varepsilon^2}{|z'|^2}\right)
\\ B(z',\barred{z}') & = \frac{k}{2}\ln\left|\frac{z'}{\varepsilon}\right|^2.
\end{align}
Then, the sum of these one-point functions is
\begin{align}
\sum_{n=0}^{\infty}\frac{(ik')^n}{n!}\left\langle\normal{X^n(z,\barred{z})}\right\rangle & = \sum_{n=0}^{\infty}\sum_{m=0}^{\lfloor n/2\rfloor}\frac{k'^n}{2^m m!(n-2m)!}[-A(z,\barred{z})]^m B(z,\barred{z})^{n-2m}
\\ & = \sum_{m=0}^{\infty}\frac{k'^{2m}}{2^m m!}\ln^m\left[\frac{1}{\varepsilon}\left(1 - \frac{\varepsilon^2}{|z|^2}\right)\right]\sum_{n=0}^{\infty}\frac{(k'k)^n}{2^n n!}\ln^n\left|\frac{z}{\varepsilon}\right|^2
\\ & = \varepsilon^{-k'^2/2}\left(1-\frac{\varepsilon^2}{|z|^2}\right)^{k'^2/2}\left|\frac{z}{\varepsilon}\right|^{k'k}.
\end{align}
Therefore, including the prefactor in \eqref{exponential 1-point function}, we have derived the general exponential primary one-point function in the presence of the Neumann-like boundary condition at $|z| = \varepsilon$ to be
\begin{equation}\label{exponential primary one-point function}
\left\langle\normal{e^{ik'X(z,\barred{z})}}\right\rangle_{\text{N}(k,\varepsilon)} = \frac{2\pi b \varepsilon^{\frac{k^2}{2}}\delta(k+k')}{|z|^{k^2}}\left(1 - \frac{\varepsilon^2}{|z|^2}\right)^{k^2/2},
\end{equation}
whose limit as $\varepsilon\rightarrow 0$ is indeed given by \eqref{exponential 1-point limit}. This result proves that the Neumann-like boundary condition shrinks to the operator $b\varepsilon^{k^2/2}\normal{e^{ikX(0)}}$ on the plane as claimed. The shrinking of the other boundary conditions written in this paper may be proven directly with only minimal modifications to these computations, so we shall not explicitly write these derivations.

\subsection{Hilbert Space and Rindler Hamiltonian}

Having determined boundary conditions that shrink to exponential primaries, it is now straightforward to construct the angular quantization Hilbert spaces and associated Rindler Hamiltonians via canonical quantization on the very wide strip with complex coordinate $y = y_1 + iy_2$. First, one determines the correct action which imposes the given boundary conditions; since we are working in Euclidean signature with $y_1$ the strip time direction, the canonical momentum conjugate to $X$ in terms of the Lagrangian is $\Pi = -i\frac{\delta L}{\delta \partial_{y_1}X}$. Then, one mode expands $X$ and $\Pi$ at a fixed time $y_1$ and imposes the usual equal-time canonical commutation relations
\begin{align}
[X(y_1,y_2),\Pi(y_1,y_2')] & = i\delta(y_2-y_2')
\\ [X(y_1,y_2),X(y_1,y_2')] & = [\Pi(y_1,y_2),\Pi(y_1,y_2')] = 0;
\end{align}
for this step it is useful to note the distributional identities
\begin{align}
\delta(y_2-y_2') & = -\frac{1}{2\ln\varepsilon}\sum_{n\in\mathds{Z}}\cos\left(\frac{\pi n (y_2-\ln\varepsilon)}{2\ln\varepsilon}\right)\cos\left(\frac{\pi n (y_2'-\ln\varepsilon)}{2\ln\varepsilon}\right)
\\ & = -\frac{1}{2\ln\varepsilon}\sum_{n\in\mathds{Z}}\sin\left(\frac{\pi n (y_2-\ln\varepsilon)}{2\ln\varepsilon}\right)\sin\left(\frac{\pi n (y_2'-\ln\varepsilon)}{2\ln\varepsilon}\right),
\end{align}
valid on the domain $\ln\varepsilon \leqslant y_2 \leqslant -\ln\varepsilon$. Next, the Rindler Hamiltonian $H_{\text{R}}$ must be computed via\footnote{The $1/2\pi$ normalization is due to the stringy convention of normalizing all currents with an extra $2\pi$ and subsequently dividing all Noether charges by $2\pi$.}
\begin{equation}\label{Rindler Hamiltonian general integral}
H_{\text{R}}(y_1) = \frac{1}{2\pi}\int_{\ln\varepsilon}^{-\ln\varepsilon}dy_2\s\s T_{y_1 y_1}(y_1,y_2);
\end{equation}
since the mode expansions obey the equation of motion and boundary conditions, this Hamiltonian will be time independent, and hence the Euclidean strip evolution operator will simply be\footnote{We are admittedly using an unfortunate definition for the direction of Euclidean time. Having defined angular quantization to evolve in angles counterclockwise in the plane, the positive direction of Euclidean time on the strip is in decreasing $y_1$ coordinate. The formula \eqref{Rindler Hamiltonian general integral} remains the same because the usual right-handed orientation corresponding to spatial normal vector $n = -\partial_{y_1}$ gives the spatial tangent vector $t = -\partial_{y_2}$. Then, evolution in Euclidean time always corresponds to $\Delta y_1$ being negative, so that $e^{\Delta y_1 H_{\text{R}}}$ is indeed a decaying operator. For example, the thermal trace corresponds to evolving from $y_1 = 0$ to $y_1 = -2\pi$ and gives the usual operator $e^{-2\pi H_{\text{R}}}$. This convention is also why the canonical momentum is $\Pi = i\frac{\delta L}{\delta\partial_{\tau}X} = -i\frac{\delta L}{\delta\partial_{y_1}X}$.} $e^{\Delta y_1 H_{\text{R}}}$. However, it is crucial that the finite-regulator stress-energy tensor is not traceless due to the explicit breaking of conformal symmetry by the boundary conditions. As such, the component appearing in the Hamiltonian is
\begin{equation}
T_{y_1 y_1}(y_1,y_2) = T_{yy}(y,\barred{y}) + T_{\barred{y}\barred{y}}(y,\barred{y}) + 2T_{y\barred{y}}(y,\barred{y}),
\end{equation}
where $T_{yy}$ and $T_{\barred{y}\barred{y}}$ are now \emph{not} holomorphic and antiholomorphic, respectively. However, the trace $4T_{y\barred{y}}$ as well as the non-(anti)holomorphic parts of $T_{yy}$ ($T_{\barred{y}\barred{y}}$) are localized to the boundary because local conformal symmetry is still preserved in the bulk. We may thus write each stress-energy component as a bulk piece plus a boundary piece, and we still have $T_{y\barred{y}}^{\text{bulk}} = 0$ as well as $T_{yy}^{\text{bulk}} \equiv T(y)$ being holomorphic and $T_{\barred{y}\barred{y}}^{\text{bulk}} \equiv \widetilde{T}(\barred{y})$ antiholomorphic. As such, we may still use the usual conformal transformations
\begin{align}
T(y) & = -z(y)^2 T\big(z(y)\big) + \frac{c}{24}
\\ \widetilde{T}(\barred{y}) & = -\barred{z}(\barred{y})^2\widetilde{T}\big(\barred{z}(\barred{y})\big) + \frac{\widetilde{c}}{24}
\end{align}
of the bulk stress-energy tensor components between the cylinder and the plane; the boundary contributions provide important boundary terms in the Rindler Hamiltonian which are necessary for the consistency of the theory. The angular quantization Hilbert space at finite regulator is thus constructed as the Fock space built up from a vacuum by the positive-energy mode operators of the canonical quantization, whose energies are immediately read off from the mode expression for the Rindler Hamiltonian. Finally, one computes the mode expansions of the exponential primaries themselves and evaluates the thermal trace over the Fock space as well as the thermal one-point functions of the bulk primaries, showing that the matching conditions \eqref{2-point matching} and \eqref{3-point matching} are indeed satisfied with specific normalization constants.

\subsubsection{Neumann-like Boundary Conditions}

Consider the Neumann-like boundary conditions \eqref{N noncompact boson} on the very wide strip, which will result in the endpoint operators $\normal{e^{ik_1 X(0)}}$ and $\normal{e^{ik_2 X(\infty)}}$ in the shrinking limit on the sphere. The free boson action which imposes these boundary conditions is given by
\begin{equation}\label{NN strip action}
S = \frac{1}{4\pi}\int_{-\infty}^{\infty}\hspace{-3pt}dy_1 \int_{\ln\varepsilon}^{-\ln\varepsilon}\hspace{-8pt}dy_2 \left[\left(\partial_{y_1} X\right)^2 + \left(\partial_{y_2} X\right)^2\right] - \frac{ik_1}{2\pi}\int_{-\infty}^{\infty}\hspace{-3pt}dy_1\s\s X\Big|_{y_2 = \ln\varepsilon} - \frac{ik_2}{2\pi}\int_{-\infty}^{\infty}\hspace{-3pt}dy_1\s\s X\Big|_{y_2 = -\ln\varepsilon}.
\end{equation}
For fixed $k_1$ and $k_2$, the field $X$ is analytically continued to be defined on an appropriate complexified contour \`{a} la Keldysh in the target space, so strictly speaking it is only a real field when $k_1$ and $k_2$ are pure imaginary. The stress-energy tensor for this theory is
\begin{equation}
T_{\alpha\beta} = -\normal{\left(\partial_{\alpha}X\partial_{\beta}X - \frac{1}{2}\delta_{\alpha\beta}\partial_{\gamma}X\partial^{\gamma}X\right)} - t_{\alpha}t_{\beta} X\left[ik_1 \delta(y_2 - \ln\varepsilon) + ik_2\delta(y_2 + \ln\varepsilon)\right],
\end{equation}
where $\delta_{\alpha\beta}$ is the flat strip metric and $t^{\alpha}$ is the unit vector tangent to the boundaries. We immediately see that the classical trace of this stress-energy tensor is\footnote{In covariant notation, this classical trace is $T^{\alpha}_{\phantom{\alpha}\alpha} = -Xn^{\alpha}\partial_{\alpha}X \delta(\Sigma-\partial\Sigma)$, where $n^{\alpha}$ is the outward-pointing unit vector normal to the boundaries, which equals the classical trace written here on account of the boundary conditions.}
\begin{equation}\label{NN noncompact classical trace}
T^{\alpha}_{\phantom{\alpha}\alpha}(y_1,y_2) = -X(y_1,y_2)\left[ik_1 \delta(y_2 - \ln\varepsilon) + ik_2 \delta(y_2 + \ln\varepsilon)\right];
\end{equation}
furthermore, $T_{yy} = T(y) + \frac{1}{4}T^{\alpha}_{\phantom{\alpha}\alpha}$ and $T_{\barred{y}\barred{y}} = \widetilde{T}(\barred{y}) + \frac{1}{4}T^{\alpha}_{\phantom{\alpha}\alpha}$, where\footnote{By convention, the variable of a holomorphic or antiholomorphic derivative is implicitly given by its argument, so that for example $\partial X(y)$ means $\partial_y X(y)$ and not $\partial_z X\big(z(y)\big)$.} 
\begin{align}
T(y) & = -\normal{\partial X(y)\partial X(y)}
\\ \widetilde{T}(\barred{y}) & = -\normal{\barred{\partial}X(\barred{y})\barred{\partial}X(\barred{y})}
\end{align}
are the usual bulk holomorphic and antiholomorphic stress-energy components, so that the integrand appearing in the Rindler Hamiltonian in this case is $T_{y_1 y_1} = T(y) + \widetilde{T}(\barred{y}) + T^{\alpha}_{\phantom{\alpha}\alpha}$. 

The mode expansions of $X$ and its canonical momentum conjugate $\Pi = -\frac{i}{2\pi}\partial_{y_1}X$ obeying the equation of motion $(\partial_{y_1}^2 + \partial_{y_2}^2)X = 0$ and the boundary conditions are
\begin{align}\label{NN noncompact strip expansion}
X(y_1,y_2) & = x + \frac{i(k_1+k_2)}{12}\ln\varepsilon - \frac{\pi i p}{\ln\varepsilon}y_1 - \frac{i(k_1 - k_2)}{2}y_2 + \frac{i(k_1+k_2)}{4\ln\varepsilon}\left(y_1^2 - y_2^2\right) 
\\ \notag & \phantom{= x } + i\sqrt{2}\sum_{\substack{n \in \mathds{Z} \\ n \neq 0}}\frac{\alpha_n}{n}e^{-\frac{\pi n }{2\ln\varepsilon}y_1}\cos\left(\frac{\pi n(y_2 - \ln\varepsilon)}{2\ln\varepsilon}\right)
\\ \Pi(y_1,y_2) & = -\frac{p}{2\ln\varepsilon} + \frac{k_1+k_2}{4\pi\ln\varepsilon}y_1 - \frac{1}{2\sqrt{2}\ln\varepsilon}\sum_{\substack{n \in \mathds{Z} \\ n \neq 0}}\alpha_n e^{-\frac{\pi n }{2\ln\varepsilon}y_1}\cos\left(\frac{\pi n (y_2 - \ln\varepsilon)}{2\ln\varepsilon}\right),
\end{align}
where the constant contribution $\frac{i}{12}(k_1+k_2)\ln\varepsilon$ in the expansion of $X$ appears so that the target-space $\text{NL}\sigma\text{M}$ center-of-mass variables are
\begin{align}
x & = -\frac{1}{2\ln\varepsilon}\int_{\ln\varepsilon}^{-\ln\varepsilon}dy_2 \s\s X(0,y_2)
\\ p & = \int_{\ln\varepsilon}^{-\ln\varepsilon}dy_2\s\s\Pi(0,y_2).
\end{align}
As a classical expression, \eqref{NN noncompact strip expansion} is indeed a saddle-point of the Euclidean path integral on the very wide \emph{strip} for all values of its parameters, but note that it is not a saddle-point of the Euclidean path integral on the very wide \emph{cylinder} (where $y_1 \sim y_1 - 2\pi$ is now periodic) for any values of its parameters unless $k_1 + k_2 = 0$ (for which a periodic saddle-point is given by $p = \alpha_n = 0$). The lack of a saddle on the cylinder means that the path integral on the two-holed plane vanishes. Thus, even at the level of the classical expansion we see that the path integral on the two-holed plane at finite regulator will already provide the $\delta(k_1+k_2)$ appearing in the two-point function of $\normal{e^{ik_1 X}}$ and $\normal{e^{ik_2 X}}$. Now, imposing the canonical commutation relations fixes the commutators between the operators $x$, $p$ and $\alpha_n$ to be
\begin{align}
[x,p] & = i
\\ [\alpha_n, \alpha_m] & = n\delta_{n,-m},
\end{align}
with all other commutators vanishing.

Next we want to compute the Hamiltonian and evolution operator so that the angular quantization Hilbert space can be identified with the Fock space of positive-energy mode operators acting on a Fock vacuum annihilated by all negative-energy mode operators. We have already seen that $T_{y_1 y_1} = T(y) + \widetilde{T}(\barred{y}) + T^{\alpha}_{\phantom{\alpha}\alpha}$, so the Rindler Hamiltonian on the very wide strip associated to Neumann-like boundary conditions on both ends is computed by
\begin{equation}\label{NN Hamiltonian integral}
H_{\text{R}}^{\text{NN}}(y_1) = \frac{1}{2\pi}\int_{\ln\varepsilon}^{-\ln\varepsilon}dy_2\left[T(y) + \widetilde{T}(\barred{y})\right] - \frac{1}{2\pi}\left[ik_1 X(y_1,y_2 = \ln\varepsilon) + ik_2 X(y_1,y_2 = -\ln\varepsilon)\right].
\end{equation}
It remains to compute the bulk stress-energy tensors $T(y)$ and $\widetilde{T}(\barred{y})$. Below we will see that the usual relations $\alpha_n^{\dagger} = \alpha_{-n}$ and $[\alpha_n,\alpha_{-n}] = n$ identify $\alpha_{n < 0}$ as creation operators and $\alpha_{n > 0}$ as annihilation operators. So define creation/annihilation normal ordering $\circnormal{\mathcal{O}}$ in canonical quantization as the mode expansion where the operators $x$ and $\alpha_{-n}$ ($n > 0$) are all placed to the left of the operators $p$ and $\alpha_n$ ($n > 0$); it is of course irrelevant whether we group $x$ with the creation operators and $p$ with the annihilation operators or \emph{vice versa}, so we make this aesthetic choice as we shall usually take a momentum basis of the non-oscillator Hilbert space. Then, the bulk stress-energy tensors on the strip are
\begin{align}
T(y) & = \frac{(k_1+k_2)^2}{16\ln^2\varepsilon}y^2 - \frac{\pi(k_1+k_2)}{4\sqrt{2}\ln^2\varepsilon}y\sum_{n\in\mathds{Z}}\alpha_n e^{\frac{\pi i n}{2}}e^{-\frac{\pi n y}{2\ln\varepsilon}} 
\\ \notag & \hspace{15pt} + \frac{\pi^2}{8\ln^2\varepsilon}\sum_{n,m\in\mathds{Z}}\circnormal{\alpha_n\alpha_m}e^{\frac{\pi i(n+m)}{2}}e^{-\frac{\pi(n+m)y}{2\ln\varepsilon}} - \frac{\pi^2}{96\ln^2\varepsilon}
\\ \widetilde{T}(\barred{y}) & = \frac{(k_1+k_2)^2}{16\ln^2\varepsilon}\barred{y}^2 - \frac{\pi(k_1+k_2)}{4\sqrt{2}\ln^2\varepsilon}\barred{y}\sum_{n\in\mathds{Z}}\widetilde{\alpha}_n e^{-\frac{\pi i n}{2}}e^{-\frac{\pi n \barred{y}}{2\ln\varepsilon}} 
\\ \notag & \hspace{15pt} + \frac{\pi^2}{8\ln^2\varepsilon}\sum_{n,m\in\mathds{Z}}\circnormal{\widetilde{\alpha}_n\widetilde{\alpha}_m}e^{-\frac{\pi i(n+m)}{2}}e^{-\frac{\pi(n+m)\barred{y}}{2\ln\varepsilon}} - \frac{\pi^2}{96\ln^2\varepsilon},
\end{align}
where we have introduced the oscillator zero-modes
\begin{align}
\alpha_0 & \equiv \sqrt{2}\left(p - \frac{i(k_1 - k_2)}{2\pi}\ln\varepsilon\right)
\\ \widetilde{\alpha}_0 & \equiv \sqrt{2}\left(p + \frac{i(k_1 - k_2)}{2\pi}\ln\varepsilon\right)
\end{align}
and defined $\widetilde{\alpha}_n = \alpha_{- n}$ for all $n\neq 0$. The finite-regulator Rindler Hamiltonian \eqref{NN Hamiltonian integral} computed from the above mode expansions is
\begin{equation}\label{NN Hamiltonian answer}
H_{\text{R}}^{\text{NN}} = -\frac{\pi p^2}{2\ln\varepsilon} - \frac{i(k_1+k_2)}{2\pi}x - \frac{\pi}{2\ln\varepsilon}\sum_{n=1}^{\infty}\alpha_{-n}\alpha_n - \frac{(k_1 - k_2)^2}{8\pi}\ln\varepsilon - \frac{(k_1+k_2)^2}{24\pi}\ln\varepsilon + \frac{\pi}{48\ln\varepsilon}.
\end{equation}
This result is manifestly independent of time because the original expansion \eqref{NN noncompact strip expansion} was chosen to obey the equation of motion, i.e.~the time-dependent operators $x(y_1) \equiv x - \frac{\pi i p(y_1)}{\ln\varepsilon}y_1 - \frac{i(k_1+k_2)}{4\ln\varepsilon}y_1^2$, $p(y_1) \equiv p - \frac{k_1+k_2}{2\pi}y_1$ and $\alpha_n(y_1)\equiv \alpha_n e^{-\frac{\pi n}{2\ln\varepsilon}y_1}$ automatically obey the Heisenberg equation\footnote{As a reminder, the Euclidean Heisenberg equation is $\partial_{\tau}\mathcal{O} = [H,\mathcal{O}]$, and the sign change is due to our time convention $\dot{\mathcal{O}} \equiv \partial_{y_1}\mathcal{O} = -\partial_{\tau}\mathcal{O}$.} $\dot{\mathcal{O}} = -[H_{\text{R}}^{\text{NN}},\mathcal{O}]$; the boundary terms arising from the classical trace of the stress-energy tensor were needed to obtain this simple result. Therefore, the evolution operator around one full $y_1 \sim y_1 - 2\pi$ cycle on the cylinder is the ordinary exponential given by
\begin{align}\label{NN evolution operator}
e^{-2\pi H_{\text{R}}^{\text{NN}}} & = e^{i(k_1+k_2)x}e^{\frac{\pi^2}{\ln\varepsilon}\left(p + \frac{k_1+k_2}{2}\right)^2}e^{\frac{\pi^2}{\ln\varepsilon}\sum_{n=1}^{\infty}\alpha_{-n}\alpha_n}e^{\frac{(k_1-k_2)^2}{4}\ln\varepsilon + \frac{(k_1+k_2)^2}{12}\ln\varepsilon + \frac{\pi^2(k_1+k_2)^2}{12\ln\varepsilon} - \frac{\pi^2}{24\ln\varepsilon}},
\end{align}
where we split the exponentials using the Zassenhaus form of the Baker-Campbell-Hausdorff formula. Note that $x$ appears in the Hamiltonian itself and is hence not a zero-mode in the strict sense of the word; more accurately it is a zero-momentum mode, so we shall refrain from using the conflicting term ``zero-mode'' as much as possible. Recall that in radial quantization of a unitary theory, real operators obey reflection positivity; see, for instance, \cite{Simmons-Duffin16}. Here, however, the free boson is defined on a complexified contour, so the appropriate notion of Hermiticity in angular quantization is
\begin{equation}
X(y_1,y_2)^{\dagger}_{k_1,k_2} = X(-y_1,y_2)_{-k_1,-k_2},
\end{equation}
where we have labeled its dependence on the boundary conditions explicitly, from which it follows that $x^{\dagger} = x$, $p^{\dagger} = p$ and $\alpha_n^{\dagger} = \alpha_{-n}$ as expected. The Rindler Hamiltonian \eqref{NN Hamiltonian answer} then also obeys
\begin{equation}
H_{\text{R}}^{\text{NN}}(k_1,k_2)^{\dagger} = H_{\text{R}}^{\text{NN}}(-k_1,-k_2).
\end{equation}
One may formally work with purely imaginary momenta and say that the free boson and the Rindler Hamiltonian are ordinary Hermitian operators in this analytically continued theory. Note that this is actually the situation that will arise in later applications to winding operators in certain theories with a linear dilaton.

For real Lorentzian momenta, note that the vertex operator $\normal{e^{i k X}}$ is itself non-Hermitian. Moreover, the only non-Hermitian part of the Hamiltonian is the constant-mode term $-\frac{i(k_1+k_2)}{2\pi}x$, whose role is to provide the correct momentum-conserving delta functions when computing thermal traces, as apparent from the appearance of $e^{i(k_1+k_2)x}$ in the evolution operator \eqref{NN evolution operator}. The non-invariance of the Hamiltonian under the shift symmetry of the free boson is not problematic and in fact is required, as it encapsulates the effects of endpoint operators which do transform nontrivially under this symmetry.

The finite-regulator angular quantization Hilbert space $\mathcal{H}_{k_1 k_2}^{\text{NN}}$ is defined as the Fock space built on top of the oscillator vacuum states $|p;\{0\}\rangle$ defined by $\alpha_n |p;\{0\}\rangle = 0$ for all $n > 0$ and $p|p;\{0\}\rangle = p|p;\{0\}\rangle$; since we shall always work in the momentum basis for the center-of-mass modes, we abuse notation and do not make a distinction between the momentum operator and its eigenvalue. Explicitly, the states of $\mathcal{H}_{k_1 k_2}^{\text{NN}}$ may be written as
\begin{equation}
|p;\{N_n\}\rangle \equiv \left(\prod_{n=1}^{\infty}\frac{1}{\sqrt{n^{N_n}N_n!}}\alpha_{-n}^{N_n}\right)|p;\{0\}\rangle
\end{equation}
where $N_n \geqslant 0$ is the integer occupation number of the $n^{\text{th}}$ mode, which are Dirac orthonormalized in the sense
\begin{equation}
\left\langle p'; \{N_n'\}| p;\{N_n\}\right\rangle = 2\pi \delta_{\{N_n'\},\{N_n\}}\delta(p'-p).
\end{equation}
This basis of $\mathcal{H}_{k_1 k_2}^{\text{NN}}$ is not an eigenbasis of the Rindler Hamiltonian \eqref{NN Hamiltonian answer} due to the fact that $x$ is not a zero-mode for $k_1 + k_2 \neq 0$, but we only ever need to compute traces over the angular quantization Hilbert space, which may be performed in any Dirac-orthonormal basis.

The finite-regulator cylinder thermal trace over \eqref{NN evolution operator} is straightforwardly computed as
\begin{equation}
\mathrm{Tr}_{\mathcal{H}_{k_1 k_2}^{\text{NN}}}\left[e^{-2\pi H_{\text{R}}^{\text{NN}}}\right]_{S^1\times \mathds{R}} = \frac{\delta(k_1+k_2)}{\sqrt{2}\s\s\eta\hspace{-2pt}\left(-\frac{2i\ln\varepsilon}{\pi}\right)}e^{\frac{\ln\varepsilon}{4}(k_1-k_2)^2 + \frac{\ln\varepsilon}{12}(k_1+k_2)^2 + \frac{\pi^2}{12\ln\varepsilon}(k_1+k_2)^2},
\end{equation}
where $\eta(\tau)$ is the Dedekind eta function (see \hyperref[formulas]{Appendix \ref{formulas}} for relevant formulas) and the delta function arose from $\langle p|e^{i(k_1+k_2)x}|p\rangle = \langle p - k_1 - k_2|p\rangle = 2\pi\delta(k_1+k_2)$. The dependence in the exponential on $(k_1+k_2)^2$ then disappears on the support of the delta function. The result in the sphere conformal frame is simply obtained by multiplying by the Weyl transformation factor $\varepsilon^{1/6}$ from \eqref{Weyl transformation}. Therefore, since the dimensions of the endpoint operators are $\Delta_1 = \frac{k_1^2}{2}$ and $\Delta_2 = \frac{k_2^2}{2}$ and we may write $\frac{1}{4}(k_1 - k_2)^2 = \Delta_1 + \Delta_2$ on the support of the delta function, the finite-regulator thermal trace on the sphere is
\begin{equation}
\mathrm{Tr}_{\mathcal{H}_{k_1 k_2}^{\text{NN}}}\left[e^{-2\pi H_{\text{R}}^{\text{NN}}}\right]_{S^2} = \frac{\delta(k_1+k_2)}{\sqrt{2}}\frac{\varepsilon^{\Delta_1+\Delta_2}}{\varepsilon^{-1/6}\s\s\eta\hspace{-2pt}\left(-\frac{2i\ln\varepsilon}{\pi}\right)}.
\end{equation}
We have written the result in this way because of the limit
\begin{equation}
\lim_{\varepsilon\rightarrow 0}\left[\varepsilon^{-1/6}\s\s\eta\hspace{-2pt}\left(-\frac{2i\ln\varepsilon}{\pi}\right)\right] = 1.
\end{equation}
Hence, we have shown that the angular quantization two-point function matching condition \eqref{2-point matching} is indeed satisfied, namely
\begin{equation}
\lim_{\varepsilon\rightarrow 0}\left(\varepsilon^{-\Delta_1-\Delta_2}\text{Tr}_{\mathcal{H}_{k_1 k_2}^{\text{NN}}}\left[e^{-2\pi H_{\text{R}}^{\text{NN}}}\right]_{S^2}\right) = 2\pi \mathcal{N}_{k_1 k_2}^{\text{NN}}\delta(k_1+k_2),
\end{equation}
with the normalization constant
\begin{equation}\label{NN normalization}
\mathcal{N}_{k_1 k_2}^{\text{NN}} = \frac{1}{2\sqrt{2}\pi}.
\end{equation}
In principle, the normalization constant appearing in \eqref{3-point matching} could be this result times $f(k_1 + k_2)$, where $f(0) = 1$; the factorization of $\mathcal{N}_{k_1 k_2}^{\text{NN}}$ still allows a nontrivial function $f(k_1 + k_2)$ given, for instance, by an exponential. The exact constant $\mathcal{N}_{k_1 k_2}^{\text{NN}}$ as a function of $k_1$ and $k_2$ is only uniquely determined by matching the thermal one-point functions computed in the following subsection; we shall find that \eqref{NN normalization} is the full answer.

\subsubsection{Dirichlet-like Boundary Conditions}

Consider the Dirichlet-like boundary conditions \eqref{D noncompact boson} on the very wide strip. These boundary conditions by themselves shrink to the identity operator. To obtain the endpoint operators $\normal{e^{ik_1 X(0)}}$ and $\normal{e^{ik_2 X(\infty)}}$ we add appropriate factors of $ikX$ to the boundary action, which will dynamically fix the operator charges. 

The free boson action which imposes the Dirichlet-like boundary conditions that shrink to $\normal{e^{ik_1 X(0)}}$ and $\normal{e^{ik_2 X(\infty)}}$ is then given by
\begin{multline}\label{DD strip action}
S = \frac{1}{4\pi}\int dy_1 dy_2 \left[\left(\partial_{y_1} X\right)^2 + \left(\partial_{y_2} X\right)^2\right] 
\\ - \frac{i}{2\pi}\int_{y_2 = \ln\varepsilon}\hspace{-8pt}dy_1\left(k_1 X - \lambda_1\partial_{y_1}X\right) - \frac{i}{2\pi}\int_{y_2 = -\ln\varepsilon}\hspace{-10pt}dy_1\left(k_2 X + \lambda_2 \partial_{y_1}X\right),
\end{multline}
where $\lambda_1(y_1)$ and $\lambda_2(y_1)$ are boundary Lagrange multipliers imposing the vanishing of $\partial_{y_1}X$ there. In addition to the bulk equation of motion $(\partial_{y_1}^2 + \partial_{y_2}^2)X = 0$ and boundary conditions $\partial_{y_1}X(y_2 = \mp \ln\varepsilon) = 0$, the equations of motion for the boundary Lagrange multipliers are
\begin{align}
\partial_{y_1}\lambda_1 & = i\partial_{y_2}X\Big|_{y_2 = +\ln\varepsilon} - k_1
\\ \partial_{y_1}\lambda_2 & = i\partial_{y_2}X\Big|_{y_2 = -\ln\varepsilon} + k_2.
\end{align}
The stress-energy tensor for this theory is
\begin{multline}
T_{\alpha\beta} = -\normal{\left(\partial_{\alpha}X\partial_{\beta}X - \frac{1}{2}\delta_{\alpha\beta}\partial_{\gamma}X\partial^{\gamma}X\right)} - t_{\alpha}t_{\beta} X\left[ik_1 \delta(y_2 - \ln\varepsilon) + ik_2\delta(y_2 + \ln\varepsilon)\right]
\\ - t_{(\alpha}n_{\beta)}\partial_{y_2}X\left[i\lambda_1\delta(y_2 - \ln\varepsilon) + i\lambda_2\delta(y_2 + \ln\varepsilon)\right],
\end{multline}
where $t^{\alpha}$ and $n^{\alpha}$ are the unit tangent and normal vectors to the boundary, respectively. Despite having different boundary conditions and a different form of the boundary stress-energy tensor, the Dirichlet-like theory has the same classical trace as the Neumann-like theory, namely
\begin{equation}
T^{\alpha}_{\phantom{\alpha}\alpha}(y_1,y_2) = -X(y_1,y_2)\left[ik_1\delta(y_2 - \ln\varepsilon) + ik_2\delta(y_2 + \ln\varepsilon)\right],
\end{equation}
as well as the same expression which computes the Rindler Hamiltonian, namely
\begin{equation}\label{DD Hamiltonian integral}
H_{\text{R}}^{\text{DD}}(y_1) = \frac{1}{2\pi}\int_{\ln\varepsilon}^{-\ln\varepsilon}dy_2\left[T(y) + \widetilde{T}(\barred{y})\right] - \frac{1}{2\pi}\left[ik_1 X(y_1,y_2 = \ln\varepsilon) + ik_2 X(y_1,y_2 = -\ln\varepsilon)\right].
\end{equation}
Now, however, the momentum density canonically conjugate to $X$ is 
\begin{equation}
\Pi(y_1,y_2) = -\frac{i}{2\pi}\partial_{y_1}X(y_1,y_2) + \frac{1}{2\pi}\left[\lambda_1(y_1)\delta(y_2 - \ln\varepsilon) - \lambda_2(y_1)\delta(y_2 + \ln\varepsilon)\right].
\end{equation}
The Dirichlet-like free boson $X$ has two continuous degrees of freedom, a spatially constant mode and a spatially linear mode, and the Lagrange multipliers $\lambda_1$ and $\lambda_2$ each have a continuous constant mode. We shall write the non-oscillator part of the $X$ expansion as $x  + (x_{\text{s}}-\frac{i(k_1-k_2)}{2})y_2$, calling $x$ the constant mode and $x_{\text{s}}$ the ``slope'' mode; the definition of the slope mode with the constant shift $-\frac{i(k_1-k_2)}{2}$ is only for later convenience, but its presence mimics the same term in the Neumann-like expansion \eqref{NN noncompact strip expansion}. Then, we see from the above form of $\Pi$ that the constant modes of $\lambda_1$ and $\lambda_2$ should be linear combinations of the conjugate momenta of $x$ and $x_{\text{s}}$. Specifically, mode expanding $X$, $\lambda_1$ and $\lambda_2$ on the strip and imposing the equal-time canonical commutation relations\footnote{Strictly speaking, since $\lambda_1$ and $\lambda_2$ do not have canonical conjugates, one should more properly perform Dirac quantization with the constraints that $\Pi_{\lambda_1}$ and $\Pi_{\lambda_2}$ both vanish and that $\lambda_1$ and $\lambda_2$ are proportional to $\Pi_X(y_2 = \ln\varepsilon)$ and $\Pi_X(y_2 = -\ln\varepsilon)$, respectively. One may explicitly check that the results of Dirac quantization are identical to those of canonical quantization here, essentially because the constraints are all compatible with a truncation of the original Poisson algebra.} between $X$ and $\Pi$ results in
\begin{align}\label{DD noncompact strip expansion}
X(y_1,y_2) & = x + \left(x_{\text{s}} - \frac{i(k_1 - k_2)}{2}\right)y_2 - \sqrt{2}\sum_{\substack{n \in \mathds{Z} \\ n\neq 0}}\frac{\alpha_n}{n}e^{-\frac{\pi n y_1}{2\ln\varepsilon}}\sin\left(\frac{\pi n(y_2-\ln\varepsilon)}{2\ln\varepsilon}\right)
\\ \lambda_1(y_1) & = +\pi\left(p + \frac{ p_{\text{s}}}{\ln\varepsilon}\right) + i\left(x_{\text{s}} + \frac{i(k_1+k_2)}{2}\right)y_1 + i\sqrt{2}\sum_{\substack{n \in \mathds{Z} \\ n\neq 0}}\frac{\alpha_n}{n}e^{-\frac{\pi n y_1}{2\ln\varepsilon}}
\\ \lambda_2(y_2) & = -\pi\left(p - \frac{ p_{\text{s}}}{\ln\varepsilon}\right) + i\left(x_{\text{s}} - \frac{i(k_1+k_2)}{2}\right)y_1 + i\sqrt{2}\sum_{\substack{n \in \mathds{Z} \\ n\neq 0}}\frac{\alpha_n}{n}(-1)^n e^{-\frac{\pi n y_1}{2\ln\varepsilon}},
\end{align}
where the nonvanishing commutators are\footnote{To obtain this result, the Fourier expansions
\begin{equation*}
1 \pm \frac{y_2}{\ln\varepsilon} = \begin{cases}
1 \pm 1 & \text{if} \ y_2 = \ln\varepsilon 
\\ \mp \frac{2}{\pi}\sum_{n\neq 0}\frac{(\pm 1)^n}{n}\sin\left(\frac{\pi n(y_2 - \ln\varepsilon)}{2\ln\varepsilon}\right) & \text{if} \ \ln\varepsilon < y_2 < -\ln\varepsilon
\\ 1 \mp 1 & \text{if} \ y_2 = -\ln\varepsilon
\end{cases}
\end{equation*}
are useful.}
\begin{align}
[x,p] & = [x_{\text{s}},p_{\text{s}}] = i
\\ [\alpha_n,\alpha_m] & = n\delta_{n,-m}.
\end{align}
As usual, we are free to define the operators in the mode expansion of $X$ however we wish, and we may make any operator redefinitions that preserve the commutators. With these definitions, the Heisenberg time evolutions of the operators $p$ and $p_{\text{s}}$ are $p(y_1) = p - \frac{k_1+k_2}{2\pi}y_1$ and $p_{\text{s}}(y_1) = p_{\text{s}} + \frac{ix_{\text{s}}}{\pi}y_1\ln\varepsilon$, which are indeed the total target space momentum and first moment thereof,
\begin{align}
\int_{\ln\varepsilon}^{-\ln\varepsilon}dy_2 \ \Pi(y_1,y_2) & = p - \frac{k_1+k_2}{2\pi}y_1
\\ \int_{\ln\varepsilon}^{-\ln\varepsilon}dy_2 \ y_2\Pi(y_1,y_2) & = p_{\text{s}} + \frac{ix_{\text{s}}}{\pi}y_1\ln\varepsilon.
\end{align}
The constant mode $x$, on the other hand, is no longer the target space center-of-mass, which now involves the oscillators as well. There is no point separating out this oscillator contribution, as the constant mode $x$ will become the center-of-mass in the shrinking limit $\varepsilon\rightarrow 0$. In the special case $k_1 = k_2 = 0$, $(-2\ln\varepsilon)x_{\text{s}}$ is the distance between the string endpoints, so $-\frac{1}{2\ln\varepsilon}p_{\text{s}}$ is the momentum of one endpoint relative to the other.

Hermitian conjugation works as before, where now in addition $\lambda_1(y_1)_{k_1,k_2}^{\dagger} = \lambda_1(-y_1)_{-k_1,-k_2}$ and $\lambda_2(y_1)_{k_1,k_2}^{\dagger} = \lambda_2(-y_1)_{-k_1,-k_2}$; formally, one may again consider the case of pure imaginary $k_1$ and $k_2$ for which the target space contour is real and Hermiticity behaves normally. Therefore, $x$, $p$, $x_{\text{s}}$ and $p_{\text{s}}$ are all Hermitian, and $\alpha_n^{\dagger} = \alpha_{-n}$. Note moreover that there exists a Euclidean saddle point on the cylinder if and only if $k_1+k_2 = 0$, where the classical saddle point obeys $x_{\text{s}} = \alpha_n = 0$ with $p$ and $p_{\text{s}}$ arbitrary; we have defined the slope mode of $X$ with the shift $-\frac{i(k_1-k_2)}{2}y_2$ precisely so that the classical saddle point occurs at $x_{\text{s}} = 0$.

The bulk stress-energy tensors on the strip are 
\begin{align}
T(y) & = \frac{\pi^2}{8\ln^2\varepsilon}\sum_{n,m\in\mathds{Z}}\circnormal{\alpha_n\alpha_m}e^{\frac{\pi i(n+m)}{2}}e^{-\frac{\pi(n+m)y}{2\ln\varepsilon}} - \frac{\pi^2}{96\ln^2\varepsilon}
\\ \widetilde{T}(\barred{y}) & = \frac{\pi^2}{8\ln^2\varepsilon}\sum_{n,m\in\mathds{Z}}\circnormal{\alpha_n\alpha_m}e^{-\frac{\pi i(n+m)}{2}}e^{-\frac{\pi(n+m)\barred{y}}{2\ln\varepsilon}} - \frac{\pi^2}{96\ln^2\varepsilon},
\end{align}
where we have defined
\begin{equation}
\alpha_0 \equiv -\frac{\sqrt{2}}{\pi}\left(x_{\text{s}} - \frac{i(k_1-k_2)}{2}\right)\ln\varepsilon;
\end{equation}
by convention we shall also group $x$ and $p_{\text{s}}$ with the creation operators and $p$ and $x_{\text{s}}$ with the annihilation operators. Therefore, the Rindler Hamiltonian on the very wide strip associated to Dirichlet-like boundary conditions on both ends, as computed by \eqref{DD Hamiltonian integral}, is
\begin{equation}\label{DD Hamiltonian answer}
H_{\text{R}}^{\text{DD}} = -\frac{i(k_1+k_2)}{2\pi}x - \frac{x_{\text{s}}^2}{2\pi}\ln\varepsilon - \frac{\pi}{2\ln\varepsilon}\sum_{n=1}^{\infty}\alpha_{-n}\alpha_n - \frac{(k_1-k_2)^2}{8\pi}\ln\varepsilon + \frac{\pi}{48\ln\varepsilon}.
\end{equation}
This Hamiltonian is time independent, and moreover the evolution operator $e^{\Delta y_1 H_{\text{R}}^{\text{DD}}}$ is trivially split as the product of the individual exponentials since all the terms in \eqref{DD Hamiltonian answer} mutually commute. The finite-regulator angular quantization Hilbert space $\mathcal{H}_{k_1 k_2}^{\text{DD}}$ on the strip thus consists of the tensor product of the oscillator Fock space and the two continuum modes; we write the general state as $|p,x_{\text{s}};\{N_n\}\rangle$, taking the momentum basis for $(x,p)$ and the position basis for $(x_{\text{s}},p_{\text{s}})$, with conventional normalization
\begin{equation}
\left\langle p',x_{\text{s}}';\{N_n'\}| p,x_{\text{s}};\{N_n\}\right\rangle = 2\pi \delta_{\{N_n'\},\{N_n\}}\delta(p'-p)\delta(x_{\text{s}}' - x_{\text{s}}).
\end{equation}
Note that $p$ and $p_{\text{s}}$ are true zero-modes of the Hamiltonian \eqref{DD Hamiltonian answer}, reflecting the conservation of $x$ and $x_{\text{s}}$. The Hermitian conjugate again obeys $H_{\text{R}}^{\text{DD}}(k_1,k_2)^{\dagger} = H_{\text{R}}^{\text{DD}}(-k_1,-k_2)$.

The next step is to compute the thermal trace and thermal one-point functions on the cylinder and transform them back to the sphere. However, we must pause to consider the path integral on the cylinder with the action \eqref{DD strip action}, which is what the angular quantization trace $\text{Tr}_{\mathcal{H}_{k_1 k_2}^{\text{DD}}}[e^{-2\pi H_{\text{R}}^{\text{DD}}}]$ is computing. The cylinder action \eqref{DD strip action} has two independent symmetries corresponding to constant shifts of the Lagrange multipliers, since $\partial_{y_1}X$ integrates along the boundary to zero. These shift symmetries would cause the path integral to diverge if left untreated. However, it is a familiar fact that any path integral with a noncompact symmetry must be divided by the volume of the symmetry group to render the partition function well-defined and finite\footnote{The most notable exception is for the constant mode of the noncompact free boson itself, which is usually left in so as to keep track of the spacetime volume dependence of quantities. Nevertheless, even for the free boson the constant mode of the path integral must be factored out, and correlators are normalized by dividing by the partition function with the zero-mode absorbed, as we have done previously.}. Note that this divergence is not present for the path integral on the strip, as the integrals over the zero-modes of $\lambda_1$ and $\lambda_2$ in that case just produce delta functions equating the boundary values of $X$ in the infinite future and the infinite past. Thus the treatment of the path integral divergence on the cylinder should be interpreted as how to define the trace in angular quantization. As such, the zero-mode sector of the trace in angular quantization must always be divided by the zero-mode volume factor
\begin{equation}
\mathrm{Vol}(\lambda_{1,0},\lambda_{2,0}) = \int_{-\infty}^{\infty}d\lambda_{1,0}\int_{-\infty}^{\infty}d\lambda_{2,0} = -\frac{8\pi^4}{\ln\varepsilon}\int_{-\infty}^{\infty}\frac{dp}{2\pi}\int_{-\infty}^{\infty}\frac{dp_{\text{s}}}{2\pi}.
\end{equation}
This infinite division also makes sense from the point of view of the angular quantization Hamiltonian \eqref{DD Hamiltonian answer}, where it is obvious that $p$ and $p_{\text{s}}$ are true continuous zero-modes, making all traces proportional to an overall unphysical infinity\footnote{Note that the Neumann-like Hamiltonian \eqref{NN Hamiltonian answer} did not share this property, as in that case $x$ is a true zero-mode if and only if $k_1 + k_2 = 0$, which is why all those traces are proportional to $2\pi\delta(k_1+k_2)$, supplying the infinity only for the Hilbert space with the actual zero-mode.}. For example, the zero-mode part of the trace of the cylinder evolution operator is
\begin{align}
\text{Tr}_{(p,x_{\text{s}})}\hspace{-3pt}\left[e^{-2\pi H_{\text{R}}^{\text{DD}}}\right] & = \frac{1}{\mathrm{Vol}(\lambda_{1,0},\lambda_{2,0})}\hspace{-2pt}\int_{-\infty}^{\infty}\hspace{-2pt}\frac{dp}{2\pi}\langle p|e^{i(k_1+k_2)x}|p\rangle \hspace{-2pt}\int_{-\infty}^{\infty}\hspace{-3pt}dx_{\text{s}}\langle x_{\text{s}}|e^{\left(x_{\text{s}}^2 + \frac{(k_1-k_2)^2}{4}\right)\ln\varepsilon}|x_{\text{s}}\rangle 
\\ & = -\frac{\ln\varepsilon}{4\pi^3}\delta(k_1+k_2)\frac{\int_{-\infty}^{\infty}\frac{dp}{2\pi}}{\int_{-\infty}^{\infty}\frac{dp}{2\pi}}\frac{\langle x_{\text{s}}|x_{\text{s}}\rangle}{\int_{-\infty}^{\infty}\frac{dp_{\text{s}}}{2\pi}}\sqrt{-\frac{\pi}{\ln\varepsilon}} \ e^{\frac{(k_1 - k_2)^2}{4}\ln\varepsilon}
\\ & = \frac{\delta(k_1+k_2)}{4\pi^3}\sqrt{-\pi\ln\varepsilon}\ \varepsilon^{(k_1-k_2)^2/4}.
\end{align}
The full finite-regulator cylinder thermal trace is therefore
\begin{equation}
\text{Tr}_{\mathcal{H}_{k_1 k_2}^{\text{DD}}}\left[e^{-2\pi H_{\text{R}}^{\text{DD}}}\right]_{S^1\times\mathds{R}} = \frac{\delta(k_1+k_2)}{4\sqrt{2}\pi^2}\frac{\varepsilon^{k_2^2}}{\eta\hspace{-2pt}\left(-\frac{2i\ln\varepsilon}{\pi}\right)}.
\end{equation}
Transforming back to the sphere again requires the Weyl transformation factor $\varepsilon^{1/6}$ from \eqref{Weyl transformation}, and therefore the finite-regulator thermal trace on the sphere is
\begin{equation}
\text{Tr}_{\mathcal{H}_{k_1 k_2}^{\text{DD}}}\left[e^{-2\pi H_{\text{R}}^{\text{DD}}}\right]_{S^2} = \frac{\delta(k_1+k_2)}{4\sqrt{2}\pi^2}\frac{\varepsilon^{\Delta_1+\Delta_2}}{\varepsilon^{-1/6}\eta\hspace{-2pt}\left(-\frac{2i\ln\varepsilon}{\pi}\right)}.
\end{equation}
Thus, the angular quantization two-point function matching condition \eqref{2-point matching} is also satisfied with these boundary conditions, namely
\begin{equation}
\lim_{\varepsilon\rightarrow 0}\left(\varepsilon^{-\Delta_1-\Delta_2}\text{Tr}_{\mathcal{H}_{k_1 k_2}^{\text{DD}}}\left[e^{-2\pi H_{\text{R}}^{\text{DD}}}\right]_{S^2}\right) = 2\pi \mathcal{N}_{k_1 k_2}^{\text{DD}}\delta(k_1+k_2),
\end{equation}
with the normalization constant
\begin{equation}\label{DD normalization}
\mathcal{N}_{k_1 k_2}^{\text{DD}} = \frac{\sqrt{2}\pi}{(2\pi)^4}.
\end{equation}
As before, matching the two-point function is not sensitive to an overall function $f(k_1+k_2)$ with $f(0) = 1$ multiplying this normalization, which can only be completely determined by matching the thermal one-point functions; we shall do this computation in the following subsection and find that \eqref{DD normalization} is the full answer.

\subsection{Thermal One-Point Function}

The last necessary computation is the matching condition \eqref{3-point matching} for the thermal one-point functions of bulk primary insertions. The angular quantization Hilbert spaces $\mathcal{H}_{k_1 k_2}^{\text{NN}}$ and $\mathcal{H}_{k_1 k_2}^{\text{DD}}$ with respective Rindler Hamiltonians $H_{\text{R}}^{\text{NN}}$ and $H_{\text{R}}^{\text{DD}}$ were derived explicitly in the previous subsection, so it remains only to compute the mode expansions of the exponential primaries $\mathcal{O}_{k_3}(z,\barred{z}) = \s\s\s\normal{e^{ik_3 X(z,\barred{z})}}$ in order to evaluate the traces. The mode expansions are most easily obtained by relating the point-split normal-ordered product $\normal{X(z_1,\barred{z}_1)X(z_2,\barred{z}_2)}$ to the creation/annihilation normal-ordered product $\circnormal{X(z_1,\barred{z}_1)X(z_2,\barred{z}_2)}$, taking the coincident limit and using Wick's Theorem to exponentiate this result. We perform these calculations directly on the two-holed sphere, since we already know that the evolution operator on the sphere differs from that on the cylinder by the Weyl factor $\varepsilon^{1/6}$ in both cases.

\subsubsection{Neumann-like Boundary Conditions}

Since $X$ transforms with weights $(h,\widetilde{h}) = (0,0)$, the mode expansion $X(z,\barred{z})$ of the free boson on the two-holed sphere is immediately obtained from the mode expansion \eqref{NN noncompact strip expansion} by setting $y = i\ln z$. The relation between conformal and creation/annihilation normal ordering on the plane is then straightforwardly computed as
\begin{multline}
\normal{X(z_1,\barred{z}_1)X(z_2,\barred{z}_2)} = \s\circnormal{X(z_1,\barred{z}_1)X(z_2,\barred{z}_2)} 
\\ - \frac{1}{2}\ln\left|\left[\frac{z_1^{-\pi i/2\ln\varepsilon} - z_2^{-\pi i/2\ln\varepsilon}}{(z_1 z_2)^{-\pi i/2\ln\varepsilon} z_{12}}\right]\left[\frac{1}{z_1^{-\pi i/2\ln\varepsilon}} + \barred{z}_2^{-\pi i/2\ln\varepsilon}\right]\right|^2.
\end{multline}
Taking the coincident limit and using Wick's Theorem, the relation between the conformal and creation/annihilation normal-ordered exponential operator on the plane is
\begin{equation}\label{NN exponential operator}
\normal{e^{ik_3 X(z,\barred{z})}} = \left[\frac{\pi^2}{\ln^2\varepsilon}\left(\frac{z}{\barred{z}}\right)^{\pi i/\ln\varepsilon}\frac{1}{|z|^2}\cos^2\left(\frac{\pi \ln|z|}{2\ln\varepsilon}\right)\right]^{k_3^2/4}\circnormal{e^{ik_3 X(z,\barred{z})}},
\end{equation}
where the creation/annihilation normal-ordered exponential is
\begin{align}\label{NN exponential mode expansion}
\circnormal{e^{ik_3 X(z,\barred{z})}} & = \exp\left[-\frac{(k_1+k_2)k_3}{12}\ln\varepsilon + \frac{(k_1-k_2)k_3}{2}\ln|z| + \frac{(k_1+k_2)k_3}{8\ln\varepsilon}\left(\ln^2 z + \ln^2\barred{z}\right)\right]
\\ \notag & \hspace{5pt} \times \exp\left[ik_3 x + \sqrt{2}k_3\sum_{n=1}^{\infty}\frac{\alpha_{-n}}{n}\left(\frac{z}{\barred{z}}\right)^{\pi i n/4\ln\varepsilon}\cos\left(\frac{\pi n}{2\ln\varepsilon}\ln\left|\frac{z}{\varepsilon}\right|\right)\right]
\\ \notag & \hspace{5pt} \times \exp\left[\frac{\pi i k_3 p}{2\ln\varepsilon}\ln\left(\frac{z}{\barred{z}}\right) - \sqrt{2}k_3\sum_{m=1}^{\infty}\frac{\alpha_m}{m}\left(\frac{z}{\barred{z}}\right)^{-\pi i m/4\ln\varepsilon}\cos\left(\frac{\pi m}{2\ln\varepsilon}\ln\left|\frac{z}{\varepsilon}\right|\right)\right].
\end{align}
It is now a straightforward, albeit slightly tedious, exercise to compute the finite-regulator thermal trace with this mode expansion inserted, using the cylinder evolution operator \eqref{NN evolution operator} and the Weyl transformation back to the sphere. The $(x,p)$ sector in the trace is still simple, and the details of the oscillator calculation are provided in \hyperref[formulas]{Appendix \ref{formulas}}; the final result is
\begin{multline}
\text{Tr}_{\mathcal{H}_{k_1 k_2}^{\text{NN}}}\hspace{-2pt}\left[\mathcal{O}_{k_3}(z,\barred{z})e^{-2\pi H_{\text{R}}^{\text{NN}}}\right]_{S^2}
\\ = \frac{\delta(k_1\hspace{-1pt}+\hspace{-1pt}k_2\hspace{-1pt}+\hspace{-1pt}k_3)}{\sqrt{2}\varepsilon^{-\frac{1}{6}}\eta\hspace{-2pt}\left(\hspace{-1pt}-\frac{2i\ln\varepsilon}{\pi}\right)}\frac{\varepsilon^{(k_1^2+k_2^2)/2}}{|z|^{(k_1^2 + k_3^2 - k_2^2)/2}}\hspace{-2pt}\left[\varepsilon^{-\frac{1}{6}}\eta\hspace{-2pt}\left(\hspace{-2pt}-\frac{2i\ln\varepsilon}{\pi}\right)\hspace{-2pt}\vartheta_4\hspace{-2pt}\left(\left.\hspace{-3pt}-\frac{i\ln|z|}{\pi}\right|\hspace{-2pt}-\hspace{-2pt}\frac{2i\ln\varepsilon}{\pi}\hspace{-1pt}\right)\hspace{-1pt}\right]^{k_3^2/2},
\end{multline}
where $\vartheta_4(\nu|\tau)$ is the fourth Jacobi elliptic theta function, obeying $\lim_{\tau\rightarrow i\infty}\vartheta_4(\nu|\tau) = 1$. Since the three-point coefficient of exponential primaries is $C_{k_1 k_2 k_3} = 2\pi \delta(k_1 + k_2 + k_3)$, we indeed find the desired matching condition \eqref{3-point matching}, namely
\begin{equation}
\lim_{\varepsilon\rightarrow 0}\left(\varepsilon^{-\frac{k_1^2}{2}-\frac{k_2^2}{2}}\text{Tr}_{\mathcal{H}_{k_1 k_2}^{\text{NN}}}\left[\mathcal{O}_{k_3}(z,\barred{z})e^{-2\pi H_{\text{R}}^{\text{NN}}}\right]_{S^2}\right) = \frac{\mathcal{N}_{k_1 k_2}^{\text{NN}}C_{k_1 k_3 k_2}}{|z|^{(k_1^2 + k_3^2 - k_2^2)/2}},
\end{equation}
with the same normalization constant \eqref{NN normalization}.

\subsubsection{Dirichlet-like Boundary Conditions}

From the mode expansion \eqref{DD noncompact strip expansion}, the relation between conformal and creation/annihilation normal ordering on the plane is 
\begin{equation}
\normal{X(z_1,\barred{z}_1)X(z_2,\barred{z}_2)} = \s\circnormal{X(z_1,\barred{z}_1)X(z_2,\barred{z}_2)} - \frac{1}{2}\ln\left|\frac{1-(z_1/z_2)^{-\pi i/2\ln\varepsilon}}{[1+(z_1\barred{z}_2)^{-\pi i/2\ln\varepsilon}]z_{12}}\right|^2,
\end{equation}
which gives the mode expansion of the bulk exponential primary as
\begin{equation}\label{DD exponential operator}
\normal{e^{ik_3 X(z,\barred{z})}} = \left[\left(-\frac{\pi}{4\ln\varepsilon}\right)\frac{1}{|z|\cos\left(\frac{\pi\ln|z|}{2\ln\varepsilon}\right)}\right]^{k_3^2/2}\circnormal{e^{ik_3 X(z,\barred{z})}},
\end{equation}
where
\begin{multline}\label{DD exponential mode expansion}
\circnormal{e^{ik_3 X(z,\barred{z})}} = e^{ik_3 x}|z|^{ik_3 x_{\text{s}} + \frac{(k_1 - k_2)k_3}{2}}\exp\left[-i\sqrt{2}k_3 \sum_{n=1}^{\infty}\frac{\alpha_{-n}}{n}\left(\frac{z}{\barred{z}}\right)^{\pi i n/4\ln\varepsilon}\sin\left(\frac{\pi n}{2\ln\varepsilon}\ln\left|\frac{z}{\varepsilon}\right|\right)\right]
\\  \times \exp\left[-i\sqrt{2}k_3 \sum_{m=1}^{\infty}\frac{\alpha_{m}}{m}\left(\frac{z}{\barred{z}}\right)^{-\pi i m/4\ln\varepsilon}\sin\left(\frac{\pi m}{2\ln\varepsilon}\ln\left|\frac{z}{\varepsilon}\right|\right)\right].
\end{multline}
The trace is computed similarly as before, remembering to divide by the volume of the Lagrange multiplier zero-modes. The result is
\begin{equation}
\text{Tr}_{\mathcal{H}_{k_1 k_2}^{\text{DD}}}\hspace{-2pt}\left[\mathcal{O}_{k_3}(z,\barred{z})e^{-2\pi H_{\text{R}}^{\text{DD}}}\right]_{S^2} = \frac{\delta(k_1\hspace{-1pt}+\hspace{-1pt}k_2\hspace{-1pt}+\hspace{-1pt}k_3)}{4\sqrt{2}\pi^2\varepsilon^{-\frac{1}{6}}\eta\hspace{-2pt}\left(\hspace{-1pt}-\frac{2i\ln\varepsilon}{\pi}\right)}\frac{\varepsilon^{(k_1^2+k_2^2)/2}}{|z|^{(k_1^2 + k_3^2 - k_2^2)/2}}\hspace{-2pt}\left[\frac{\varepsilon^{-\frac{1}{2}}\eta^3\hspace{-2pt}\left(\hspace{-2pt}-\frac{2i\ln\varepsilon}{\pi}\right)}{\vartheta_4\hspace{-2pt}\left(\left.\hspace{-2pt}-\frac{i\ln|z|}{\pi}\right|\hspace{-2pt}-\hspace{-2pt}\frac{2i\ln\varepsilon}{\pi}\hspace{-1pt}\right)}\hspace{-1pt}\right]^{k_3^2/2}.
\end{equation}
We again find the desired matching condition \eqref{3-point matching}, namely
\begin{equation}
\lim_{\varepsilon\rightarrow 0}\left(\varepsilon^{-\frac{k_1^2}{2}-\frac{k_2^2}{2}}\text{Tr}_{\mathcal{H}_{k_1 k_2}^{\text{DD}}}\left[\mathcal{O}_{k_3}(z,\barred{z})e^{-2\pi H_{\text{R}}^{\text{DD}}}\right]_{S^2}\right) = \frac{\mathcal{N}_{k_1 k_2}^{\text{DD}}C_{k_1 k_3 k_2}}{|z|^{(k_1^2 + k_3^2 - k_2^2)/2}},
\end{equation}
with the same normalization constant \eqref{DD normalization} as before.

As a final comment, it may not be surprising that the normalization constants $\mathcal{N}_{k_1 k_2}^{\text{NN}}$ and $\mathcal{N}_{k_1 k_2}^{\text{DD}}$ are essentially those appearing in the Cardy states associated to the conformally-invariant Neumann and Dirichlet boundary conditions. Recall that the latter are non-normalizable states determined by the Cardy condition \cite{Cardy89}, i.e.~they are constructed from the usual Hilbert space on $S^1$ so that their annulus matrix elements reproduce the cylinder traces over the strip Hilbert spaces (with conventional spatial length of $\pi$) with Neumann or Dirichlet boundary conditions at the endpoints. The usual Cardy states associated to $\partial_n X = 0$ and to $X = x_0$ are then determined to be
\begin{align}\label{Neumann Cardy state}
|\text{N}\rrangle & = \frac{1}{(8\pi^2)^{1/4}}e^{-\sum_{n=1}^{\infty}\frac{1}{n}\alpha_{-n}\widetilde{\alpha}_{-n}}|0;\{0\},\{0\}\rangle
\\ \label{Dirichlet Cardy state} |\text{D},x_0\rrangle & = (2\pi^2)^{1/4}\int_{-\infty}^{\infty}\frac{dk}{2\pi}e^{ikx_0}e^{\sum_{n=1}^{\infty}\frac{1}{n}\alpha_{-n}\widetilde{\alpha}_{-n}}|k;\{0\},\{0\}\rangle,
\end{align}
where $|k;\{0\},\{0\}\rangle$ is the oscillator vacuum in $\mathcal{H}_{S^1}$ with target space momentum $k$. The normalizations $\mathcal{N}_{|\text{N}\rrangle} = \frac{1}{(8\pi^2)^{1/4}}$ and $\mathcal{N}_{|\text{D}\rrangle} = (2\pi^2)^{1/4}$ of the above Cardy states are related to the angular quantization normalizations here by
\begin{align}
\mathcal{N}_{k_1 k_2}^{\text{NN}} & = \mathcal{N}_{|\text{N}\rrangle}^2
\\ \mathcal{N}_{k_1 k_2}^{\text{DD}} & = \left(\frac{\mathcal{N}_{|\text{D}\rrangle}}{(2\pi)^2}\right)^2,
\end{align}
which is to be expected since the oscillator parts of the thermal trace matching computations above are nearly identical to those appearing in the Cardy condition\footnote{The extra factor of $(2\pi)^2$ accompanying the Dirichlet normalization is simply due to the fact that the normalization of a continuous basis is arbitrary, and the normalization we chose here differs slightly from that conventionally used in the Cardy states.}. We should also point out that the conformally-invariant boundary condition on which the Dirichlet-like approach here is based is $\partial_t X = 0$, whose associated boundary state is
\begin{equation}
|\text{D}\rrangle = \int_{-\infty}^{\infty} dx_0 \ |\text{D},x_0\rrangle = (2\pi^2)^{1/4}e^{\sum_{n=1}^{\infty}\frac{1}{n}\alpha_{-n}\widetilde{\alpha}_{-n}}|0;\{0\},\{0\}\rangle.
\end{equation}
It is this boundary state which shrinks to the identity operator. Even though the lowest dimension state appearing in the expression \eqref{Dirichlet Cardy state} is that corresponding to the identity, $|\text{D},x_0\rrangle$ also contains a continuum of states with infinitesimal dimension all with the same coefficient, and the contribution from the identity operator cannot be separated from this tail of exponential primaries. It is in this sense that pure Dirichlet boundary conditions are pathological for the noncompact free boson, a problem which is obviously rectified in the compact theory.

\section{Free CFT Examples}\label{free CFTs}

Having presented the angular quantization of the noncompact free boson for the two most natural boundary conditions in great detail, in this section we present the results of angular quantization for three other free CFTs --- the linear dilaton, the compact free boson and the $bc$ ghost system. As these CFTs are free, we again analyze them in canonical quantization, leading to explicit Fock space constructions of the associated Hilbert spaces. Many of the details are only slight modifications of those presented for the noncompact free boson in Section \ref{free boson}, so after deriving the relevant boundary conditions we shall focus primarily on presenting the results and noting where they differ; some computational details are again relegated to an Appendix.

\subsection{Linear Dilaton}

Even though the linear dilaton CFT is also somewhat pathological --- having a non-closed OPE\footnote{Alternatively, one could take its closure under the OPE at the cost of unitarity.} and being logarithmic --- its appearance in string theory makes it relevant for our later applications in \cite{Agia22}. The linear dilaton $\phi$ with background charge $Q$ has central charges $c = \widetilde{c} = 1 + 6Q^2$; we normalize the linear dilaton on the plane in the same way as the free boson, so that its OPE with itself is
\begin{equation}
\phi(z_1,\barred{z}_1)\phi(z_2,\barred{z}_2) = -\frac{1}{2}\ln|z_{12}|^2 + \normal{\phi(z_1,\barred{z}_1)\phi(z_2,\barred{z}_2)}.
\end{equation}
The exponential primaries are $\mathcal{O}_{\alpha} \equiv \s\s\s\normal{e^{2\alpha \phi}}$ with conformal weights $h_{\alpha} = \widetilde{h}_{\alpha} = \alpha(Q-\alpha)$. We shall perform the angular quantization for the linear dilaton with exponential endpoint operators analogously to the Neumann-like approach for the noncompact free boson, but an important distinction is that the linear dilaton field $\phi(z,\barred{z})$ has different conformal transformations.

Under a conformal transformation associated to $z \mapsto z'(z)$, the field transforms as $\phi(z,\barred{z}) \mapsto \phi'(z,\barred{z})$, where
\begin{equation}\label{linear dilaton general transformation}
\phi'(z',\barred{z}') = \phi(z,\barred{z}) - \frac{Q}{2}\ln\left|\partial_z z'\right|^2.
\end{equation}
The anomalous holomorphic current $\partial\phi(z)$ does transform as a weight-$(1,0)$ operator but is not a primary; its anomalous conformal transformation is instead given by
\begin{equation}
(\partial\phi)'(z') = (\partial_z z')^{-1}\partial\phi(z) - \frac{Q}{2}\frac{\partial_z^2 z'}{(\partial_z z')^2}.
\end{equation}
The most important conformal transformation for us is that associated with the map $z = e^{-iy}$ between the flat plane and the flat cylinder; for ease of notation, we shall drop the prime on the transformation of the operator itself, instead letting the argument determine whether the operator is the one on the plane with coordinates $(z,\barred{z})$ or on the strip with coordinates $(y,\barred{y})$ or equivalently $(y_1,y_2)$. Then, the transformations of the linear dilaton and its derivatives from the plane to the cylinder are
\begin{align}\label{anomalous linear dilaton transformation}
\phi(y,\barred{y}) & = \phi\big(z(y),\barred{z}(\barred{y})\big) + Qy_2
\\ \partial\phi(y) & = -iz(y)\partial\phi\big(z(y)\big) - \frac{iQ}{2}
\\ \barred{\partial}\phi(\barred{y}) & = +i\barred{z}(\barred{y})\barred{\partial}\phi\big(\barred{z}(\barred{y})\big) + \frac{iQ}{2}.
\end{align}
There will thus be a shift in the boundary conditions shrinking to exponential operators on the plane versus on the cylinder.

We now wish to write Neumann-like boundary conditions on the plane at $|z| = \varepsilon$ and at $|z| = \frac{1}{\varepsilon}$ which shrink to $\mathcal{O}_{\alpha_1} = \s\s\s\normal{e^{2\alpha_1\phi(0)}}$ and to $\mathcal{O}_{\alpha_2} = \s\s\s\normal{e^{2\alpha_2\phi(\infty)}}$, respectively. Na\"{i}vely, the appropriate boundary conditions might seem to be obtained from the free boson ones with the replacement $k \rightarrow -2i\alpha$, but this is not quite true. Recall that the covariant linear dilaton action on a closed Riemann surface $\Sigma$ with metric $g_{\alpha\beta}$ is
\begin{equation}
S = \frac{1}{4\pi}\int_{\Sigma} d^2\sigma\sqrt{g}\left(g^{\alpha\beta}\partial_{\alpha}\phi\partial_{\beta}\phi + QR\phi\right),
\end{equation}
with equation of motion $\nabla^2\phi = \frac{1}{2}QR$. Since the sphere has Euler characteristic $\chi_{S^2} = 2$, working in the ``flat'' sphere conformal frame really means using the flat metric in the $z$-patch stereographic plane with a curvature singularity at infinity such that $R = \frac{8\pi}{\sqrt{g}}\delta^2(z-\infty)$. Therefore, when inserting the operators $\normal{e^{2\alpha_1\phi(0)}}$ and $\normal{e^{2\alpha_2\phi(\infty)}}$ into the path integral on the sphere, the effective action (in the sense of absorbing the contributions of the operator insertions into the action up to some renormalization terms\footnote{Or, more precisely, one may consider separating out the zero-mode from the exponential operators like we did in the path integral derivation for the shrinking of the free boson Neumann-like boundary condition. Then just the zero-modes are absorbed into the action, with the bulk insertions of the nonzero-mode exponentials still present. This is what one would do to explicitly evaluate the path integral by expanding the exponentials, for example.}) in the flat $z$-coordinates is
\begin{equation}
S = \frac{1}{2\pi}\int d^2 z \s\s \partial\phi\barred{\partial}\phi - 2\alpha_1\phi(|z|=0) + 2(Q-\alpha_2)\phi(|z|=\infty).
\end{equation}
Thus the Neumann-like boundary conditions are obtained from those of the free boson via the replacements $k_1\rightarrow -2i\alpha_1$ and $k_2\rightarrow -2i(\alpha_2-Q)$. That is, on the two-holed sphere we impose the boundary conditions
\begin{align}
r\partial_r \phi\Big|_{|z|=\varepsilon} & = -2\alpha_1
\\ r\partial_r \phi\Big|_{|z|=\frac{1}{\varepsilon}} & = +2(\alpha_2 - Q).
\end{align}
Another way to obtain the latter result is to apply the inversion $z\mapsto z' = 1/\barred{z}$ using the parity-reversed version of the anomalous transformation law \eqref{linear dilaton general transformation}, which says that $\phi'(\frac{1}{\barred{z}},\frac{1}{z}) = \phi(z,\barred{z}) + 2Q\ln|z|$. Since the inverted operator obeys the boundary condition near the origin $r'\partial_{r'}\phi'(r' = \varepsilon) = -2\alpha_2$ without subtleties, the boundary condition in the original $z$-frame is $r\partial_r\phi(r = \frac{1}{\varepsilon}) = -r'\partial_{r'}\phi'(r' = \varepsilon) - 2Q = 2(\alpha_2 - Q)$. These statements are equivalent to the familiar fact that the linear dilaton theory on the stereographically-projected sphere behaves like a free boson with a ``background charge operator'' $\normal{e^{-2Q\phi}}$ placed at infinity. Using \eqref{anomalous linear dilaton transformation}, the appropriate boundary conditions on the very wide strip are
\begin{align}\label{LD bc 1}
\partial_{y_2}\phi\Big|_{y_2 = +\ln\varepsilon} & = -2\alpha_1 + Q
\\ \label{LD bc 2}\partial_{y_2}\phi\Big|_{y_2 = -\ln\varepsilon} & = +2\alpha_2 - Q.
\end{align}
The boundary conditions on the strip are more symmetric, corresponding to the fact that the infinite cylinder without boundaries is typically described in the flat $y$-patch frame with curvature singularities equally distributed between $y_2 = \pm\infty$. Note also that there can only be a Euclidean saddle on the two-holed sphere or on the very wide cylinder if the Neumann-like boundary conditions at $\ln|z| = y_2 = \pm \ln\varepsilon$ are equal, meaning that the corresponding Euclidean path integrals will only be nonvanishing when $\alpha_1 + \alpha_2 = Q$, reproducing the well-known anomalous conservation law of the linear dilaton translation current.

The strip action imposing the boundary conditions \eqref{LD bc 1} and \eqref{LD bc 2} is
\begin{equation}\label{LD strip action}
S = \frac{1}{4\pi}\int dy_1 dy_2\left[\left(\partial_{y_1}\phi\right)^2 + \left(\partial_{y_2}\phi\right)^2\right] - \frac{2\alpha_1\hspace{-2pt}-\hspace{-2pt}Q}{2\pi}\int_{y_2 = \ln\varepsilon}\hspace{-5pt}dy_1 \ \phi - \frac{2\alpha_2\hspace{-2pt}-\hspace{-2pt}Q}{2\pi}\int_{y_2 = -\ln\varepsilon}\hspace{-7pt}dy_1 \ \phi.
\end{equation}
One must be careful in computing the stress-energy tensor, however, as even the ``pure'' linear dilaton on a Riemann surface with boundary has a boundary contribution to the stress-energy tensor, which is not often written in the literature since one typically works with a boundary condition in a frame where this term vanishes. It is useful first to write the fully covariant stress-energy tensor for the ``pure'' linear dilaton theory on a manifold with boundary, i.e.~without including the extra boundary terms yet which will impose our choice of boundary conditions. The covariant pure action is
\begin{equation}\label{LD pure}
S_{\text{pure}} = \frac{1}{4\pi}\int_{\Sigma}d^2\sigma\sqrt{g}\left(g^{\alpha\beta}\partial_{\alpha}\phi\partial_{\beta}\phi + QR\phi\right) + \frac{1}{2\pi}\int_{\partial\Sigma}d\ell\sqrt{f^* g}\s\s QK\phi,
\end{equation}
whose covariant stress-energy tensor is
\begin{equation}
T_{\alpha\beta}^{\text{pure}} \hspace{-2pt}= \hspace{-2pt}-\hspace{-0.5pt}\normal{\hspace{-1pt}\left(\hspace{-2pt}\nabla_{\alpha}\phi\nabla_{\beta}\phi \hspace{-2pt}-\hspace{-2pt} \frac{1}{2}g_{\alpha\beta}\nabla^{\gamma}\phi\nabla_{\gamma}\phi \hspace{-2pt}-\hspace{-2pt} Q\left(\nabla_{\alpha}\nabla_{\beta}\phi \hspace{-2pt}-\hspace{-2pt} g_{\alpha\beta}\nabla^2\phi\right)\hspace{-2pt}\right)\hspace{-1pt}} \hspace{-2pt}+ Q\delta(\Sigma-\partial\Sigma)(g_{\alpha\beta} - n_{\alpha}n_{\beta})n^{\gamma}\nabla_{\gamma}\phi.
\end{equation}
The boundary contribution to $T_{\alpha\beta}^{\text{pure}}$ only vanishes identically if one chooses exactly Neumann boundary conditions $n^{\alpha}\nabla_{\alpha}\phi\big|_{\partial\Sigma} = 0$, which is not the case here (and is not even the conformally-invariant choice). The full boundary action in a general frame is the strip boundary action in \eqref{LD strip action} plus the extrinsic curvature term in \eqref{LD pure}. Including the boundary contribution from the ``pure'' stress-energy tensor with the normal derivatives \eqref{LD bc 1} and \eqref{LD bc 2} as well as the boundary terms from the additional boundary action in \eqref{LD strip action} on the strip (for which $R=0$ and $K=0$), the correct stress-energy tensor for this theory in the flat cylinder frame is
\begin{multline}
T_{\alpha\beta}(y_1,y_2) = -\normal{\left(\partial_{\alpha}\phi\partial_{\beta}\phi - \frac{1}{2}\delta_{\alpha\beta}\partial^{\gamma}\phi\partial_{\gamma}\phi - Q\partial_{\alpha}\partial_{\beta}\phi\right)}
\\ - t_{\alpha}t_{\beta}(\phi-Q)\left[(2\alpha_1 - Q)\delta(y_2 - \ln\varepsilon) + (2\alpha_2 - Q)\delta(y_2 + \ln\varepsilon)\right].
\end{multline}
The Rindler Hamiltonian with these boundary conditions is therefore computed by
\begin{equation}\label{LD Hamiltonian integral}
H_{\text{R}}^{\text{LD}} = \frac{1}{2\pi}\int_{\ln\varepsilon}^{-\ln\varepsilon}\hspace{-5pt}dy_2\hspace{-2pt}\left[T(y) + \widetilde{T}(\barred{y})\right] - \frac{1}{2\pi}\left[(2\alpha_1\hspace{-2pt}-Q)(\phi-Q)\Big|_{y_2 = \ln\varepsilon}\hspace{-5pt} + (2\alpha_2\hspace{-2pt}-Q)(\phi-Q)\Big|_{y_2=-\ln\varepsilon}\right],
\end{equation}
where
\begin{align}
T(y) & = -\normal{\partial\phi(y)\partial\phi(y)} \hspace{-2pt}+\s\s Q\partial^2\phi(y)
\\ \widetilde{T}(\barred{y}) & = -\normal{\barred{\partial}\phi(\barred{y})\barred{\partial}\phi(\barred{y})} \hspace{-2pt}+\s\s Q\barred{\partial}^2\phi(\barred{y})
\end{align}
are the usual bulk holomorphic and antiholomorphic components of the stress-energy tensor.

Before proceeding, let us briefly comment on conformal invariance in boundary linear dilaton theory. A linear boundary condition for a linear dilaton is frame dependent since $\phi$ does not have definite weights. The boundary condition imposed by the pure linear dilaton action on its own is $n^{\alpha}\nabla_{\alpha}\phi = -QK$, which is exactly Neumann only in frames where the boundary has no extrinsic curvature. The presence of the boundary stress-energy tensor even in the pure theory is a consequence of the fact that the background charge $Q$ already gives rise to a ``classical central charge''. In particular, using the bulk equation of motion $\nabla^2\phi = \frac{1}{2}QR$ and boundary condition $n^{\alpha}\nabla_{\alpha}\phi = -QK$ arising from the action $S_{\text{pure}}$ with no additional boundary terms, the classical trace of the pure stress-energy tensor is $(T^{\alpha}_{\phantom{\alpha}\alpha})^{\text{pure}} = -\frac{1}{2}Q^2 R - Q^2 \delta(\Sigma-\partial\Sigma)K$, which is indeed the form of the trace dictated by diffeomorphism invariance for ``classical central charge'' $c_{\text{cl.}} = 6Q^2$. The difference between the actual quantum stress-energy tensor and the classical one is identical to that of the free boson, since such effects arise from coincident operator renormalization which is due only to the quadratic kinetic term in the action. The action $S_{\text{pure}}$ is hence conformally invariant with central charge $c = 1 + 6Q^2$ because the trace of the stress-energy tensor is exactly zero on a flat space with flat boundaries. Thus, the correct conformally-invariant boundary condition for the linear dilaton is actually $n^{\alpha}\nabla_{\alpha}\phi = -QK$, which we still call ``Neumann'' even though the normal derivative only vanishes at a flat boundary, corresponding to the construction here on the strip/cylinder with $\alpha_1 = \alpha_2 = \frac{Q}{2}$. Therefore, the conformally-invariant linear dilaton Neumann boundary condition actually shrinks to the scalar primary $\normal{e^{Q\phi}}$ of dimension $\Delta = \frac{Q^2}{2}$ on the sphere. This is a simple example of a conformally-invariant local boundary condition that does not shrink to the identity.

Back to the action \eqref{LD strip action}, the strip mode expansions of the dilaton and its conjugate momentum $\Pi_{\phi} = -\frac{i}{2\pi}\partial_{y_1}\phi$ satisfying the bulk equation of motion and the boundary conditions are
\begin{align}
\phi(y_1,y_2) & = \varphi + \frac{\alpha_1+\alpha_2-Q}{6}\ln\varepsilon - \frac{\pi i p}{\ln\varepsilon}y_1 - (\alpha_1 - \alpha_2)y_2 + \frac{\alpha_1 + \alpha_2 - Q}{2\ln\varepsilon}\left(y_1^2 - y_2^2\right)
\\ \notag & \hspace{15pt} + i\sqrt{2}\sum_{\substack{n \in\mathds{Z} \\ n \neq 0}}\frac{\phi_n}{n}e^{-\frac{\pi n}{2\ln\varepsilon}y_1}\cos\left(\frac{\pi n(y_2 - \ln\varepsilon)}{2\ln\varepsilon}\right)
\\ \Pi_{\phi}(y_1,y_2) & = -\frac{p_{\varphi}}{2\ln\varepsilon} \hspace{-2pt}-\hspace{-2pt} \frac{i(\alpha_1\hspace{-2pt}+\hspace{-2pt}\alpha_2\hspace{-2pt}-\hspace{-2pt}Q)}{2\pi\ln\varepsilon}y_1 \hspace{-2pt}-\hspace{-2pt} \frac{1}{2\sqrt{2}\ln\varepsilon}\sum_{\substack{n\in\mathds{Z} \\ n \neq 0}}\phi_n e^{-\frac{\pi n}{2\ln\varepsilon}y_1}\cos\left(\frac{\pi n(y_2 - \ln\varepsilon)}{2\ln\varepsilon}\right),
\end{align}
where $\varphi$ and $p_{\varphi}$ are again the center-of-mass variables; the nontrivial commutators are the usual $[\varphi,p_{\varphi}] = i$ and $[\phi_n,\phi_m] = n\delta_{n,-m}$. The mode expansions of the bulk holomorphic and antiholomorphic stress-energy tensor components are
\begin{align}
T(y) & = -\frac{(\alpha_1+\alpha_2-Q)^2}{4\ln^2\varepsilon}y^2 + \frac{\pi i(\alpha_1+\alpha_2-Q)}{2\sqrt{2}\ln^2\varepsilon}y\sum_{n\in\mathds{Z}}\phi_n e^{\frac{\pi i n}{2}}e^{-\frac{\pi n y}{2\ln\varepsilon}} 
\\ \notag & \hspace{15pt} + \frac{\pi^2}{8\ln^2\varepsilon}\sum_{n,m\in\mathds{Z}}\circnormal{\phi_n\phi_m}e^{\frac{\pi i(n+m)}{2}}e^{-\frac{\pi(n+m)y}{2\ln\varepsilon}} - \frac{\pi^2}{96\ln^2\varepsilon}
\\ \notag & \hspace{15pt} + Q\partial^2\phi(y) + iQ\partial\phi(y) - \frac{Q^2}{4}
\\ \widetilde{T}(\barred{y}) & = -\frac{(\alpha_1+\alpha_2-Q)^2}{4\ln^2\varepsilon}\barred{y}^2 + \frac{\pi i(\alpha_1+\alpha_2-Q)}{2\sqrt{2}\ln^2\varepsilon}\barred{y}\sum_{n\in\mathds{Z}}\widetilde{\phi}_n e^{-\frac{\pi i n}{2}}e^{-\frac{\pi n \barred{y}}{2\ln\varepsilon}} 
\\ \notag & \hspace{15pt} + \frac{\pi^2}{8\ln^2\varepsilon}\sum_{n,m\in\mathds{Z}}\circnormal{\widetilde{\phi}_n\widetilde{\phi}_m}e^{-\frac{\pi i(n+m)}{2}}e^{-\frac{\pi(n+m)\barred{y}}{2\ln\varepsilon}} - \frac{\pi^2}{96\ln^2\varepsilon}
\\ \notag & \hspace{15pt} + Q\barred{\partial}^2\phi(\barred{y}) - iQ\barred{\partial}\phi(\barred{y}) - \frac{Q^2}{4},
\end{align}
where we have defined
\begin{align}
\phi_0 & \equiv \sqrt{2}\left(p_{\varphi} - \frac{\alpha_1 - \alpha_2 + Q}{\pi}\ln\varepsilon\right)
\\ \widetilde{\phi}_0 & \equiv \sqrt{2}\left(p_{\varphi} + \frac{\alpha_1 - \alpha_2 + Q}{\pi}\ln\varepsilon\right)
\end{align}
as well as $\widetilde{\phi}_n = \phi_n$ for all $n\neq 0$. Using these expansions in \eqref{LD Hamiltonian integral} results in the Rindler Hamiltonian on the very wide strip as
\begin{equation}\label{LD Hamiltonian answer}
H_{\text{R}}^{\text{LD}} = -\frac{\pi p_{\varphi}^2}{2\ln\varepsilon} - \frac{(\alpha_1\hspace{-2pt}+\hspace{-2pt}\alpha_2\hspace{-2pt}-\hspace{-2pt}Q)\varphi}{\pi} - \frac{\pi}{2\ln\varepsilon}\sum_{n=1}^{\infty}\phi_{-n}\phi_n + \frac{(\alpha_1\hspace{-2pt}-\hspace{-2pt}\alpha_2)^2}{2\pi}\ln\varepsilon + \frac{(\alpha_1\hspace{-2pt}+\hspace{-2pt}\alpha_2\hspace{-2pt}-\hspace{-2pt}Q)^2}{6\pi}\ln\varepsilon + \frac{\pi}{48\ln\varepsilon}.
\end{equation}
The finite-regulator angular quantization Hilbert space $\mathcal{H}_{\alpha_1\alpha_2}^{\text{LD}}$ is isomorphic as a vector space to that of the noncompact free boson with Neumann-like boundary conditions, being spanned by $|p_{\varphi};\{N_n\}\rangle$. The mode expansion of the bulk exponential primary $\mathcal{O}_{\alpha_3}(z,\barred{z}) = \normal{e^{2\alpha_3\phi(z,\barred{z})}}$ on the plane is
\begin{align}
\normal{e^{2\alpha_3\phi(z,\barred{z})}} & = \left[-\frac{\pi}{\ln\varepsilon}\hspace{-2pt}\left(\frac{z}{\barred{z}}\right)^{\frac{\pi i}{2\ln\varepsilon}}\hspace{-0pt}\cos\hspace{-1pt}\left(\hspace{-1pt}\frac{\pi\ln|z|}{2\ln\varepsilon}\hspace{-1pt}\right)\hspace{-1pt}\right]^{-2\alpha_3^2}
\\ \notag & \hspace{5pt} \times \exp\hspace{-2pt}\left[\hspace{-2pt}\frac{(\hspace{-1pt}\alpha_1\hspace{-3pt}+\hspace{-3pt}\alpha_2\hspace{-3pt}-\hspace{-3pt}Q\hspace{-1pt})\hspace{-1pt}\alpha_3}{3}\hspace{-2pt}\ln\hspace{-1pt}\varepsilon \hspace{-2pt}-\hspace{-2pt} 2\hspace{-1pt}(\hspace{-1pt}\alpha_1\hspace{-3pt}-\hspace{-3pt}\alpha_2\hspace{-3pt}-\hspace{-3pt}\alpha_3\hspace{-3pt}+\hspace{-3pt}Q\hspace{-1pt})\hspace{-1pt}\alpha_3\hspace{-1pt}\ln\hspace{-1pt}|z| \hspace{-2pt}-\hspace{-2pt} \frac{(\hspace{-1pt}\alpha_1\hspace{-3pt}+\hspace{-3pt}\alpha_2\hspace{-3pt}-\hspace{-3pt}Q\hspace{-1pt})\hspace{-1pt}\alpha_3}{2\ln\varepsilon}\hspace{-3pt}\left(\ln^2 \hspace{-2pt}z \hspace{-2pt}+\hspace{-2pt} \ln^2\hspace{-2pt}\barred{z}\right)\hspace{-2pt}\right]
\\ \notag & \hspace{5pt} \times \exp\left[2\alpha_3\varphi - 2i\sqrt{2}\alpha_3\sum_{n=1}^{\infty}\frac{\phi_{-n}}{n}\left(\frac{z}{\barred{z}}\right)^{\frac{\pi i n}{4\ln\varepsilon}}\cos\left(\frac{\pi n}{2\ln\varepsilon}\ln\left|\frac{z}{\varepsilon}\right|\right)\right]
\\ \notag & \hspace{5pt} \times \exp\left[\frac{\pi \alpha_3}{\ln\varepsilon}p_{\varphi}\ln\left(\frac{z}{\barred{z}}\right) + 2i\sqrt{2}\alpha_3\sum_{m=1}^{\infty}\frac{\phi_{m}}{m}\left(\frac{z}{\barred{z}}\right)^{-\frac{\pi i m}{4\ln\varepsilon}}\cos\left(\frac{\pi m}{2\ln\varepsilon}\ln\left|\frac{z}{\varepsilon}\right|\right)\right].
\end{align}
The thermal trace and bulk thermal one-point function on the two-holed sphere are
\begin{align}
\text{Tr}_{\mathcal{H}_{\alpha_1\alpha_2}^{\text{LD}}}\hspace{-2pt}\left[e^{-2\pi H_{\text{R}}^{\text{LD}}}\right]_{S^2} & = \frac{\delta(\alpha_1\hspace{-2pt}+\hspace{-2pt}\alpha_2\hspace{-2pt}-\hspace{-2pt}Q)}{2\sqrt{2}}\frac{\varepsilon^{2\alpha_1(Q-\alpha_1)+2\alpha_2(Q-\alpha_2)}}{\varepsilon^{-1/6}\eta\hspace{-2pt}\left(-\frac{2i\ln\varepsilon}{\pi}\right)}
\\ \text{Tr}_{\mathcal{H}_{\alpha_1\alpha_2}^{\text{LD}}}\hspace{-2pt}\left[\mathcal{O}_{\alpha_3}(z,\barred{z})e^{-2\pi H_{\text{R}}^{\text{LD}}}\right]_{S^2} & = \frac{\delta(\alpha_1\hspace{-2pt}+\hspace{-2pt}\alpha_2\hspace{-2pt}+\hspace{-2pt}\alpha_3\hspace{-2pt}-\hspace{-2pt}Q)}{2\sqrt{2}}\frac{\varepsilon^{2\alpha_1(Q-\alpha_1)+2\alpha_2(Q-\alpha_2)}}{|z|^{2[\alpha_1(Q-\alpha_1)+\alpha_3(Q-\alpha_3)-\alpha_2(Q-\alpha_2)]}}
\\ \notag & \hspace{10pt} \times\hspace{-2pt} \frac{1}{\varepsilon^{-\frac{1}{6}}\eta\hspace{-2pt}\left(\hspace{-1pt}-\hspace{-0pt}\frac{2i\ln\hspace{-1pt}\varepsilon}{\pi}\hspace{-1pt}\right)}\hspace{-3pt}\left[\varepsilon^{-\frac{1}{6}}\eta\hspace{-2pt}\left(\hspace{-3pt}-\hspace{-1pt}\frac{2i\hspace{-0.5pt}\ln\hspace{-1pt}\varepsilon}{\pi}\hspace{-2pt}\right)\hspace{-2pt}\vartheta_4\hspace{-3pt}\left(\left.\hspace{-4pt}-\frac{i\hspace{-0.5pt}\ln\hspace{-1pt}|z|}{\pi}\right|\hspace{-2pt}-\hspace{-2pt}\frac{2i\hspace{-0.5pt}\ln\hspace{-1pt}\varepsilon}{\pi}\hspace{-2pt}\right)\hspace{-2pt}\right]^{\hspace{-1pt}-\hspace{-1pt}2\alpha_3^2}.
\end{align}
Since the canonically normalized linear dilaton sphere two- and three-point coefficients are $\langle\mathcal{O}_{\alpha_1}(0)\mathcal{O}_{\alpha_2}(1)\rangle_{S^2} = \pi\delta(\alpha_1+\alpha_2-Q)$ and $C_{\alpha_1\alpha_2\alpha_3} = \pi\delta(\alpha_1+\alpha_2+\alpha_3-Q)$, we again find the required matching conditions \eqref{2-point matching} and \eqref{3-point matching}, namely
\begin{align}
\lim_{\varepsilon\rightarrow 0}\left(\varepsilon^{-2\alpha_1(Q-\alpha_1)-2\alpha_2(Q-\alpha_2)}\text{Tr}_{\mathcal{H}_{\alpha_1\alpha_2}^{\text{LD}}}\left[e^{-2\pi H_{\text{R}}^{\text{LD}}}\right]_{S^2}\right) & = \pi \mathcal{N}_{\alpha_1\alpha_2}^{\text{LD}}\delta(\alpha_1+\alpha_2-Q)
\\ \lim_{\varepsilon\rightarrow 0}\left(\varepsilon^{-2\alpha_1(Q-\alpha_1)-2\alpha_2(Q-\alpha_2)}\text{Tr}_{\mathcal{H}_{\alpha_1\alpha_2}^{\text{LD}}}\left[\mathcal{O}_{\alpha_3}(z,\barred{z})e^{-2\pi H_{\text{R}}^{\text{LD}}}\right]_{S^2}\right) & = \frac{\pi\mathcal{N}_{\alpha_1\alpha_2}^{\text{LD}}\delta(\alpha_1+\alpha_2+\alpha_3-Q)}{|z|^{2[\alpha_1(Q-\alpha_1) + \alpha_3(Q-\alpha_3) - \alpha_2(Q-\alpha_2)]}},
\end{align}
where the normalization
\begin{equation}
\mathcal{N}_{\alpha_1\alpha_2}^{\text{LD}} = \frac{1}{2\sqrt{2}\pi}
\end{equation}
is the same as in the Neumann-like angular quantization of the noncompact free boson.

\subsection{Compact Free Boson}

Now we turn to the free boson CFT compactified at radius $R$, so that $X \sim X + 2\pi R$, rendering the spectrum discrete. The exponential primaries $\mathcal{O}_{n,w}$ are characterized by integer momentum $n$ and integer winding $w$; as usual on the plane we write $\mathcal{O}_{n,w}(z,\barred{z}) \equiv \s\s\s \normal{e^{i[k_{\text{L}}X_{\text{L}}(z) + k_{\text{R}}X_{\text{R}}(\barred{z})]}}$, where $k_{\text{L}} \equiv \frac{n}{R} + wR$, $k_{\text{R}} \equiv \frac{n}{R} - wR$ and $X(z,\barred{z}) = X_{\text{L}}(z) + X_{\text{R}}(\barred{z})$. The momentum and winding exponential primary $\mathcal{O}_{n,w}$ has conformal weights $(h,\widetilde{h}) = (\frac{k_{\text{L}}^2}{4},\frac{k_{\text{R}}^2}{4})$, i.e.~scaling dimension $\Delta = \frac{1}{2}(\frac{n^2}{R^2} + w^2 R^2)$ and spin $s = nw$, but the expression $\normal{e^{i(k_{\text{L}}X_{\text{L}} + k_{\text{R}}X_{\text{R}})}}$ is really only a slogan until one defines the chiral splitting more precisely, as it leaves the zero-mode ambiguous. In radial quantization, one often just says that the choice of zero-mode splitting is tantamount to a choice of branch cut, and at the end of the day the choice will be irrelevant for correlators of well-defined local operators. In more modern parlance, we would say that $X_{\text{L}}$ and $X_{\text{R}}$ are chiral defect operators and the monodromy from a branch cut is just the effect of passing an operator through the defect line \cite{Chang18}. When mode expanding the compact boson in radial quantization, the zero-mode ambiguity is thus left unfixed and requires appending 2-cocycles to the oscillator exponentials in order to recover the correct operator algebra \cite{Polchinski981} (just like with bosonization). In angular quantization, on the other hand, we shall find that the zero-mode splitting is actually determined by construction and moreover automatically incorporates the 2-cocycles that seemed \emph{ad hoc} in radial quantization.

The example of most importance for us in future applications is the angular quantization associated to Euclidean time-winding operators. Using the results laid out here, in \cite{Agia22} we shall perform the BRST quantization associated to pairs of worldsheet vertex operators which wind Euclidean target-space time, describing a distinct class of string states which are neither open nor closed.

The chiral OPEs of the compact boson are
\begin{align}
X_{\text{L}}(z_1)X_{\text{L}}(z_2) & = -\frac{1}{2}\ln z_{12} + \normal{X_{\text{L}}(z_1)X_{\text{L}}(z_2)}
\\ X_{\text{R}}(\barred{z}_1)X_{\text{R}}(\barred{z}_2) & = -\frac{1}{2}\ln\barred{z}_{12} + \normal{X_{\text{R}}(\barred{z}_1)X_{\text{R}}(\barred{z}_2)},
\end{align}
which are well-defined as soon as we give a proper definition of the chiral splitting below. The OPEs of the radial and angular derivatives in the presence of an exponential primary are
\begin{align}
\left[z\partial X_{\text{L}}(z) + \barred{z}\barred{\partial}X_{\text{R}}(\barred{z})\right]\normal{e^{i[k_{\text{L}}X_{\text{L}}(0) + k_{\text{R}}X_{\text{R}}(0)]}} & = -\frac{in}{R}\normal{e^{i[k_{\text{L}}X_{\text{L}}(0) + k_{\text{R}}X_{\text{R}}(0)]}} + \dotsc
\\ \left[iz\partial X_{\text{L}}(z) - i\barred{z}\barred{\partial}X_{\text{R}}(\barred{z})\right]\normal{e^{i[k_{\text{L}}X_{\text{L}}(0) + k_{\text{R}}X_{\text{R}}(0)]}} & = wR\normal{e^{i[k_{\text{L}}X_{\text{L}}(0) + k_{\text{R}}X_{\text{R}}(0)]}} + \dotsc;
\end{align}
the first OPE suggests a boundary condition $r\partial_r X = -\frac{in}{R}$ to fix the momentum charge, while the second OPE suggests a boundary condition $\partial_{\theta}X = wR$ to fix the winding charge. However, it is incompatible with the equal-time canonical commutation relation to impose both of these boundary conditions, as one fixes the coordinate and the other the conjugate momentum. Instead we may use the boundary condition to fix one of the charges and construct the action so that the other charge is fixed dynamically; this is really the same situation that we saw for the Dirichlet-like approach for the noncompact free boson, so we follow this Dirichlet-like approach again here\footnote{One may try to work out the Neumann-like approach here directly, though it is more awkward to impose the winding condition dynamically on the strip. It it easier to first perform the Dirichlet-like quantization here, but with momentum $w$ and winding $n$, and then T-dualize everything.}. That is, for endpoint operators $\mathcal{O}_{n_1,w_1}(0)$ and $\mathcal{O}_{n_2,w_2}(\infty)$, we take the boundary conditions on the plane to be 
\begin{align}
\partial_{\theta}X\Big|_{|z| = \varepsilon} & = +w_1 R
\\ \partial_{\theta}X\Big|_{|z| = \frac{1}{\varepsilon}} & = -w_2 R,
\end{align}
with the relative minus sign in the latter again due to the reversed orientation of the boundary near infinity\footnote{This can be seen in two equivalent ways. One may define the operator at infinity by first transitioning to the $z' = 1/z$ coordinate patch, for which $\partial_{\theta'} = -\partial_{\theta}$. If one instead maps the operator at infinity to the origin by an actual conformal inversion via $z \mapsto z' = 1/\barred{z}$, one has $\partial_{\theta'} = \partial_{\theta}$ but also $k_{\text{L}} \leftrightharpoons k_{\text{R}}$ because the inversion flips chirality.}. We therefore wish to impose the very wide strip boundary conditions
\begin{align}
\partial_{y_1}X\Big|_{y_2 = +\ln\varepsilon} & = -w_1 R
\\ \partial_{y_1}X\Big|_{y_2 = -\ln\varepsilon} & = +w_2 R
\end{align}
via Lagrange multipliers and impose the momentum conservation dynamically as before. Therefore, the strip action is
\begin{multline}\label{compact boson action}
S = \frac{1}{4\pi}\int dy_1 dy_2\left[\left(\partial_{y_1}X\right)^2 + \left(\partial_{y_2}X\right)^2\right] - \frac{i}{2\pi}\int_{y_2 = \ln\varepsilon}\hspace{-5pt}dy_1\left[\frac{n_1}{R}X - \lambda_1\left(\partial_{y_1}X + w_1 R\right)\right]
\\ -\frac{i}{2\pi}\int_{y_2 = -\ln\varepsilon}dy_1\left[\frac{n_2}{R}X + \lambda_2\left(\partial_{y_1}X - w_2 R\right)\right],
\end{multline}
whose equations of motion in addition to the usual bulk $(\partial_{y_1}^2 + \partial_{y_2}^2)X = 0$ together with the boundary conditions are
\begin{align}
\partial_{y_1}\lambda_1(y_1) & = i\partial_{y_2}X\Big|_{y_2 = +\ln\varepsilon} - \frac{n_1}{R}
\\ \partial_{y_1}\lambda_2(y_1) & = i\partial_{y_2}X\Big|_{y_2 = -\ln\varepsilon} + \frac{n_2}{R}.
\end{align}
It should be noted that the action \eqref{compact boson action} might look strange because it contains an undifferentiated $X$ which is circle-valued. Nevertheless, the quantity that actually needs to be a well-defined functional is $e^{-S}$ on the cylinder, which it is because shifting $X$ by $2\pi R$ shifts the cylinder action by $-2\pi i(n_1+n_2)$. On the very wide strip where the quantization is actually performed, there is also no issue because $X$, while circle-valued, is forced to be single-valued since the strip has trivial homology, and again a global shift of $2\pi R$ does not change $e^{-S}$. Furthermore, the boundary Lagrange multipliers are defined to be noncompact. Another way to view the compact boson CFT is from the orbifold $\mathds{R}/2\pi R\mathds{Z}$, in which non-single-valued operators in the compact theory are viewed as defect operators in the noncompact theory; in this way functionals such as \eqref{compact boson action} are always well-defined in the covering theory.

Proceeding with the canonical quantization on the very wide strip to determine the Hilbert space and Rindler Hamiltonian is nearly identical to the Dirichlet-like approach to the noncompact free boson; we just need to put in the winding dependence. The general solution to the equations of motion and boundary conditions on the very wide strip is
\begin{align}
X(y_1,y_2) & = x + \left(x_{\text{s}} - \frac{i(n_1 - n_2)}{2R}\right)y_2 - \frac{(w_1 - w_2)R}{2}y_1 - \frac{(w_1+w_2)R}{2\ln\varepsilon}y_1 y_2 
\\ \notag & \hspace{10pt} - \sqrt{2}\sum_{\substack{n\in\mathds{Z} \\ n\neq 0}}\frac{\alpha_n}{n}e^{-\frac{\pi n}{2\ln\varepsilon}y_1}\sin\left(\frac{\pi n(y_2-\ln\varepsilon)}{2\ln\varepsilon}\right)
\\ \lambda_1(y_1) & = \lambda_{1,0} + i\left(x_{\text{s}} + \frac{i(n_1 \hspace{-2pt}+\hspace{-2pt} n_2)}{2R}\right)y_1 - \frac{i(w_1\hspace{-2pt}+\hspace{-2pt}w_2)R}{4\ln\varepsilon}y_1^2 + i\sqrt{2}\sum_{\substack{n\in\mathds{Z} \\ n\neq 0}}\frac{\alpha_n}{n}e^{-\frac{\pi n}{2\ln\varepsilon}y_1}
\\ \lambda_2(y_1) & = \lambda_{2,0} + i\left(x_{\text{s}} - \frac{i(n_1 \hspace{-2pt}+\hspace{-2pt} n_2)}{2R}\right)y_1 - \frac{i(w_1\hspace{-2pt}+\hspace{-2pt}w_2)R}{4\ln\varepsilon}y_1^2 + i\sqrt{2}\sum_{\substack{n\in\mathds{Z} \\ n\neq 0}}\frac{\alpha_n}{n}(-1)^n e^{-\frac{\pi n}{2\ln\varepsilon}y_1},
\end{align}
where canonical quantization determines that $\lambda_{1,0}$ equals $\pi(p+p_{\text{s}}/\ln\varepsilon)$ plus a constant and that $\lambda_{2,0}$ equals $-\pi(p - p_{\text{s}}/\ln\varepsilon)$ plus a constant. As before, the slope mode $x_{\text{s}}$ is defined with the same shift $-\frac{i(n_1 - n_2)}{2R}$ so that the classical Euclidean saddle point exists when $n_1 + n_2 = w_1 + w_2 = 0$ with $x_{\text{s}} = 0$, and we define the momenta $p(y_1)$ and $p_{\text{s}}(y_1)$ canonically conjugate to $x(y_1) = x - \frac{(w_1 - w_2)R}{2}y_1$ and $x_{\text{s}}(y_1) = x_{\text{s}} - \frac{(w_1+w_2)R}{2\ln\varepsilon}y_1$ again as the total momentum and first moment thereof. As $\Pi$ has the same relation to $X$, $\lambda_1$ and $\lambda_2$ as in the noncompact case, these integrals are trivially computed as
\begin{align}
\int_{\ln\varepsilon}^{-\ln\varepsilon}\hspace{-10pt}dy_2 \ \Pi(y_1,y_2) & = -\frac{i(w_1-w_2)R}{2\pi}\ln\varepsilon + \frac{\lambda_{1,0} - \lambda_{2,0}}{2\pi} -\frac{n_1+n_2}{2\pi R}y_1
\\ & \stackrel{!}{=} p(y_1) = p -\frac{n_1+n_2}{2\pi R}y_1
\\ \int_{\ln\varepsilon}^{-\ln\varepsilon}\hspace{-10pt}dy_2 \ y_2\Pi(y_1,y_2) & = -\frac{i(w_1\hspace{-2pt}+\hspace{-2pt}w_2)R}{6\pi}\ln^2\varepsilon + \frac{ix_{\text{s}}}{\pi}y_1\ln\varepsilon - \frac{i(w_1\hspace{-2pt}+\hspace{-2pt}w_2)R}{4\pi}y_1^2 +\frac{\lambda_{1,0} \hspace{-2pt}+\hspace{-2pt} \lambda_{2,0}}{2\pi}\ln\varepsilon
\\ & \stackrel{!}{=} p_{\text{s}}(y_1) = p_{\text{s}} + \frac{ix_{\text{s}}}{\pi}y_1\ln\varepsilon - \frac{i(w_1+w_2)R}{4\pi}y_1^2,
\end{align}
which determines the precise relations
\begin{align}\label{compact Lagrange multiplier zero-mode}
\lambda_{1,0} & = +\pi\left(p+\frac{p_{\text{s}}}{\ln\varepsilon}\right) + \frac{i}{2}\left(\frac{w_1+w_2}{3} + (w_1-w_2)\right)R\ln\varepsilon
\\ \lambda_{2,0} & = -\pi\left(p - \frac{p_{\text{s}}}{\ln\varepsilon}\right) + \frac{i}{2}\left(\frac{w_1+w_2}{3}-(w_1-w_2)\right)R\ln\varepsilon.
\end{align}
We emphasize that one is free to redefine the mode operators with any normalization and constant offsets preserving the commutators. The definition here that $p$ is the total momentum conjugate to the compact constant mode $x$ means that the spectrum of this operator will be $\frac{1}{R}\mathds{Z}$; one of course will obtain the same results if one alternatively defined $\lambda_{1,0} - \lambda_{2,0} = 2\pi p$, but would instead have to work with the spectrum of $p$ being $\frac{i(w_1-w_2)R}{2\pi}\ln\varepsilon + \frac{1}{R}\mathds{Z}$. The offset of $p_{\text{s}}$ does not change its continuous spectrum, but it is still a convenient calculational choice in the same spirit of the chosen offset of $x_{\text{s}}$.

The stress-energy tensor associated to the action \eqref{compact boson action} is
\begin{multline}
T_{\alpha\beta} = -\normal{\left(\partial_{\alpha}X\partial_{\beta}X - \frac{1}{2}\delta_{\alpha\beta}\partial_{\gamma}X\partial^{\gamma}X\right)} 
\\ - t_{\alpha}t_{\beta}\left[i\left(\frac{n_1}{R}X + w_1 R \lambda_1\right) \delta(y_2 - \ln\varepsilon) + i\left(\frac{n_2}{R}X + w_2 R \lambda_2\right)\delta(y_2 + \ln\varepsilon)\right]
\\ - t_{(\alpha}n_{\beta)}\partial_{y_2}X\left[i\lambda_1\delta(y_2 - \ln\varepsilon) + i\lambda_2\delta(y_2 + \ln\varepsilon)\right],
\end{multline}
so that the Rindler Hamiltonian is computed to be
\begin{align}
H_{\text{R}}^{\text{DD}} & = \frac{1}{2\pi}\hspace{-2pt}\int_{\ln\varepsilon}^{-\ln\varepsilon}\hspace{-10pt}dy_2\hspace{-2pt}\left[T(y) \hspace{-2pt}+\hspace{-2pt} \widetilde{T}(\barred{y})\right] \hspace{-2pt}-\hspace{-2pt}\frac{1}{2\pi}\hspace{-3pt}\left[i\hspace{-2pt}\left(\frac{n_1}{R}\hspace{-1pt}X \hspace{-3pt}+\hspace{-2pt} w_1 R\lambda_1\hspace{-2pt}\right)\hspace{-3pt}\Big|_{y_2=\ln\varepsilon} \hspace{-8pt}+\hspace{-2pt} i\hspace{-2pt}\left(\frac{n_2}{R}\hspace{-1pt}X \hspace{-3pt}+\hspace{-2pt} w_2 R\lambda_2\hspace{-2pt}\right)\hspace{-3pt}\Big|_{y_2 = -\hspace{-1pt}\ln\varepsilon}\right]
\\ \label{DD compact Hamiltonian} & = -\frac{i(n_1+n_2)}{2\pi R}x + \frac{i(w_1-w_2)R}{2}p + \frac{i(w_1+w_2)R}{2\ln\varepsilon}p_{\text{s}} -\frac{x_{\text{s}}^2}{2\pi}\ln\varepsilon - \frac{\pi}{2\ln\varepsilon}\sum_{n=1}^{\infty}\alpha_{-n}\alpha_n
\\ \notag & \hspace{12pt} - \frac{(n_1 - n_2)^2}{8\pi R^2}\ln\varepsilon - \frac{(w_1+w_2)^2 R^2}{24\pi}\ln\varepsilon - \frac{(w_1-w_2)^2 R^2}{8\pi}\ln\varepsilon + \frac{\pi}{48\ln\varepsilon}.
\end{align}
The finite-regulator angular quantization Hilbert space $\mathcal{H}_{n_1 w_1;n_2 w_2}^{\text{DD}}$ is spanned by the states $|m,x_{\text{s}};\{N_n\}\rangle$, where $p = \frac{m}{R}$ with $m\in\mathds{Z}$ are the discrete momentum eigenvalues. The explicit mode expansions of the stress-energy tensors and the resulting cylinder evolution operator are written in \hyperref[formulas]{Appendix \ref{formulas}}.

Note that the generic finite-regulator Hilbert space $\mathcal{H}_{n_1 w_1;n_2 w_2}^{\text{DD}}$ contains no true zero-modes because all operators and their conjugates now appear in \eqref{DD compact Hamiltonian}, unlike in the noncompact case \eqref{DD Hamiltonian answer}, though $x$ and $p_{\text{s}}$ both become zero-modes in the special case $n_1 + n_2 = w_1 + w_2 = 0$ for which the two endpoint operators are BPZ conjugate. Nevertheless, the path integral on the cylinder still has the shift symmetries in the Lagrange multiplier constant modes\footnote{Technically the path integral on the cylinder sums over all winding sectors of the free boson around the nontrivial homology cycle, and it is only one of these winding sectors that has the Lagrange multiplier shift symmetries. However, not coincidentally, the single winding sector with this symmetry is also the single winding sector that gives a nonvanishing contribution to the path integral (i.e.~it is the one which satisfies winding conservation), so the net effect is that the symmetry is still present.}, so the angular quantization prescription to compute the sphere correlators still involves taking the thermal trace divided by the volume of this symmetry group. The only difference from the noncompact case is that $p = \frac{m}{R}$ is now discrete, so the relevant volume factor is
\begin{equation}
\mathrm{Vol}(\lambda_{1,0},\lambda_{2,0}) = -\frac{4\pi^3}{R\ln\varepsilon}\left(\sum_{m\in\mathds{Z}}1\right)\int_{-\infty}^{\infty}\frac{dp_{\text{s}}}{2\pi}.
\end{equation}

The last ingredient we need before computing the thermal trace and thermal one-point function is the chiral splitting of the mode expansion $X(y_1,y_2)$ into $X_{\text{L}}(y)$ and $X_{\text{R}}(\barred{y})$. The chiral expansions are immediately written as
\begin{align}\label{compact chiral L expansion}
X_{\text{L}}(y) & = x_{\text{L}} \hspace{-2pt}-\hspace{-2pt}\frac{i}{2}\hspace{-3pt}\left(\hspace{-2pt}x_{\text{s}} \hspace{-2pt}-\hspace{-2pt} \frac{i(n_1 \hspace{-2pt}-\hspace{-2pt} n_2)}{2R} \hspace{-2pt}-\hspace{-2pt} \frac{i(w_1\hspace{-2pt}-\hspace{-2pt}w_2)R}{2}\hspace{-2pt}\right)\hspace{-2pt}y \hspace{-2pt}+\hspace{-2pt} \frac{i(w_1\hspace{-2pt}+\hspace{-2pt}w_2)R}{8\ln\varepsilon}y^2 \hspace{-2pt}-\hspace{-2pt} \frac{i}{\sqrt{2}}\sum_{\substack{n\in\mathds{Z} \\ n\neq 0}}\frac{\alpha_n}{n}e^{\frac{\pi i n}{2}}e^{-\frac{\pi n y}{2\ln\varepsilon}}
\\ \label{compact chiral R expansion} X_{\text{R}}(\barred{y}) & = x_{\text{R}} \hspace{-2pt}+\hspace{-2pt} \frac{i}{2}\hspace{-3pt}\left(\hspace{-2pt}x_{\text{s}} \hspace{-2pt}-\hspace{-2pt} \frac{i(n_1 \hspace{-2pt}-\hspace{-2pt} n_2)}{2R} \hspace{-2pt}+\hspace{-2pt} \frac{i(w_1\hspace{-2pt}-\hspace{-2pt}w_2)R}{2}\hspace{-2pt}\right)\hspace{-2pt}\barred{y} \hspace{-2pt}-\hspace{-2pt} \frac{i(w_1\hspace{-2pt}+\hspace{-2pt}w_2)R}{8\ln\varepsilon}\barred{y}^2 \hspace{-2pt}+\hspace{-2pt} \frac{i}{\sqrt{2}}\hspace{-2pt}\sum_{\substack{n\in\mathds{Z} \\ n\neq 0}}\hspace{-2pt}\frac{\alpha_n}{n}e^{\hspace{-0.5pt}-\hspace{-0.5pt}\frac{\pi in}{2}}\hspace{-1pt}e^{\hspace{-0.5pt}-\hspace{-0.5pt}\frac{\pi n\barred{y}}{2\ln\varepsilon}}\hspace{-1pt},
\end{align}
where we must determine $x_{\text{L}}$ and $x_{\text{R}}$. Their sum is of course $x_{\text{L}} + x_{\text{R}} = x$, so it remains to determine their difference, which may be inferred directly from the constant modes in the path integral boundary terms, as these must capture the corresponding constant modes of the endpoint operators $\mathcal{O}_{n_1,w_1}(0)$ and $\mathcal{O}_{n_2,w_2}(\infty)$ in the shrinking limit. This derivation is performed in \hyperref[chiral splitting]{Appendix \ref{chiral splitting}}, with the result being
\begin{align}
x_{\text{L}} & = \frac{x}{2} - \frac{\pi}{2}\left(p+\frac{p_{\text{s}}}{\ln\varepsilon}\right) + \frac{i(w_1+w_2)R}{24}\ln\varepsilon
\\ x_{\text{R}} & = \frac{x}{2} + \frac{\pi}{2}\left(p + \frac{p_{\text{s}}}{\ln\varepsilon}\right) - \frac{i(w_1+w_2)R}{24}\ln\varepsilon,
\end{align}
so that the exponential primary operator expression $\mathcal{O}_{n_3,w_3} \equiv \normal{e^{i(\frac{n_3}{R} + w_3 R)X_{\text{L}} + i(\frac{n_3}{R} - w_3 R)X_{\text{R}}}}$ is finally well-defined; all the freedom in redefining the mode operators by multiplicative and additive constants is gone, so the chiral splitting is uniquely determined by the consistency of the definition $\normal{e^{i(\frac{n_3}{R} + w_3 R)X_{\text{L}} + i(\frac{n_3}{R} - w_3 R)X_{\text{R}}}}$ with the path integral. It is very important that bulk winding operators therefore do actually depend on the momenta $p$ and $p_{\text{s}}$, which is necessary to obtain conservation of winding in traces with bulk operator insertions. It might seem strange that a winding operator, which is a perfectly well-defined bulk local operator, could depend on the Lagrange multipliers which are defined only on the boundary. However, this construction of the compact boson theory really is an orbifold of the noncompact theory, so the action \eqref{compact boson action} and all mode expansions are really constructed in the covering theory before the quotient is taken. In the noncompact theory, winding operators are defect operators and hence not local; its attached topological defect line can run all the way out to the boundary and give a dependence on $\lambda$.

The rest of the story proceeds without surprises. Let $k_{\text{L}} = \frac{n_3}{R} + w_3 R$ and $k_{\text{R}} = \frac{n_3}{R} - w_3 R$ denote the chiral charges of a bulk exponential primary. The mode expansion of this primary on the plane is
\begin{multline}\label{compact exponential expansion}
\normal{e^{i[k_{\text{L}}X_{\text{L}}(z) + k_{\text{R}}X_{\text{R}}(\barred{z})]}} 
\\ = \left(-\frac{\pi}{\ln\varepsilon}\right)^{\frac{k_{\text{L}}^2 + k_{\text{R}}^2}{4}}\frac{e^{-\frac{\pi i}{4}(k_{\text{L}}^2 - k_{\text{R}}^2)}|z|^{-\frac{\pi i}{8\ln\varepsilon}(k_{\text{L}}^2 - k_{\text{R}}^2)}\big(\frac{z}{\barred{z}}\big)^{-\frac{\pi i}{16\ln\varepsilon}(k_{\text{L}}-k_{\text{R}})^2}}{2^{\frac{(k_{\text{L}} + k_{\text{R}})^2}{4}}z^{\frac{k_{\text{L}}^2}{4}}\barred{z}^{\frac{k_{\text{R}}^2}{4}}\cos^{\frac{k_{\text{L}}k_{\text{R}}}{2}}\left(\frac{\pi\ln|z|}{2\ln\varepsilon}\right)}\circnormal{e^{i[k_{\text{L}}X_{\text{L}}(z) + k_{\text{R}}X_{\text{R}}(\barred{z})]}},
\end{multline}
where
\begin{align}
\circnormal{e^{i[k_{\text{L}}X_{\text{L}}(z) + k_{\text{R}}X_{\text{R}}(\barred{z})]}} & = \varepsilon^{-\frac{(w_1+w_2)w_3 R^2}{12}}|z|^{\frac{(n_1-n_2)n_3}{2R^2}+\frac{(w_1-w_2)w_3 R^2}{2}}\left(\frac{z}{\barred{z}}\right)^{\frac{(n_1-n_2)w_3+n_3(w_1-w_2)}{4}}
\\ \notag & \hspace{-60pt} \times e^{\frac{(w_1+w_2)R}{8\ln\varepsilon}[(\frac{n_3}{R} + w_3 R)\ln^2 z - (\frac{n_3}{R} - w_3 R)\ln^2\barred{z}]}e^{\frac{in_3}{R}x}e^{-\pi i w_3 R p}|z|^{\frac{in_3}{R}x_{\text{s}}}\hspace{-2pt}\left(\frac{z}{\barred{z}}\right)^{\frac{iw_3 R}{2}x_{\text{s}}}\hspace{-4pt}e^{-\frac{\pi i w_3 R}{\ln\varepsilon} p_{\text{s}}}
\\ \notag & \hspace{-60pt} \times \exp\left\{-i\sqrt{2}\sum_{n=1}^{\infty}\frac{\alpha_{-n}}{n}\left(\frac{z}{\barred{z}}\right)^{\frac{\pi i n}{4\ln\varepsilon}}\left[\frac{n_3}{R}\sin\left(\frac{\pi n}{2\ln\varepsilon}\ln\left|\frac{z}{\varepsilon}\right|\right) - iw_3 R\cos\left(\frac{\pi n}{2\ln\varepsilon}\ln\left|\frac{z}{\varepsilon}\right|\right)\right]\right\}
\\ \notag & \hspace{-60pt} \times \exp\left\{-i\sqrt{2}\sum_{\ell=1}^{\infty}\frac{\alpha_{\ell}}{\ell}\left(\frac{z}{\barred{z}}\right)^{-\frac{\pi i \ell}{4\ln\varepsilon}}\left[\frac{n_3}{R}\sin\left(\frac{\pi \ell}{2\ln\varepsilon}\ln\left|\frac{z}{\varepsilon}\right|\right) + iw_3 R\cos\left(\frac{\pi \ell}{2\ln\varepsilon}\ln\left|\frac{z}{\varepsilon}\right|\right)\right]\right\}.
\end{align}
It is worth pointing out that these exponential operators $\normal{e^{i(k_{\text{L}}X_{\text{L}} + k_{\text{R}}X_{\text{R}})}}$ automatically commute at equal times, a property which in radial quantization is only achieved by appending 2-cocycles by hand. This commutativity is also derived in \hyperref[chiral splitting]{Appendix \ref{chiral splitting}}, and it holds because angular quantization determines the chiral splitting for us, essentially accounting for the defect lines attached to $X_{\text{L}}$ and $X_{\text{R}}$.

The thermal trace on the sphere is then computed to be
\begin{equation}
\text{Tr}_{\mathcal{H}_{n_1 w_1;n_2 w_2}^{\text{DD}}}\left[e^{-2\pi H_{\text{R}}^{\text{DD}}}\right]_{S^2} = \frac{R\delta_{n_1+n_2,0}\delta_{w_1+w_2,0}}{4\sqrt{2}\pi^2}\frac{e^{-\frac{\pi i}{2}(n_1 w_1-n_2 w_2)}\varepsilon^{\frac{1}{2}\big(\frac{n_1^2}{R^2} + w_1^2 R^2\big)+\frac{1}{2}\big(\frac{n_2^2}{R^2} + w_2^2 R^2\big)}}{\varepsilon^{-1/6}\eta\hspace{-2pt}\left(-\frac{2i\ln\varepsilon}{\pi}\right)},
\end{equation}
and the thermal one-point function of an exponential primary is computed to be
\begin{multline}
\text{Tr}_{\mathcal{H}_{n_1 w_1;n_2 w_2}^{\text{DD}}}\hspace{-3pt}\left[\mathcal{O}_{n_3,w_3}(z,\barred{z})e^{-2\pi H_{\text{R}}^{\text{DD}}}\right]_{\hspace{-1pt}S^2} \hspace{-2pt}= \frac{R\delta_{n_1+n_2+n_3,0}\delta_{w_1+w_2+w_3,0}}{4\sqrt{2}\pi^2}\frac{e^{\hspace{-1pt}-\hspace{-1pt}\frac{\pi i}{2}(n_1 w_1\hspace{-1pt}-n_2 w_2)}\varepsilon^{h_1+\widetilde{h}_1+h_2+\widetilde{h}_2}}{z^{h_1+h_3-h_2}\barred{z}^{\widetilde{h}_1+\widetilde{h}_3-\widetilde{h}_2}}
\\ \times \frac{e^{\frac{\pi i}{2}(n_1 w_2 - n_2 w_1)}}{\varepsilon^{-1/6}\eta\hspace{-2pt}\left(-\frac{2i\ln\varepsilon}{\pi}\right)}\frac{[\varepsilon^{-1/6}\eta\hspace{-2pt}\left(-\frac{2i\ln\varepsilon}{\pi}\right)]^{\frac{1}{2}(\frac{3n_3^2}{R^2} + w_3^2 R^2)}}{\vartheta_4\hspace{-2pt}\left(\left.\hspace{-2pt}-\frac{i\ln|z|}{\pi}\right|-\frac{2i\ln\varepsilon}{\pi}\right)^{\frac{1}{2}(\frac{n_3^2}{R^2} - w_3^2 R^2)}}.
\end{multline}
Recalling that the usual normalization of the compact boson exponential operators gives the two- and three-point coefficients as\footnote{The phase factor in the three-point coefficient is easily seen to be what is required for the three-point function to be symmetric under operator exchange. There is no problem for this three-point coefficient in a unitary theory to be complex, because $\mathcal{O}_{n,w}$ is a complex observable; it is only in the ``good'' self-conjugate basis of operators with a positive two-point coefficient that the three-point coefficient must be real, which does indeed happen for the sine and cosine basis.}
\begin{align}
\left\langle \mathcal{O}_{n_1,w_1}(0)\mathcal{O}_{n_2,w_2}(1)\right\rangle_{S^2} & = 2\pi R \delta_{n_1+n_2,0}\delta_{w_1+w_2,0}
\\ C_{n_1 w_1;n_2 w_2;n_3 w_3} & = 2\pi R e^{\frac{\pi i}{2}(n_1 w_2 - n_2 w_1)}\delta_{n_1+n_2+n_3,0}\delta_{w_1+w_2+w_3,0},
\end{align}
we see that the requisite matching conditions \eqref{2-point matching} and \eqref{3-point matching} are again satisfied. Explicitly, we have found
\begin{align}
\lim_{\varepsilon\rightarrow 0}\hspace{-2pt}\left(\hspace{-2pt}e^{\frac{\pi i}{2}(s_1 - s_2)}\varepsilon^{-\Delta_1-\Delta_2}\text{Tr}_{\mathcal{H}_{1;2}^{\text{DD}}}\hspace{-2pt}\left[e^{-2\pi H_{\text{R}}^{\text{DD}}}\right]_{\hspace{-2pt}S^2}\hspace{-2pt}\right) & \hspace{-3pt}=\hspace{-2pt} \mathcal{N}_{\hspace{-1pt}n_1 w_1;n_2 w_2}^{\text{DD}}\hspace{-3pt}\left\langle\hspace{-1pt} \mathcal{O}_{\hspace{-1pt}n_1,w_1}\hspace{-1.5pt}(0)\mathcal{O}_{\hspace{-1pt}n_2,w_2}\hspace{-1.5pt}(1)\hspace{-1pt}\right\rangle_{\hspace{-2pt}S^2}
\\ \lim_{\varepsilon\rightarrow 0}\hspace{-3pt}\left(\hspace{-3pt}e^{\frac{\pi i}{2}(s_1 - s_2)}\varepsilon^{-\Delta_1-\Delta_2}\hspace{-1pt}\text{Tr}_{\hspace{-1pt}\mathcal{H}_{1;2}^{\hspace{-0.5pt}\text{DD}}}\hspace{-3pt}\left[\hspace{-1pt}\mathcal{O}_{\hspace{-0.5pt}n_3,w_3}\hspace{-1pt}(z,\barred{z})e^{\hspace{-0.5pt}-\hspace{-0.5pt}2\pi H_{\text{R}}^{\text{DD}}}\hspace{-1pt}\right]_{\hspace{-2pt}S^2}\hspace{-2pt}\right) & \hspace{-3pt}=\hspace{-2pt} \frac{\mathcal{N}_{n_1 w_1;n_2 w_2}^{\text{DD}}C_{n_1 w_1;n_3 w_3;n_2 w_2}}{z^{h_1 + h_3 - h_2}\barred{z}^{\widetilde{h}_1 + \widetilde{h}_3 - \widetilde{h}_2}}
\end{align}
with normalization constant
\begin{equation}
\mathcal{N}_{n_1 w_1;n_2 w_2}^{\text{DD}} = \frac{\sqrt{2}\pi}{(2\pi)^4},
\end{equation}
which is the same as before. Note that no phase has been generated in the normalization constant for spinning operators precisely because the spin-dependent factors in the matching conditions \eqref{2-point matching} and \eqref{3-point matching} have been included. Obviously it does not matter where these spin factors are included, but our choice of having no phases in the shrinking on the cylinder is the natural choice for which the normalization constant $\mathcal{N}^{\text{DD}}$ is always positive.

\subsection{\texorpdfstring{$bc$}{bc} Ghost System}

As a final explicit free-field example, we perform the angular quantization associated to the $bc$ ghost system. As our original motivation is to describe the angular quantization associated to certain BRST-invariant vertex operators in string theory, we want to consider the case where both endpoints operators are $c\widetilde{c}$. Even though the ghost fields $b_{\alpha\beta}$ and $c^{\alpha}$ are free, the fact that they are Grassmann variables with a first-order action makes finding a simple boundary condition shrinking to $c\widetilde{c}$ problematic. Indeed, the previous free field boundary conditions were all imposing the charge under a current, so an appropriate boundary condition here would impose the correct charge under the total ghost number current $j_{\alpha} = -\normal{b_{\alpha\beta}c^{\beta}}$. Since this current is bilinear in the fields, there will not be a linear and local boundary action which can fix it at finite regulator. The easiest way to proceed is to bosonize the $bc$ ghost system to a compact linear dilaton \cite{Friedan85}. In this language, the angular quantization of the $bc$ ghost system is nearly identical to the previous examples, essentially combining the technique of the compact boson with the background charge effects of the linear dilaton.

The bosonization of the ghost system amounts to the identifications
\begin{alignat}{2}
b(z) & \cong \normal{e^{i\sqrt{2}H_{\text{L}}(z)}} \qquad \qquad & \widetilde{b} & \cong \normal{e^{i\sqrt{2}H_{\text{R}}(\barred{z})}}
\\ c(z) & \cong \normal{e^{-i\sqrt{2}H_{\text{L}}(z)}} \qquad \qquad & \widetilde{c} & \cong \normal{e^{-i\sqrt{2}H_{\text{R}}(\barred{z})}},
\end{alignat}
where\footnote{We apologize for the horrible notation of using $H_{\text{R}}$ to denote the antichiral linear dilaton for the bosonized ghost, despite the same symbol denoting the Rindler Hamiltonian. The latter will always be denoted $H_{\text{R}}^{\text{gh}}$, so there should not be any confusion.} $H = H_{\text{L}} + H_{\text{R}}$ is a canonically normalized linear dilaton (with chiral splitting defined below) with background charge
\begin{equation}
Q_{\text{gh}} = -\frac{3i}{\sqrt{2}},
\end{equation}
which reproduces the bulk $bc$ stress-energy tensor with central charges $c = \widetilde{c} = -26$. The holomorphic and antiholomorphic ghost number currents bosonize to
\begin{align}
-\normal{b(z)c(z)} & \cong -i\sqrt{2}\partial H(z)
\\ -\normal{\widetilde{b}(\barred{z})\widetilde{c}(\barred{z})} & \cong -i\sqrt{2}\barred{\partial}H(\barred{z}).
\end{align}
That the bosonized current is linear is why this description possesses simple linear local boundary conditions shrinking to the primaries. The bosonized $bc$ primaries generically take the form of the linear dilaton exponentials $\normal{e^{-i\sqrt{2}(n_{\text{gh}}H_{\text{L}} + \widetilde{n}_{\text{gh}}H_{\text{R}})}}$ for a pair of integers $(n_{\text{gh}},\widetilde{n}_{\text{gh}}) \in \mathds{Z}^2$, with conformal weights $(h,\widetilde{h}) = (\frac{n_{\text{gh}}(n_{\text{gh}}+3)}{2},\frac{\widetilde{n}_{\text{gh}}(\widetilde{n}_{\text{gh}}+3)}{2})$ and ghost numbers $(n_{\text{gh}},\widetilde{n}_{\text{gh}})$. The ghost number anomaly coincides with the linear dilaton shift anomaly, so the ghost number anomalous conservation law (e.g.~that nonzero sphere correlators have $\sum n_{\text{gh}} = \sum \widetilde{n}_{\text{gh}} = 3$) is simply reproduced by the linear dilaton anomalous conservation law (e.g.~that nonzero sphere correlators have $-\frac{i}{\sqrt{2}}\sum n_{\text{gh}} = -\frac{i}{\sqrt{2}}\sum \widetilde{n}_{\text{gh}} = Q_{\text{gh}}$). We do not use momentum and winding notations since they are frame-dependent quantities due to the background charge. The only place where the ``radius'' of the linear dilaton $H$ matters is in the periodic identification of its constant mode $h$, whose minimal choice\footnote{The identification must be such that $\normal{e^{-i\sqrt{2}(n_{\text{gh}}H_{\text{L}} + \widetilde{n}_{\text{gh}}H_{\text{R}})}}$ is a well-defined local operator for all pairs $(n_{\text{gh}},\widetilde{n}_{\text{gh}}) \in \mathds{Z}^2$. The part of this operator involving $h$ is $e^{-\frac{i}{\sqrt{2}}(n_{\text{gh}}+\widetilde{n}_{\text{gh}})h}$, which is single-valued only if $h$ is identified with a shift of $2\sqrt{2}\pi$ times some integer. It is irrelevant whether we choose $2\sqrt{2}\pi$ as the period or any integer multiple of it, because the momentum eigenvalues will be divided by that integer but the Hamiltonian does not depend on this momentum at all, so we might as well make the minimal choice.} is $h \sim h + 2\sqrt{2}\pi$; the bosonized field $H$ should not be thought of as a target space dimension compactified at radius $R = \sqrt{2}$.

Since the linear dilaton is compact, we choose the Dirichlet-like boundary conditions so that the compact boson results may be borrowed directly, even though the endpoint operators $c\widetilde{c} \cong\s\s\s\normal{e^{-i\sqrt{2}H}}$ themselves have no ``winding'' (but $c$ and $\widetilde{c}$ individually do). That is, we take the very wide strip boundary conditions to be
\begin{equation}
\partial_{y_1}H\Big|_{y_2 = \pm\ln\varepsilon} = 0
\end{equation}
and impose the ghost number charges dynamically via the action
\begin{multline}
S = \frac{1}{4\pi}\int dy_1 dy_2\left[\left(\partial_{y_1}H\right)^2 + \left(\partial_{y_2}H\right)^2\right]
\\ - \frac{i}{2\pi}\int_{y_2=\ln\varepsilon}\hspace{-6pt}dy_1\left[\frac{1}{\sqrt{2}}H - \lambda_1\partial_{y_1}H\right] - \frac{i}{2\pi}\int_{y_2 = -\ln\varepsilon}\hspace{-8pt}dy_1\left[\frac{1}{\sqrt{2}}H + \lambda_2\partial_{y_1}H\right],
\end{multline}
where the boundary action terms linear in $H$ are obtained from the linear dilaton result \eqref{LD strip action} by setting $\alpha_1 = \alpha_2 = -\frac{i}{\sqrt{2}}$ and $Q = Q_{\text{gh}} = -\frac{3i}{\sqrt{2}}$. When going back to the plane, we must remember the anomalous transformation $H(z,\barred{z}) = H\big(y(z),\barred{y}(\barred{z})\big) - Q_{\text{gh}}\ln|z|$ and its derivatives.

The mode expansions of $H(y_1,y_2) = H_{\text{L}}(y) + H_{\text{R}}(\barred{y})$ and of the Lagrange multipliers on the very wide strip are
\begin{align}
H_{\text{L}}(y) & = \frac{h}{2} - \frac{\pi(p_h + \frac{p_{\text{s}}}{\ln\varepsilon})}{2} - \frac{ih_{\text{s}}}{2}y - \frac{i}{\sqrt{2}}\sum_{\substack{n\in\mathds{Z} \\ n\neq 0}}\frac{h_n}{n}e^{\frac{\pi i n}{2}}e^{-\frac{\pi n y}{2\ln\varepsilon}}
\\ H_{\text{R}}(\barred{y}) & = \frac{h}{2} + \frac{\pi(p_h + \frac{p_{\text{s}}}{\ln\varepsilon})}{2} + \frac{ih_{\text{s}}}{2}\barred{y} + \frac{i}{\sqrt{2}}\sum_{\substack{n\in\mathds{Z} \\ n\neq 0}}\frac{h_n}{n}e^{-\frac{\pi i n}{2}}e^{-\frac{\pi n\barred{y}}{2\ln\varepsilon}}
\\ \lambda_1(y_1) & = +\pi\left(p_h + \frac{p_{\text{s}}}{\ln\varepsilon}\right) + i\left(h_{\text{s}} + \frac{i}{\sqrt{2}}\right)y_1 + i\sqrt{2}\sum_{\substack{n\in\mathds{Z} \\ n\neq 0}}\frac{h_n}{n}e^{-\frac{\pi n}{2\ln\varepsilon}y_1}
\\ \lambda_2(y_1) & = -\pi\left(p_h - \frac{p_{\text{s}}}{\ln\varepsilon}\right) + i\left(h_{\text{s}} - \frac{i}{\sqrt{2}}\right)y_1 + i\sqrt{2}\sum_{\substack{n\in\mathds{Z} \\ n\neq 0}}\frac{h_n}{n}(-1)^n e^{-\frac{\pi n}{2\ln\varepsilon}y_1}.
\end{align}
The rest of the cylinder computations are identical for that of the compact boson, so we may directly import the Rindler Hamiltonian as
\begin{equation}\label{ghost Hamiltonian}
H_{\text{R}}^{\text{gh}} = -\frac{i}{\sqrt{2}\pi}h - \frac{h_{\text{s}}^2}{2\pi}\ln\varepsilon - \frac{\pi}{2\ln\varepsilon}\sum_{n=1}^{\infty}h_{-n}h_n + \frac{\pi}{48\ln\varepsilon},
\end{equation}
and we take the basis of the finite-regulator angular quantization Hilbert space $\mathcal{H}_{c\widetilde{c}}^{\text{gh}}$ to consist of the states $|m_h,h_{\text{s}};\{N_n\}\rangle$ with eigenvalues $h_{\text{s}} \in \mathds{R}$ and $p_h = \frac{m_h}{\sqrt{2}}$, $m_h\in\mathds{Z}$; as usual, the thermal traces are defined in this Dirac orthonormalized basis together with dividing by the Lagrange multiplier zero-mode volume
\begin{equation}
\mathrm{Vol}(\lambda_{1,0},\lambda_{2,0}) = -\frac{2\sqrt{2}\pi^3}{\ln\varepsilon}\Bigg(\sum_{m_h\in\mathds{Z}}1\Bigg)\int_{-\infty}^{\infty}\frac{dp_{\text{s}}}{2\pi}.
\end{equation}
When computing the trace over the evolution operator $e^{-2\pi H_{\text{R}}^{\text{gh}}}$, the fact that $\langle m_h| e^{i\sqrt{2}h}|m_h\rangle = \langle m_h-2|m_h\rangle = 0$ immediately says that
\begin{equation}
\text{Tr}_{\mathcal{H}_{c\widetilde{c}}^{\text{gh}}}\left[e^{-2\pi H_{\text{R}}^{\text{gh}}}\right]_{S^2} = 0,
\end{equation}
which should vanish (at least in the shrinking limit) because $\langle cc\widetilde{c}\widetilde{c}\rangle_{S^2} = 0$. The first nonvanishing correlator on the sphere must come from the thermal two-point function of $c(z_1)\widetilde{c}(\barred{z}_2)$ and must reproduce the $\varepsilon\rightarrow 0$ limit of the correlator
\begin{equation}
\left\langle (c\widetilde{c})(\varepsilon)c(z_1)\widetilde{c}(\barred{z}_2)(c\widetilde{c})(\varepsilon^{-1})\right\rangle_{S^2} \stackrel{\varepsilon\rightarrow 0}{\longrightarrow} \frac{1}{\varepsilon^4}z_1\barred{z}_2;
\end{equation}
we define the $bc$ ghost system so that this first nonzero correlator has this normalization and sign. Of course, thanks to holomorphic factorization the thermal two-point function here is equivalent to a thermal one-point function via
\begin{equation}
\normal{e^{-i\sqrt{2}H_{\text{L}}(z_1)}}\normal{e^{-i\sqrt{2}H_{\text{R}}(\barred{z}_2)}} = e^{-\barred{z}_{12}\partial_{\barred{z}_1}}\normal{e^{-i\sqrt{2}H(z_1,\barred{z}_1)}}.
\end{equation}
It is no more difficult to compute the mode expansion and thermal one-point function of the general exponential $\normal{e^{-i\sqrt{2}[n_{\text{gh}}H_{\text{L}}(z) + \widetilde{n}_{\text{gh}}H_{\text{R}}(\barred{z})]}}$, but we already know that the linear dilaton anomalous conservation law will immediately cause all thermal one-point functions to vanish except that of $c\widetilde{c}$ with $n_{\text{gh}} = \widetilde{n}_{\text{gh}} = 1$. So here we only record the explicit mode expansions of the $c$ ghosts themselves on the plane, which are
\begin{align}
c(z) & = -i\sqrt{-\frac{\pi}{2\ln\varepsilon}} \ z^{1-\frac{\pi i}{4\ln\varepsilon}}e^{-\frac{i}{\sqrt{2}}h}e^{\frac{\pi i}{\sqrt{2}}p_h}e^{-\frac{i\ln z}{\sqrt{2}}h_{\text{s}}}e^{\frac{\pi i}{\sqrt{2}\ln\varepsilon}p_{\text{s}}}
\\ \notag & \hspace{10pt} \times \exp\left[\sum_{n=1}^{\infty}\frac{h_{-n}}{n}e^{-\frac{\pi i n}{2}}z^{\frac{\pi i n}{2\ln\varepsilon}}\right]\exp\left[-\sum_{n'=1}^{\infty}\frac{h_{n'}}{n'}e^{\frac{\pi i n'}{2}}z^{-\frac{\pi i n'}{2\ln\varepsilon}}\right]
\\ \widetilde{c}(\barred{z}) & = +i\sqrt{-\frac{\pi}{2\ln\varepsilon}} \ \barred{z}^{1+\frac{\pi i}{4\ln\varepsilon}}e^{-\frac{i}{\sqrt{2}}h}e^{-\frac{\pi i}{\sqrt{2}}p_h}e^{-\frac{i\ln \barred{z}}{\sqrt{2}}h_{\text{s}}}e^{-\frac{\pi i}{\sqrt{2}\ln\varepsilon}p_{\text{s}}}
\\ \notag & \hspace{10pt} \times \exp\left[-\sum_{n=1}^{\infty}\frac{h_{-n}}{n}e^{\frac{\pi i n}{2}}\barred{z}^{-\frac{\pi i n}{2\ln\varepsilon}}\right]\exp\left[\sum_{n'=1}^{\infty}\frac{h_{n'}}{n'}e^{-\frac{\pi i n'}{2}}\barred{z}^{\frac{\pi i n'}{2\ln\varepsilon}}\right].
\end{align}
Even though these expansions individually are not needed here, they are required for example in the computation of the worldsheet BRST charge in angular quantization, which we present in \cite{Agia22}. The finite-regulator thermal two-point function on the sphere is then easily shown to be
\begin{equation}
\text{Tr}_{\mathcal{H}_{c\widetilde{c}}^{\text{gh}}}\left[c(z_1)\widetilde{c}(\barred{z}_2)e^{-2\pi H_{\text{R}}^{\text{gh}}}\right]_{S^2} = \frac{z_1\barred{z}_2}{4\pi^2\varepsilon^4}\left[\frac{\varepsilon^{-1/3}\eta^2\hspace{-2pt}\left(-\frac{2i\ln\varepsilon}{\pi}\right)}{\vartheta_4\hspace{-2pt}\left(\left.\hspace{-2pt}-\frac{i\ln(z_1\barred{z}_2)}{2\pi}\right|\hspace{-2pt}-\hspace{-2pt}\frac{2i\ln\varepsilon}{\pi}\right)}\right],
\end{equation}
which indeed satisfies the angular quantization matching condition
\begin{equation}
\lim_{\varepsilon\rightarrow 0}\left(\varepsilon^{4}\text{Tr}_{\mathcal{H}_{c\widetilde{c}}^{\text{gh}}}\left[c(z_1)\widetilde{c}(\barred{z}_2)e^{-2\pi H_{\text{R}}^{\text{gh}}}\right]_{S^2}\right) = \mathcal{N}_{c\widetilde{c}}^{\text{gh}}z_1\barred{z}_2
\end{equation}
with normalization
\begin{equation}
\mathcal{N}_{c\widetilde{c}}^{\text{gh}} = \frac{1}{(2\pi)^2}.
\end{equation}
Note that this differs from the compact boson normalization constant because the correlators of the bosonized exponential primaries are not normalized to have two-point coefficient $2\pi R$, since there is no target space interpretation.

Even though we did not need the explicit mode expansions of the general exponential primaries $\mathcal{O}_{n_{\text{gh}},\widetilde{n}_{\text{gh}}} = \normal{e^{-i\sqrt{2}(n_{\text{gh}}H_{\text{L}} + \widetilde{n}_{\text{gh}}H_{\text{R}})}}$ here, we should point out that they also automatically obey the correct operator algebra, meaning that they anticommute at equal times if and only if both $bc$ primaries are Grassman odd and commute at equal times otherwise. This derivation is also performed in \hyperref[chiral splitting]{Appendix \ref{chiral splitting}}, with the result given in \eqref{ghost exponential (anti)commutator}.

It should be clear that the procedure laid out in this section immediately generalizes to the free fermion CFT as well as to the $\beta\gamma$ superghost system of the RNS formalism, as the former can be written as a twisted $bc$ ghost system (where the twist ensures the correct weights and central charges) and the latter may be bosonized into another twisted $bc$ ghost system plus linear dilaton (the latter being the picture field).

\section{Connection to Entanglement Entropy}\label{EE}

Finally, we mention the connection to the computation of 2d entanglement entropy. Most of the set-up described below is well-known in the literature; for a review, see for example \cite{Casini09} and references therein. The idea that the simplest CFT entanglement scenarios are those for which there exists a map to the annulus was expounded in \cite{Cardy16}. Using this idea, the computation of vacuum entanglement entropy of an interval for the 2d free boson with different boundary conditions via oscillator methods was already performed in \cite{Michel16} by utilizing the properties of the modular Hamiltonian. However, the interpretation of the methodology here is different. In particular, the role of angular quantization in the general set-up is manifest, and the incorporation of states other than the vacuum is straightforward. This construction also emphasizes that the associated ``entanglement spectrum'' is not merely a mathematical tool but is used to define a Minkowski Hilbert space, and it highlights the need to enumerate the full class of boundary condition regulators for which the physical part of the entanglement entropy is actually independent of boundary condition.

Consider the CFT on the plane in a state $\rho$, where we wish to compute the entanglement entropy between an interval $A$ of length $L$ and its complement $\barred{A}$, which is
\begin{equation}
S_{\text{EE}} \ \text{``$=$''} -\text{Tr}_{\mathcal{H}_{A}}\left[\rho_A\ln\rho_A\right],
\end{equation}
where $\rho_A = \text{Tr}_{\mathcal{H}_{\barred{A}}}\rho$ is the reduced density matrix on the interval $A$. The equality sign is in quotes because the expression on the right side is not well-defined, due to the lack of a tensor product decomposition of the Hilbert space of states in continuum quantum field theory, $\mathcal{H} \neq \mathcal{H}_A\otimes\mathcal{H}_{\barred{A}}$. 

In the context of algebraic quantum field theory, a rigorously defined quantity is the relative entropy, from which mutual information between subregions can be defined. However, there are applications in which various finite derivatives of the entanglement entropy itself play an important role, and in general those are a distinct quantity from mutual informations. For example, in (2+1)-dimensional theories that flow to a gapped phase described by Chern-Simons theory, entanglement entropies are nontrivial, while mutual informations vanish \cite{Casini15}. It remains an open problem to give a precise definition of entanglement entropy and its unambiguous, regulator-independent content in QFT. 

It is of course straightforward to define entanglement entropy in finite-dimensional lattice Hilbert spaces, and one may introduce a UV regulator in a QFT in order to compute a regulated reduced density matrix and the von Neumann entropy thereof; the real challenge is finding the necessary and sufficient conditions guaranteeing that the physical quantities extracted from the entanglement entropy are actually finite and independent of regulator. 

Of particular interest is the study of RG monotones, beginning with the famous Zamolodchikov $c$-theorem in 2d \cite{Zamolodchikov86} and its later generalizations to the 3d $F$-theorem and the 4d $a$-theorem, which have been proposed to be constructed in $d$ dimensions from certain entanglement entropy quantities \cite{Myers10}. While there has been strong evidence in support of the generalized $d$-dimensional $F$-theorem, there is still work to be done in proving the universality of its entanglement entropy construction. 

One notable roadblock in proceeding beyond formal arguments is the computational difficulty even for free theories. The most widely-adopted prescription is the replica trick, wherein one defines the $n^{\text{th}}$ R\'{e}nyi entropy
\begin{equation}\label{Renyi entropy def}
S_n = -\frac{1}{n-1}\ln\left(\frac{\text{Tr}_{\mathcal{H}_A} \rho_A^n}{(\text{Tr}_{\mathcal{H}_A}\rho_A)^n}\right),
\end{equation}
analytically continues in $n$ (assuming some boundedness properties) and constructs the entanglement entropy as
\begin{equation}
S_{\text{EE}} = \lim_{n\rightarrow 1}S_n = -\frac{\partial}{\partial n}\left.\ln\left(\frac{\text{Tr}_{\mathcal{H}_A} \rho_A^n}{(\text{Tr}_{\mathcal{H}_A}\rho_A)^n}\right)\right|_{n=1}.
\end{equation}

\begin{figure}
\centering
\begin{tikzpicture}
\draw[|-|,gray] (-6,1) -- node[midway,fill=white,scale=1]{$L$} (-3,1);
\draw[|-|,gray] (-6.5,0.25) -- node[midway,fill=white,scale=0.9]{$\hspace{-2pt}2\varepsilon\hspace{-2pt}$} (-5.5,0.25);
\draw[|-|,gray] (-3.5,0.25) -- node[midway,fill=white,scale=0.9]{$\hspace{-2pt}2\varepsilon\hspace{-2pt}$} (-2.5,0.25);
\begin{scope}
\clip (-8,0) -- (-6.5,0) arc (-180:-115:0.5) -- ++(0,-0.6) -- ++({2*0.5*sin(25)},0) -- ++(0,0.6) arc (-65:0:0.5) -- (-3.5,0) arc (-180:-115:0.5) -- ++(0,-0.6) -- ++({2*0.5*sin(25)},0) -- ++(0,0.6) arc (-70:0:0.5) -- (-1,0) -- (-1,-2) -- (-8,-2) -- (-8,0);
\foreach \x in {-20,...,20}{
\draw[gray] ({-4.5-5*(1)},{-1-5*(1)+0.25*(\x)}) -- ({-4.5+5*(1)},{-1+5*(1)+0.25*(\x)});
}
\end{scope}
\draw (-8,0) -- node[midway,above,scale=1]{$\barred{A}$} (-6.5,0);
\draw (-5.5,0) -- node[midway,above,scale=1]{$A$} (-3.5,0);
\draw (-2.5,0) -- node[midway,above,scale=1]{$\barred{A}$} (-1,0);
\draw[very thick] (-6.5,0) arc (-180:0:0.5) node[midway,below,scale=1]{$B$};
\draw[very thick] (-3.5,0) arc (-180:0:0.5) node[midway,below,scale=1]{$B$};
\draw[very thick] (1.5,0) node[left,scale=1]{$B$} arc (180:5:0.5);
\draw[very thick] (1.5,0) arc (-180:-5:0.5);
\draw ({2+0.5*cos(5)},{0+0.5*sin(5)}) -- node[midway,above,scale=1]{$\psi'_A$} ({5-0.5*cos(5)},{0+0.5*sin(5)});
\draw ({2+0.5*cos(5)},{0-0.5*sin(5)}) -- node[midway,below,scale=1]{$\psi_A$} ({5-0.5*cos(5)},{0-0.5*sin(5)});
\draw[very thick] (5.5,0) node[right,scale=1]{$B$} arc (0:175:0.5);
\draw[very thick] (5.5,0) arc (0:-175:0.5);
\end{tikzpicture}
\caption{Left: The regularized splitting map $\imath_B(\varepsilon):\mathcal{H}\hookrightarrow \mathcal{H}_A^B\otimes\mathcal{H}_{\barred{A}}^B$ as defined by the Euclidean path integral over the shaded region. Right: The regularized reduced density matrix element $\langle \psi_A'|\rho_A|\psi_A\rangle$ as defined by the Euclidean path integral over everywhere outside the two excised holes.}
\label{EE splitting map}
\end{figure}
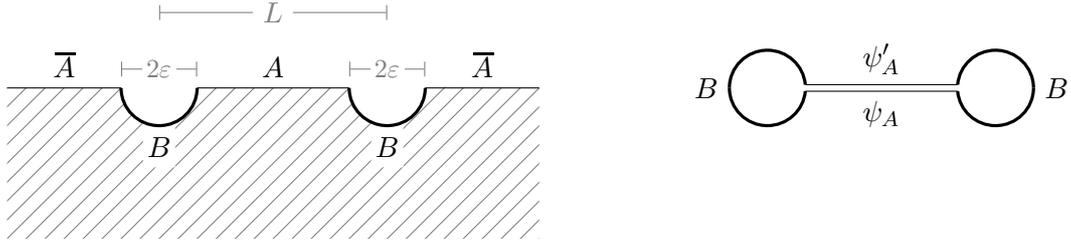

To proceed, one must define $\rho_A$ and its integer powers $\rho_A^n$ by introducing a UV regulator $\varepsilon$ and defining a map which splits the Hilbert space into a tensor product. A natural methodology is supplied by the Euclidean path integral \cite{Ohmori14}; we consider only the simplest case where $A$ is a single line segment of length $L$.
A state in the Hilbert space $\mathcal{H}$ at any fixed $\mathrm{Im}\s\s z < 0$ in the plane is prepared by performing the Euclidean path integral up from $\mathrm{Im}\s\s z = -\infty$ with appropriate initial conditions. One then performs the path integral all the way up to $\mathrm{Im}\s\s z = 0$ except for excised half-disks of radius $\varepsilon$ around the endpoints of $A$, on which one places some fixed boundary condition $B$ as depicted on the left in Figure \ref{EE splitting map}.

This procedure defines a regularized splitting map
\begin{equation}
\imath_B(\varepsilon): \mathcal{H} \lhook\joinrel\longrightarrow \mathcal{H}_A^B\otimes\mathcal{H}_{\barred{A}}^B,
\end{equation}
where $\mathcal{H}_A^B$ is the Hilbert space of states on a finite line segment of length $L-2\varepsilon$ flanked by boundary conditions $B$; we must demand that in the $\varepsilon \rightarrow 0$ limit, the splitting map is isometric, meaning that $\lim_{\varepsilon \rightarrow 0} \imath_B(\varepsilon)^{\dagger}\imath_B(\varepsilon) = \mathds{1}_{\mathcal{H}}$. The split state $\rho$ is then defined so that its matrix element between any $|\psi_A\rangle\otimes|\barred{\psi}_{\barred{A}}\rangle \in \mathcal{H}_A^B\otimes\mathcal{H}_{\barred{A}}^B$ and (the CPT conjugate of) any $|\psi'_A\rangle\otimes|\barred{\psi}'_{\barred{A}}\rangle \in \mathcal{H}_A^B\otimes\mathcal{H}_{\barred{A}}^B$ equals the Euclidean path integral over the entire two-holed plane obtained by gluing the lower-half-plane with future boundary conditions $(\psi_A,\barred{\psi}_{\barred{A}})$ at $\mathrm{Im}\s\s z = 0$ to the reversed upper-half-plane with past boundary conditions $(\psi'_A,\barred{\psi}'_{\barred{A}})$ at $\mathrm{Im}\s\s z = 0$. The finite-regulator reduced density matrix $\rho_A$ is obtained by tracing over $\mathcal{H}_{\barred{A}}^B$, which in the path integral language just produces the completeness relation on $\barred{A}$ at $\mathrm{Im}\s\s z = 0$ and is hence equivalent to erasing the boundary conditions on $\barred{A}$. That is, the matrix elements $\langle \psi'_A|\rho_A|\psi_A\rangle$ of the reduced density matrix are computed by the path integral shown on the right in Figure \ref{EE splitting map}; the trace $\text{Tr}_{\mathcal{H}_A} \rho_A$ is therefore just the partition function on the two-holed plane, which we keep explicit here instead of normalizing it to unity. 

The integer powers $\rho_A^n$ are then defined by inserting completeness relations between each factor of $\rho_A$ so that the matrix element $\langle \psi'_A|\rho_A^n|\psi_A\rangle$ is given by the path integral on the $n$-fold copy of the diagram on the right in Figure \ref{EE splitting map} with the boundary condition on the upper branch on the $i^{\text{th}}$ copy identified and summed over with the boundary condition on the lower branch of the $(i+1)^{\text{th}}$ copy for $1\leqslant i \leqslant n-1$. The trace $\text{Tr}_{\mathcal{H}_A}\rho_A^n$ is therefore just the partition function on the $n$-fold cover of the two-holed plane. Note that this $n$-fold cover does not have a conical singularity; topologically, the two-holed plane is a cylinder, and the $n$-fold cover of a cylinder is again a cylinder whose compact direction is $n$ times as large.

So far our discussion has been completely standard. The main open question is --- what are the admissible boundary conditions $B$ such that one may extract a universal finite quantity from the entanglement entropy which is independent of $B$? The most obvious condition is that $B$ must be a boundary condition which shrinks to the identity operator, since the right side of Figure \ref{EE splitting map} is supposed to be the regularized version of the plane branched at the two endpoints of the interval, and there are no nontrivial operator insertions at these branching points; if $B$ were to shrink to a nontrivial operator, then the splitting map would not be isometric in the limit.

As we have discussed in \hyperref[general construction]{Section \ref{general construction}}, this shrinkability condition is easy to satisfy in a 2d CFT since most boundary conditions will shrink to the identity operator. However, shrinkability is far less obvious for non-conformal 2d QFTs as well as for higher-dimensional field theories. In the latter, one regularizes the reduced density matrix by excising a tubular neighborhood of the boundary subregion, which will shrink to a codimension-2 surface operator. Lack of shrinkability is the culprit behind earlier results in \cite{Dowker10} for 4d free Maxwell theory seemingly not agreeing with the general result from the $a$-anomaly. This issue was later rectified in \cite{Donnelly15} which obtained the correct entanglement entropy in agreement with the $a$-anomaly. In the latter, the discrepancy was found to be from so-called ``edge modes'' which had to be present for the boundary condition to be admissible. However, we emphasize that all that matters is shrinkability of the boundary condition, which is the common ground which must unite all different approaches.

It is also natural to demand that an admissible $B$ be a local boundary condition. Locality in Euclidean time is required to express the path integral in the Hilbert space formulation and give a Lorentzian entropy interpretation. Furthermore, one might require the boundary conditions to respect various symmetries of the problem, such as local Lorentz covariance. 

Angular quantization applied to non-conformal 2d QFTs can provide oscillator methods of computing entanglement entropy quantities that may prove useful. Here we just comment on the well-understood conformal case; the hope is that more general situations are amenable to similar methods. The regulated prescription of computing $\text{Tr}_{\mathcal{H}_A}\rho_A^n$ above is simply angular quantization after performing an appropriate special conformal transformation. Since $\mathrm{PSL}(2,\mathds{C})$ transformations map circles to circles but do not map their centers to each other, we slightly modify the boundary locations drawn in Figure \ref{EE splitting map} so that after this transformation the boundary circles are symmetrically placed at $|z'| = \varepsilon'$ and $|z'| = \frac{1}{\varepsilon'}$ like before. 

\begin{figure}
\centering
\begin{tikzpicture}
\draw[very thick] (-7,0) arc (180:5:0.5);
\draw[very thick] (-7,0) arc (-180:-5:0.5);
\draw[dashed] ({-6.5+0.5*cos(5)},{0+0.5*sin(5)}) -- ({-2.5-0.5*cos(5)},{0+0.5*sin(5)});
\draw[dashed] ({-6.5+0.5*cos(5)},{0-0.5*sin(5)}) -- ({-2.5-0.5*cos(5)},{0-0.5*sin(5)});
\draw[very thick] (-2,0) arc (0:175:0.5);
\draw[very thick] (-2,0) arc (0:-175:0.5);
\draw[gray,fill=gray] (-6.3,0) circle (0.05);
\draw[gray,fill=gray] (-2.7,0) circle (0.05);
\draw[gray,|-|] (-6.3,1) -- node[midway,scale=1,fill=white]{$L$} (-2.7,1);
\draw[gray,|-|] (-6.5,-1) -- node[midway,scale=1,fill=white]{$\sqrt{L^2+4\varepsilon^2}$} (-2.5,-1);
\draw[very thick] (5,0) arc (0:175:0.5);
\draw[very thick] (5,0) arc (0:-175:0.5);
\draw[dashed] ({4.5-0.5*cos(5)},{0+0.5*sin(5)}) -- ({4.5-2.5*cos(asin(sin(5)/5))},{0+0.5*sin(5)});
\draw[dashed] ({4.5-0.5*cos(5)},{0-0.5*sin(5)}) -- ({4.5-2.5*cos(asin(sin(5)/5))},{0-0.5*sin(5)});
\draw[very thick] (7,0) arc (0:{180-asin(sin(5)/5)}:2.5);
\draw[very thick] (7,0) arc (0:{-180+asin(sin(5)/5)}:2.5);
\node[scale=1] (1) at (5.75,0) {$\times$};
\node[scale=2] (arrow) at (0,0) {$\Longrightarrow$};
\node[scale=0.8] (PSL) at (0,0.5) {$\mathrm{PSL}(2,\mathds{C})$};
\node[scale=0.9,gray] (inner) at (4.5,0.8) {$|z'|=\varepsilon'$};
\node[scale=0.9,gray] (outer) at (4.5,-2) {$\displaystyle |z'| = \frac{1}{\varepsilon'}$};
\node[scale=0.9,gray] (op) at (5.75,-0.3) {$z'=1$};
\end{tikzpicture}
\caption{The $\mathrm{PSL}(2,\mathds{C})$ transformation mapping the two-holed plane used in computing entanglement entropy to the symmetric two-holed plane used to construct angular quantization. The point at infinity is mapped to $z' = 1$.}
\label{EE AQ}
\end{figure}
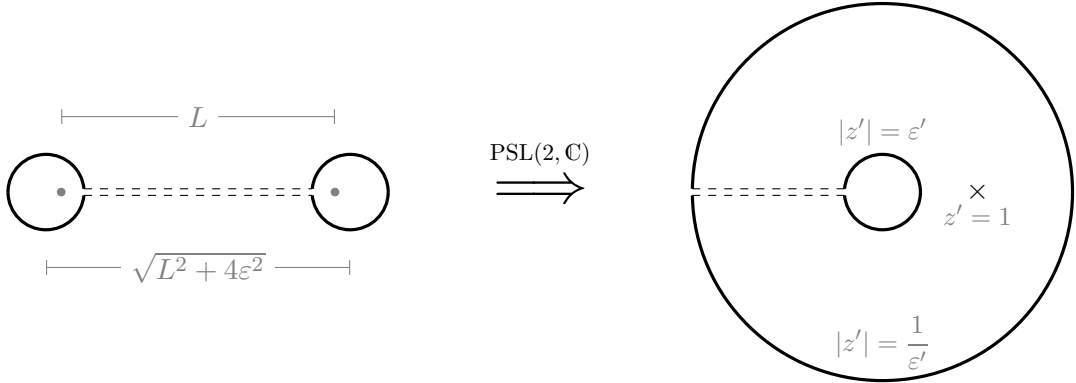

Taking the original interval $A$ to have endpoints at $z = 0$ and $z = L$, we define the boundary circles to have radius $\varepsilon$ but centered at the points $z = \frac{1}{2}(L \pm  \sqrt{L^2 + 4\varepsilon^2})$. Then, the special conformal transformation
\begin{equation}\label{EE special conformal transformation}
z \longmapsto z' = \frac{z}{z-L}
\end{equation}
maps the boundaries to the circles at $\ln |z'| = \pm\ln\varepsilon'$, where 
\begin{equation}\label{EE primed regulator}
\varepsilon' = \frac{\sqrt{L^2 + 4\varepsilon^2}-L}{2\varepsilon};
\end{equation}
see Figure \ref{EE AQ}.

The R\'{e}nyi entropy traces $\text{Tr}_{\mathcal{H}_A}\rho_A^n$ are then computed as sphere thermal traces over $e^{-2\pi n H_{\text{R}}}$ in the appropriate angular quantization Hilbert space. For a chosen shrinkable boundary condition, the latter depends only on the regulator $\varepsilon'$, which itself depends only on the ratio $\frac{\varepsilon}{L}$; at very small $\varepsilon$, the relation is $\varepsilon' = \frac{\varepsilon}{L} + \dotsc$. The only difference between the trace needed for the replica trick and the associated sphere thermal trace over $2\pi n$ angular evolution is the Weyl anomaly induced by the transformation \eqref{EE special conformal transformation}. Fortunately, this Weyl anomaly is almost inconsequential as it cancels in the R\'{e}nyi entropy definition \eqref{Renyi entropy def} because the curvatures and Weyl transformation factor are all single-valued functions and so the Weyl anomaly for $\mathrm{Tr}_{\mathcal{H}_A}\rho_A^n$ is exactly the $n^{\text{th}}$ power of the Weyl anomaly for $\mathrm{Tr}_{\mathcal{H}_A}\rho_A$.

One must be slightly careful because the flat $z'$-plane metric is obtained from the flat $z$-plane metric by applying the Weyl transformation $g_{\alpha\beta} \mapsto g'_{\alpha\beta} = e^{-2\omega}g_{\alpha\beta}$ with $\omega(z',\barred{z}') = \ln(\frac{L}{|z'-1|^2})$, which diverges at $z' = 1$. One should really cut out a hole of some small radius $\delta'$ centered around $z' = 1$ and place an appropriate boundary condition there. The Weyl anomaly still cancels in \eqref{Renyi entropy def} for the same reason as before, and now the new boundary condition on the $z$-plane fixes the asymptotic path integral behavior near the point at infinity and hence defines the total state $\rho$ in which we wish to compute the entanglement entropy. For example, if we place a shrinkable boundary condition there, then the thermal trace over the $z'$-plane will compute the entanglement entropy of an interval in the vacuum on $S^1$. Similarly, the entanglement entropy of an interval in any pure state $\rho = |\Psi_{\mathcal{O}}\rangle\langle\Psi_{\mathcal{O}}|$ simply corresponds to the thermal trace over the $z'$-plane with the operator $\mathcal{O}$ inserted at $z' = 1$ by the usual state/operator correspondence; in particular, $\text{Tr}_{\mathcal{H}_A}\rho_A^n$ involves $n$ insertions of $\mathcal{O}$, one at each pre-image of $z' = 1$ in the $n$-fold cover. Finally, the $z'$ thermal traces are easily computed by transforming to the $y'$ cylinder frame, where the Weyl anomaly action is again simply multiplied by $n$ for the $n$-fold cover. The $n$-fold cover cylinder trace is then evaluated by changing coordinates to $y'' = y'/n$. In these new coordinates, the cylinder is covered exactly once with the $n$ insertions of $\mathcal{O}(y_1'',y_2'')$ at $y_1'' = 0$ separated by angle $\Delta y_2'' = \frac{2\pi}{n}$; the only change is that the boundaries are now located at $y_2'' = \pm \frac{\ln\varepsilon'}{n}$. 

Therefore, the end result is that the traces involved in the regularized R\'{e}nyi entropy \eqref{Renyi entropy def} of an interval in any pure state $\rho = |\Psi_{\mathcal{O}}\rangle\langle\Psi_{\mathcal{O}}|$ are computed in angular quantization by
\begin{equation}
\text{Tr}_{\mathcal{H}_A}\rho_A^n = e^{-nS_{\text{W}}(\varepsilon'^{1/n})}\text{Tr}_{\mathcal{H}_{\mathds{1}}}\hspace{-3pt}\left[\mathcal{O}(0,0)\mathcal{O}\left(0,\tfrac{2\pi}{n}\right)\cdots\mathcal{O}\left(0,\tfrac{2\pi(n-1)}{n}\right)\hspace{-2pt}e^{-2\pi H_{\text{R}}^{\mathds{1}}(\varepsilon'^{1/n})}\right]_{S^1\times\mathds{R}},
\end{equation}
where $S_{\text{W}}$ is the Weyl anomaly action to go from the cylinder frame to the sphere frame, $\mathcal{H}_{\mathds{1}}$ is the angular quantization Hilbert space associated to two local boundary conditions which shrink to the identity with associated Rindler Hamiltonian $H_{\text{R}}^{\mathds{1}}$, and we have explicitly indicated that all thermal traces are to be evaluated at regulator $\varepsilon'^{1/n}$ where $\varepsilon'$ is given by \eqref{EE primed regulator}. The benefit of performing the calculation at finite regulator here is that the replica direction is an isometry (i.e.~non-contractible) and hence there is no conical deficit in the resulting geometry.

Using the above methodology and the explicit expressions given in \hyperref[free boson]{Section \ref{free boson}}, it is simple to obtain the exact finite-regulator vacuum entanglement entropy for, say, the noncompact free boson by setting $k_1 = k_2 = 0$ and writing $V = 2\pi\delta(k=0)$ for the volume of target space. The boundary conditions considered in \hyperref[free boson]{Section \ref{free boson}} then correspond to ordinary Neumann and (indefinite) Dirichlet boundary conditions, which are the familiar conformally-invariant ones. For Neumann boundary conditions, the trace over $\rho_A^n$ is therefore
\begin{equation}
\text{NN:} \qquad \text{Tr}_{\mathcal{H}_A}\rho_A^n = \frac{V}{2\sqrt{2}\pi}\frac{\varepsilon'^{\frac{1}{6}(n-\frac{1}{n})}}{\varepsilon'^{-\frac{1}{6n}}\eta\hspace{-2pt}\left(\hspace{-2pt}-\frac{2i\ln\varepsilon'}{\pi n}\right)},
\end{equation}
from which it follows that the exact finite-regulator vacuum entanglement entropy of an interval of length $L$ is
\begin{equation}\label{NN EE exact}
S_{\text{EE}}^{\text{NN}}(L,\varepsilon) = \frac{\ln\hspace{-3pt}\left(\hspace{-3pt}\frac{L}{2\varepsilon}\hspace{-3pt}+\hspace{-3pt}\sqrt{\hspace{-1pt}1\hspace{-3pt}+\hspace{-3pt}\big(\hspace{-2pt}\frac{L}{2\varepsilon}\hspace{-2pt}\big)^{\hspace{-2pt}2}}\hspace{-2pt}\right)}{6}E_2\hspace{-3pt}\left(\hspace{-2pt}\frac{2i}{\pi}\hspace{-2pt}\ln\hspace{-3pt}\left(\hspace{-3pt}\tfrac{L}{2\varepsilon}\hspace{-3pt}+\hspace{-3pt}\sqrt{\hspace{-1pt}1\hspace{-3pt}+\hspace{-3pt}\big(\hspace{-2pt}\tfrac{L}{2\varepsilon}\hspace{-2pt}\big)^{\hspace{-2pt}2}}\hspace{-2pt}\right)\hspace{-3pt}\right) \hspace{-2pt}-\hspace{-2pt} \ln\hspace{-1pt}\eta\hspace{-2pt}\left(\hspace{-2pt}\frac{2i}{\pi}\hspace{-2pt}\ln\hspace{-3pt}\left(\hspace{-3pt}\tfrac{L}{2\varepsilon}\hspace{-3pt}+\hspace{-3pt}\sqrt{\hspace{-1pt}1\hspace{-3pt}+\hspace{-3pt}\big(\hspace{-2pt}\tfrac{L}{2\varepsilon}\hspace{-2pt}\big)^{\hspace{-2pt}2}}\hspace{-2pt}\right)\hspace{-3pt}\right) \hspace{-2pt}+\hspace{-2pt} \ln\hspace{-3pt}\left(\hspace{-3pt}\frac{V}{2\hspace{-1pt}\sqrt{2}\pi}\hspace{-3pt}\right)\hspace{-2pt},
\end{equation}
where $E_2(\tau)$ is the quasi-modular Eisenstein series of weight $2$ defined in \eqref{Eisenstein}. For (indefinite) Dirichlet boundary conditions, the trace over $\rho_A^n$ is
\begin{equation}
\text{DD:} \qquad \text{Tr}_{\mathcal{H}_A}\rho_A^n = \frac{V}{8\sqrt{2}\pi^3}\frac{\varepsilon'^{\frac{1}{6}(n-\frac{1}{n})}}{\varepsilon'^{-\frac{1}{6n}}\eta\hspace{-2pt}\left(\hspace{-2pt}-\frac{2i\ln\varepsilon'}{\pi n}\right)},
\end{equation}
and hence the entanglement entropy from Dirichlet boundary conditions differs from that from Neumann boundary conditions by a pure constant, namely
\begin{equation}\label{N and D EE difference}
S_{\text{EE}}^{\text{NN}}(L,\varepsilon) = S_{\text{EE}}^{\text{DD}}(L,\varepsilon) + \ln(2\pi)^2.
\end{equation}
Indeed, it is a familiar fact in 2d CFTs that the constant part in the entanglement entropy as computed from the modular Hamiltonian with different conformally-invariant boundary conditions differs precisely by (the logarithm of) the ratio of their corresponding boundary $g$-functions \cite{Affleck91}. From the explicit form of the Neumann and Dirichlet Cardy states in \eqref{Neumann Cardy state} and \eqref{Dirichlet Cardy state}, the boundary $g$-functions obey\footnote{The shift in \eqref{N and D EE difference} actually has the wrong sign compared to the logarithm of the ratio of the Neumann and Dirichlet boundary $g$-functions, but this is due to the extra factor of $1/(2\pi)^4$ in $\mathcal{N}_{k_1 k_2}^{\text{DD}}$ compared to $\mathcal{N}_{|\text{D}\rrangle}^2$.} $\mathcal{N}^2_{|\text{D}\rrangle}/\mathcal{N}^2_{|\text{N}\rrangle} = (2\pi)^2$. However, \eqref{N and D EE difference} expresses the much stronger fact that the constant logarithm of the ratio of the boundary $g$-functions is the only difference between the exact entanglement entropies even at finite regulator. Of course, we are free to choose other boundary conditions that shrink to the identity which do differ nontrivially at finite regulator.

The leading expansion of \eqref{NN EE exact}, subtracting the contribution from the boundary $g$-function, as $\varepsilon\rightarrow 0$ is
\begin{equation}
S_{\text{EE}}(L,\varepsilon) - \ln(\mathcal{N}V) \stackrel{\varepsilon\rightarrow 0}{\longrightarrow} \left[\frac{1}{3} - \frac{1}{2}\left(\frac{\varepsilon}{L}\right)^4 + \dotsc\right]\ln\left(\frac{L}{\varepsilon}\right) + \frac{1}{3}\left(\frac{\varepsilon}{L}\right)^2 + \frac{1}{2}\left(\frac{\varepsilon}{L}\right)^4 + \dotsc,
\end{equation}
which holds for the Dirichlet result as well, agreeing with the famous result that the universal part of the vacuum entanglement entropy of an interval of length $L$ in a 2d CFT is
\begin{equation}\label{2d universal EE}
\lim_{\varepsilon\rightarrow 0}\left(\frac{\partial S_{\text{EE}}(L,\varepsilon)}{\partial\ln L}\right) = \frac{c}{3};
\end{equation}
see, for example, \cite{Calabrese09} and references therein. Indeed, the angular quantization methodology also immediately produces this result in general. By the matching condition \eqref{2-point matching}, the leading contribution to $\text{Tr}_{\mathcal{H}_A}\rho_A$ is the normalization constant $\mathcal{N}$, and hence the leading contribution to $\text{Tr}_{\mathcal{H}_{\mathds{1}}}[e^{-2\pi H_{\text{R}}(\varepsilon')}]_{S^1\times\mathds{R}}$ is $\varepsilon'^{-c/3}$. Therefore, the leading term in the $n^{\text{th}}$ R\'{e}nyi entropy is 
\begin{align}
S_n & = -\frac{1}{n-1}\ln\left(\varepsilon'^{(1-n)c/3}\frac{\mathcal{N}\varepsilon'^{-c/3n}}{\mathcal{N}^n\varepsilon'^{-nc/3}} + \dotsc\right)
\\ & = -\frac{1}{n-1}\ln \left(\frac{\varepsilon'^{(1-1/n)c/3}}{\mathcal{N}^{n-1}}\right) + \dotsc 
\\ & = \ln\left(\mathcal{N}\varepsilon'^{-c/3n}\right) + \dotsc,
\end{align}
from which \eqref{2d universal EE} immediately follows due to $\varepsilon' = \frac{\varepsilon}{L} + \dotsc$. Of course, relating the trace to the sphere partition function via the Weyl anomaly is one way \eqref{2d universal EE} is derived in general. What we gain here is not new in terms of this result, but provides a solid foundation for the independence of boundary conditions given shrinkability. For example, this perspective makes it clear why pure Dirichlet boundary conditions, $X = x_0$ for fixed $x_0$, are actually inadmissible for the noncompact free boson and so would not compute the correct entanglement entropy quantities. It is clear from the general construction in \hyperref[general construction]{Section \ref{general construction}} that pure Dirichlet boundary conditions fail the shrinkability requirement because the finite-regulator state $|\text{D},x_0\rrangle$ contains infinitely many primaries of arbitrarily small dimension with equal coefficients, as mentioned before. Pure Dirichlet boundary conditions do however shrink to the identity operator for the compact boson because of the discrete spectrum of exponential primaries, indeed leading to the correct entanglement entropy as shown in \cite{Michel16}.

\section{Summary}

In this paper we have described the framework of angular quantization as it applies to 2d CFTs. While our primary motivation is to use these results to describe string theory in the presence of Euclidean time-winding operators, there are other applications of interest where this formalism can be useful. At the computational level, angular quantization can be considered as a study of non-conformal boundary conditions in bulk CFTs, thus being a generalization of the familiar Cardy state construction for conformally-invariant boundary conditions. In this paper we have focused on explicit examples of free CFTs, namely the noncompact and compact free bosons, the linear dilaton and the $bc$ ghost system. For a generic CFT, one may consider the bootstrap equations for the angular quantization Hilbert spaces associated to slicing correlators in different ways; these conditions encode locality of the boundary conditions, but it is difficult to provide explicit realizations of the Hilbert spaces because the lack of a state/operator correspondence means that the states are not organized into conformal primaries and descendants. 

From a more practical point of view, analytically continuing angular quantization on the sphere results in Rindler quantization, thus providing an explicit construction of the CFT on Minkowski space $\mathds{R}^{1,1}$. While a great deal is known about the structure of CFTs living on a circle, far less is known about the nonperturbative structure of CFTs living on an infinite line. Angular quantization as described here provides a connection between the two and thus can help learn more about Minkowski space CFTs. In particular, a Hilbert space of a CFT living on a line $\mathds{R}$ is labeled by two asymptotic conditions which in the Euclidean language determine the endpoint operators to which the associated boundary conditions shrink. That the Hilbert space should be labeled by these asymptotics conditions is perhaps not so surprising, as it is by now a common viewpoint that theories in asymptotically flat or asymptotically AdS spacetimes should have the chosen asymptotic conditions as part of their intrinsic data, since the Hilbert space consists of normalizable modes in the interior none of which can alter the asymptotics. 

Lastly, we briefly described the connection to entanglement entropy; although we have not obtained any new results in \hyperref[EE]{Section \ref{EE}}, the perspective of angular quantization allows us to be more careful about the shrinkability of boundary conditions. A general construction of entanglement entropy in QFT that is well-defined and independent of regulator for an appropriate class of admissible boundary conditions is still lacking. It is clear that a necessary requirement for admissibility is that the boundary condition placed on a tubular neighborhood around the subregion boundary must shrink to the trivial surface operator (the identity local operator in 2d). Demonstrating the equivalence of entanglement entropy for any boundary condition which shrinks to the identity remains a formidable task even for the free massive scalar. We hope to be able to say more about this matter in the future.

\section*{Acknowledgments}

We are grateful to Thomas Dumitrescu, Elliot Schneider and Xi Yin for discussions. This work is supported in part by DOE grant DE-SC0007870.

\appendix
\section{Oscillator Traces}\label{formulas}

The Dedekind eta function and Jacobi elliptic theta functions appearing in this paper are given by
\begin{align}
\eta(\tau) & = q^{1/24}\prod_{n=1}^{\infty}\left(1-q^n\right)
\\ \vartheta_2(\nu|\tau) & = 2q^{1/8}\cos \pi\nu \prod_{n=1}^{\infty}\left(1-q^n\right)\left(1+e^{2\pi i\nu}q^n\right)\left(1+e^{-2\pi i\nu}q^n\right)
\\ \vartheta_4(\nu|\tau) & = \prod_{n=1}^{\infty}\left(1-q^n\right)\left(1-e^{2\pi i\nu}q^{n-1/2}\right)\left(1-e^{-2\pi i\nu}q^{n-1/2}\right),
\end{align}
where the elliptic nome is $q\equiv e^{2\pi i \tau}$ with $\tau$ living in the upper-half-plane. They obey the modular-$S$ transformations
\begin{align}
\eta\left(-\frac{1}{\tau}\right) & = \sqrt{-i\tau} \ \eta(\tau)
\\ \vartheta_2(\nu|\tau) & = \sqrt{\frac{i}{\tau}}\ e^{-\pi i\nu^2/\tau}\vartheta_4\left(\left.\frac{\nu}{\tau}\right|-\frac{1}{\tau}\right).
\end{align}
The Eisenstein series $E_{2n}(\tau)$, defined to be\footnote{This sum does define the Eisenstein series, usually denoted $G_{2n}(\tau)$, which conventionally equals $2\zeta(2n)$ times $E_{2n}(\tau)$. We shall stick to calling $E_{2n}(\tau)$ the Eisenstein series, for the painfully obvious reason that it starts with `E'.} the sum of $1/\lambda^{2n}$ for all lattice vectors $\lambda \in \Lambda(\tau)$ associated to the torus $T^2(\tau) \cong \mathds{C}/\Lambda(\tau)$, is a holomorphic modular form of weight $2n$ for any integer $n\geqslant 2$. For $n=1$, the Eisenstein series $E_2(\tau)$ is instead a quasi-modular form of weight $2$ and is given explicitly by
\begin{equation}\label{Eisenstein}
E_2(\tau) = 1 - 24\sum_{n=1}^{\infty}\frac{nq^n}{1-q^n} = \frac{12}{\pi i} \ \partial_{\tau}\ln\eta(\tau).
\end{equation}
The quasi-modularity here means that it is $E_2(\tau) - \frac{3}{\pi\s\mathrm{Im}\s\tau}$ which is the actual modular form, with the non-holomorphic offset needed to cancel the part of the modular transformation where $\partial_{\tau}$ acts on $\sqrt{-i\tau}$.

Let us also record here the mode expansions of the stress-energy tensors and evolution operator for the compact boson angular quantization with endpoint operators $\mathcal{O}_{n_1,w_1}(0)$ and $\mathcal{O}_{n_2,w_2}(\infty)$. The stress-energy tensors on the plane are
\begin{align}
T(z) & = \frac{(w_1+w_2)^2 R^2}{16\ln^2\varepsilon}\frac{\ln^2 z}{z^2} - \frac{\pi i(w_1+w_2)R}{4\sqrt{2}\ln^2\varepsilon}\frac{\ln z}{z^2}\sum_{n\in\mathds{Z}}\alpha_n e^{\frac{\pi i n}{2}}z^{-\frac{\pi i n}{2\ln\varepsilon}}
\\ \notag & \hspace{10pt} - \frac{\pi^2}{8z^2\ln^2\varepsilon}\sum_{n,m\in\mathds{Z}}\circnormal{\alpha_n\alpha_m}e^{\frac{\pi i(n+m)}{2}}z^{-\frac{\pi i(n+m)}{2\ln\varepsilon}} + \frac{1}{24z^2}\left(1 + \frac{\pi^2}{4\ln^2\varepsilon}\right)
\\ \widetilde{T}(\barred{z}) & = \frac{(w_1+w_2)^2 R^2}{16\ln^2\varepsilon}\frac{\ln^2\barred{z}}{\barred{z}^2} + \frac{\pi i(w_1+w_2)R}{4\sqrt{2}\ln^2\varepsilon}\frac{\ln\barred{z}}{\barred{z}^2}\sum_{n\in\mathds{Z}}\widetilde{\alpha}_n e^{-\frac{\pi i n}{2}}\barred{z}^{\frac{\pi i n}{2\ln\varepsilon}}
\\ \notag & \hspace{10pt} - \frac{\pi^2}{8\barred{z}^2\ln^2\varepsilon}\sum_{n,m\in\mathds{Z}}\circnormal{\widetilde{\alpha}_n\widetilde{\alpha}_m}e^{-\frac{\pi i(n+m)}{2}}\barred{z}^{\frac{\pi i(n+m)}{2\ln\varepsilon}} + \frac{1}{24\barred{z}^2}\left(1 + \frac{\pi^2}{4\ln^2\varepsilon}\right),
\end{align}
where we have defined the zero-momentum modes
\begin{align}
\alpha_0 & \equiv -\frac{\sqrt{2}}{\pi}\left(x_{\text{s}} - \frac{i(n_1-n_2)}{2R} - \frac{i(w_1-w_2)R}{2}\right)\ln\varepsilon
\\ \widetilde{\alpha}_0 & \equiv -\frac{\sqrt{2}}{\pi}\left(x_{\text{s}} - \frac{i(n_1-n_2)}{2R} + \frac{i(w_1-w_2)R}{2}\right)\ln\varepsilon,
\end{align}
with $\widetilde{\alpha}_{n\neq 0} = \alpha_{n\neq 0}$. The cylinder evolution operator determined by the Hamiltonian \eqref{DD compact Hamiltonian} is
\begin{multline}
e^{-2\pi H_{\text{R}}^{\text{DD}}} = \varepsilon^{\frac{(n_1-n_2)^2}{4R^2} + [3(w_1-w_2)^2 + (w_1+w_2)^2]\frac{R^2}{12}}e^{-\frac{\pi i}{2}(n_1+n_2)(w_1-w_2)}e^{\frac{\pi^2(w_1+w_2)^2 R^2}{3\ln\varepsilon}}
\\ \times e^{\frac{i(n_1+n_2)}{R}x}e^{-\pi i(w_1-w_2)Rp}e^{-\frac{\pi i(w_1+w_2)R}{\ln\varepsilon}p_{\text{s}}}e^{x_{\text{s}}^2 \ln\varepsilon + \pi(w_1+w_2)R x_{\text{s}}}e^{\frac{\pi^2}{\ln\varepsilon}\sum_{n=1}^{\infty}\alpha_{-n}\alpha_n - \frac{\pi^2}{24\ln\varepsilon}}.
\end{multline}

The oscillator traces that appear in the matching of the two-point function are all of the form
\begin{align}
e^{-\frac{\pi^2}{24\ln\varepsilon}}\hspace{-3pt}\sum_{\{N_n\}}\hspace{-2pt}\langle\{N_n\}|e^{\frac{\pi^2}{\ln\varepsilon}\sum_{m=1}^{\infty}\alpha_{-m}\alpha_m}|\{N_n\}\rangle & = e^{-\frac{\pi^2}{24\ln\varepsilon}}\prod_{n=1}^{\infty}\sum_{N=0}^{\infty}\langle N| e^{\frac{\pi^2}{\ln\varepsilon}nN}|N\rangle 
\\ & = e^{-\frac{\pi^2}{24\ln\varepsilon}}\prod_{n=1}^{\infty}\frac{1}{1-e^{\frac{\pi^2}{\ln\varepsilon} n}}
\\ & = \frac{1}{\eta\left(-\frac{\pi i}{2\ln\varepsilon}\right)} = \sqrt{-\frac{\pi}{2\ln\varepsilon}}\frac{1}{\eta\left(-\frac{2i\ln\varepsilon}{\pi}\right)}.
\end{align}
The oscillator traces that appear in the thermal one-point functions are only slightly more annoying. We perform the calculation here explicitly for the compact boson, as the other traces are obtained from it as special cases. Define the quantity
\begin{equation}\label{oscillator coefficient}
f_n(z,\barred{z}) \equiv \left(\frac{z}{\barred{z}}\right)^{-\frac{\pi i n}{4\ln\varepsilon}}\left[\frac{n_3}{R}\sin\left(\frac{\pi n}{2\ln\varepsilon}\ln\left|\frac{z}{\varepsilon}\right|\right) + iw_3 R\cos\left(\frac{\pi n}{2\ln\varepsilon}\ln\left|\frac{z}{\varepsilon}\right|\right)\right].
\end{equation}
The oscillator trace needed for the thermal one-point function of $\mathcal{O}_{n_3,w_3}$ for the compact boson is
\begin{align}
\hspace{120pt} & \hspace{-120pt} \text{Tr}_{\text{osc.}}\left[e^{i\sqrt{2}\sum_{\ell=1}^{\infty}\frac{\alpha_{-\ell}}{\ell}f_{-\ell}}e^{-i\sqrt{2}\sum_{m=1}^{\infty}\frac{\alpha_m}{m}f_m}e^{\frac{\pi^2}{\ln\varepsilon}\sum_{n=1}^{\infty}\alpha_{-n}\alpha_n -\frac{\pi^2}{24\ln\varepsilon}}\right]
\\ & = e^{-\frac{\pi^2}{24\ln\varepsilon}}\prod_{n=1}^{\infty}\sum_{N=0}^{\infty}e^{\frac{\pi^2}{\ln\varepsilon}nN}\langle N|e^{i\sqrt{2}\frac{\alpha_{-n}}{n}f_{-n}}e^{-i\sqrt{2}\frac{\alpha_n}{n}f_n}|N\rangle
\\ & = e^{-\frac{\pi^2}{24\ln\varepsilon}}\prod_{n=1}^{\infty}\sum_{N=0}^{\infty}N! e^{\frac{\pi^2}{\ln\varepsilon}nN}\sum_{m=0}^N\frac{(2f_{-n}f_n)^m}{m!^2(N-m)!n^m}
\\ & = e^{-\frac{\pi^2}{24\ln\varepsilon}}\prod_{n=1}^{\infty}\sum_{N=0}^{\infty} e^{\frac{\pi^2}{\ln\varepsilon}nN}L_N\left(-\frac{2}{n}f_{-n}f_n\right)
\\ & = e^{-\frac{\pi^2}{24\ln\varepsilon}}\prod_{n=1}^{\infty}\frac{1}{1-e^{\frac{\pi^2 n}{\ln\varepsilon}}}\exp\left[\frac{2f_{-n}f_n}{n\left(e^{-\frac{\pi^2 n}{\ln\varepsilon}}-1\right)}\right],
\end{align}
where $L_N(x)$ is the Laguerre polynomial of degree $N$, whose generating function is
\begin{equation}
\sum_{N=0}^{\infty}L_N(x)t^N = \frac{e^{-tx/(1-t)}}{1-t}.
\end{equation}
The first infinite product is the same as above, given by the Dedekind eta function. For the second infinite product, we use
\begin{equation}
f_{-n}f_n = -\frac{1}{2}\left(\frac{n_3^2}{R^2} + w_3^2 R^2\right) + \frac{(-1)^n}{2}\left(\frac{n_3^2}{R^2} - w_3^2 R^2\right)\cos\left(\frac{\pi n\ln|z|}{\ln\varepsilon}\right)
\end{equation}
as well as the double summations
\begin{align}
\sum_{n,m=1}^{\infty}\frac{1}{n}e^{\frac{\pi^2 nm}{\ln\varepsilon}} & = \ln\left(e^{-\frac{\pi^2}{24\ln\varepsilon}}\prod_{m=1}^{\infty}\frac{1}{1-e^{\frac{\pi^2 m}{\ln\varepsilon}}}\right) + \frac{\pi^2}{24\ln\varepsilon}
\\ & = -\ln\left(\sqrt{-\frac{2\ln\varepsilon}{\pi}}\eta\left(-\frac{2i\ln\varepsilon}{\pi}\right)\right) + \frac{\pi^2}{24\ln\varepsilon}
\\ \sum_{n,m=1}^{\infty}\hspace{-3pt}\frac{(-1)^n}{n}e^{\frac{\pi^2 nm}{\ln\varepsilon}}\hspace{-2pt}\cos\hspace{-2pt}\left(\hspace{-2pt}\frac{\pi n\ln\hspace{-1pt}|z|}{\ln\varepsilon}\hspace{-2pt}\right) & = -\frac{1}{2}\ln\hspace{-2pt}\left[\prod_{m=1}^{\infty}\hspace{-3pt}\left(1 + e^{-\frac{\pi i\ln|z|}{\ln\varepsilon}}e^{\frac{\pi^2 m}{\ln\varepsilon}}\right)\left(1 + e^{\frac{\pi i\ln|z|}{\ln\varepsilon}}e^{\frac{\pi^2 m}{\ln\varepsilon}}\right)\right]
\\ & = -\frac{1}{2}\ln\left[\frac{1}{2\cos\left(\frac{\pi\ln|z|}{2\ln\varepsilon}\right)}\frac{\vartheta_2\hspace{-2pt}\left(\left.\hspace{-2pt}-\frac{\ln|z|}{2\ln\varepsilon}\right|\hspace{-2pt}-\hspace{-2pt}\frac{\pi i}{2\ln\varepsilon}\right)}{\eta\left(-\frac{\pi i}{2\ln\varepsilon}\right)}\right] \hspace{-2pt}+\hspace{-2pt} \frac{\pi^2}{24\ln\varepsilon}
\\ & = -\frac{1}{2}\ln\left[\frac{\vartheta_4\left(\left.-\frac{i\ln|z|}{\pi}\right|-\frac{2i\ln\varepsilon}{\pi}\right)}{\eta\left(-\frac{2i\ln\varepsilon}{\pi}\right)}\right] - \frac{\ln^2|z|}{4\ln\varepsilon} + \frac{\pi^2}{24\ln\varepsilon}
\end{align}
to compute
\begin{multline}
\prod_{n=1}^{\infty}\exp\left[\frac{2f_{-n}f_n}{n\left(e^{-\frac{\pi^2 n}{\ln\varepsilon}}-1\right)}\right] = \left(-\frac{\ln\varepsilon}{\pi}\right)^{\frac{1}{2}(\frac{n_3^2}{R^2} + w_3^2 R^2)}2^{\frac{n_3^2}{R^2}}\cos\left(\frac{\pi\ln|z|}{2\ln\varepsilon}\right)^{\frac{1}{2}(\frac{n_3^2}{R^2} - w_3^2 R^2)}
\\ \times \eta\hspace{-2pt}\left(\hspace{-2pt}-\frac{2i\ln\hspace{-1pt}\varepsilon}{\pi}\hspace{-2pt}\right)^{\hspace{-2pt}\frac{1}{2}\hspace{-1pt}\big(\hspace{-1pt}\frac{3n_3^2}{R^2} \hspace{-1pt}+\hspace{-1pt} w_3^2 R^2\hspace{-1pt}\big)}\hspace{-5pt}\vartheta_4\hspace{-2pt}\left(\left.\hspace{-4pt}-\hspace{-1pt}\frac{i\ln\hspace{-1pt}|z|}{\pi}\right|\hspace{-2pt}-\hspace{-2pt}\frac{2i\hspace{-1pt}\ln\hspace{-1pt}\varepsilon}{\pi}\hspace{-2pt}\right)^{\hspace{-3pt}-\hspace{-1pt}\frac{1}{2}\hspace{-1pt}\big(\hspace{-1pt}\frac{n_3^2}{R^2} \hspace{-1pt}-\hspace{-1pt} w_3^2 R^2\hspace{-1pt}\big)}\hspace{-4pt}e^{\hspace{-1pt}-\hspace{-1pt}\frac{\ln^2\hspace{-1pt}|z|}{4\ln\varepsilon}\hspace{-1pt}\big(\hspace{-1pt}\frac{n_3^2}{R^2} \hspace{-1pt}-\hspace{-1pt} w_3^2 R^2\hspace{-1pt}\big)}e^{\hspace{-1pt}-\hspace{-1pt}\frac{\pi^2 w_3^2 R^2}{12\ln\varepsilon}}.
\end{multline}
Therefore, the original oscillator trace is
\begin{align}
\hspace{30pt} & \hspace{-30pt} \text{Tr}_{\text{osc.}}\left[e^{i\sqrt{2}\sum_{\ell=1}^{\infty}\frac{\alpha_{-\ell}}{\ell}f_{-\ell}}e^{-i\sqrt{2}\sum_{m=1}^{\infty}\frac{\alpha_m}{m}f_m}e^{\frac{\pi^2}{\ln\varepsilon}\sum_{n=1}^{\infty}\alpha_{-n}\alpha_n -\frac{\pi^2}{24\ln\varepsilon}}\right]
\\ \notag & = \sqrt{-\frac{\pi}{2\ln\varepsilon}}\frac{1}{\eta\left(-\frac{2i\ln\varepsilon}{\pi}\right)}\left[-\frac{4\ln\varepsilon}{\pi}\cos\left(\frac{\pi\ln|z|}{2\ln\varepsilon}\right)\frac{\eta^3\hspace{-2pt}\left(-\frac{2i\ln\varepsilon}{\pi}\right)}{\vartheta_4\hspace{-2pt}\left(\left.\hspace{-2pt}-\frac{i\ln|z|}{\pi}\right|\hspace{-2pt}-\hspace{-2pt}\frac{2i\ln\varepsilon}{\pi}\hspace{-2pt}\right)}\right]^{\frac{n_3^2}{2R^2}}
\\ \notag & \hspace{10pt} \times \hspace{-2pt}\left[-\frac{\ln\varepsilon}{\pi\hspace{-1pt}\cos\hspace{-2pt}\left(\hspace{-2pt}\frac{\pi\ln\hspace{-1pt}|z|}{2\ln\varepsilon}\hspace{-2pt}\right)}\eta\hspace{-2pt}\left(\hspace{-2pt}-\frac{2i\hspace{-1pt}\ln\hspace{-1pt}\varepsilon}{\pi}\hspace{-2pt}\right)\hspace{-2pt}\vartheta_4\hspace{-2pt}\left(\left.\hspace{-4pt}-\hspace{-1pt}\frac{i\ln\hspace{-2pt}|z|}{\pi}\right|\hspace{-2pt}-\hspace{-2pt}\frac{2i\hspace{-1pt}\ln\hspace{-1pt}\varepsilon}{\pi}\hspace{-2pt}\right)\hspace{-2pt}\right]^{\hspace{-2pt}\frac{w_3^2 R^2}{2}}\hspace{-6pt}e^{-\frac{\ln^2|z|}{4\ln\varepsilon}\hspace{-1pt}\big(\hspace{-1pt}\frac{n_3^2}{R^2} \hspace{-1pt}-\hspace{-1pt} w_3^2 R^2\hspace{-1pt}\big)}e^{\hspace{-1pt}-\hspace{-1pt}\frac{\pi^2 w_3^2 R^2}{12\ln\varepsilon}}.
\end{align}
The noncompact Neumann-like free boson trace is obtained by setting $\frac{n_3}{R} = 0$ and $w_3 R = -k_3$, that for the noncompact Dirichlet-like free boson by setting $\frac{n_3}{R} = k_3$ and $w_3 R = 0$, that for the linear dilaton by setting $\frac{n_3}{R} = 0$ and $w_3 R = 2i\alpha_3$ and finally that for the bosonized $bc$ ghost system by setting $\frac{n_3}{R} = -\sqrt{2}$ and $w_3 R = 0$ (for the thermal one-point function of $c\widetilde{c}$). The elliptic functions are all evaluated at the nome $q = \varepsilon^4$, from which it is obvious that $\varepsilon^{-1/6}\eta(\tau)$ and $\vartheta_4(\nu|\tau)$ both approach unity as $\varepsilon\rightarrow 0$ for any fixed $\nu$.

\section{Chiral Splitting and the Operator Algebra}\label{chiral splitting}

In this appendix, we obtain the chiral splitting of the free compact boson $X(y,\barred{y}) = X_{\text{L}}(y) + X_{\text{L}}(\barred{y})$ in angular quantization as dictated by the path integral and demonstrate that the resulting mode expansions of the exponential primaries $\mathcal{O}_{n,w}$ automatically commute at equal times. 

The Euclidean path integral is performed by decomposing the field into eigenmodes of the Laplacian on the surface in question, separating out any ``zero-modes'' which appear only linearly or not at all in the action. For the path integral on the sphere, such terms are necessarily zero-modes of the Laplacian and hence satisfy the bulk equation of motion. For the compact boson path integral with action \eqref{compact boson action}, one decomposes $dX = dX_0 + 2\pi wR\omega$ into the single-valued part $dX_0$ and the winding part where $\omega$ generates $H^1(\Sigma;\mathds{Z})$ and subsequently sums over winding sectors. Only the single-valued part of $X$ can have a zero-mode since the bulk kinetic term involves a term proportional to $w^2$. This zero-mode is thus obtained by setting $X = x_0$ to a constant, for which the exponentiated boundary action at $|z| = \varepsilon$ reads
\begin{equation}
\left.\exp\left[\frac{i}{2\pi}\int_{|z|=\varepsilon}\hspace{-6pt}d\theta\left(\frac{n_1}{R}X + \lambda_1(\partial_{\theta}X - w_1 R)\right)\right]\right|_{\text{zero-mode}} = e^{\frac{in_1}{R}x_0 - iw_1 R\lambda_{1,0}},
\end{equation}
since the remaining integral over $\lambda_1$ just yields its constant Fourier mode. The regulated path integral is constructed to reproduce the path integral with the insertions $\mathcal{O}_{n_1,w_1}(\varepsilon)\mathcal{O}_{n_2,w_2}(\varepsilon^{-1})$ in the $\varepsilon\rightarrow 0$ limit, so in particular the above zero-mode must match that of the mode expansion $\mathcal{O}_{n_1,w_1} = \s\s\s\normal{e^{i[\frac{n_1}{R}X + w_1 R(X_{\text{L}}-X_{\text{R}})]}}$. From the chiral mode expansions \eqref{compact chiral L expansion} and \eqref{compact chiral R expansion} with $x_{\text{s}} = \alpha_n = 0$, we evaluate
\begin{multline}
\frac{in_1}{R}X(\varepsilon) + iw_1 R[X_{\text{L}}(\varepsilon) - X_{\text{R}}(\varepsilon)]\Big|_{\text{zero-mode}} 
\\ = \frac{in_1}{R}x + \frac{n_1(n_1-n_2)}{2R^2}\ln\varepsilon + iw_1 R(x_{\text{L}} - x_{\text{R}}) + \frac{w_1(w_1-w_2)R^2}{2}\ln\varepsilon + \frac{w_1(w_1+w_2)R^2}{4}\ln\varepsilon.
\end{multline}
Matching of these two zero-modes using \eqref{compact Lagrange multiplier zero-mode} then requires
\begin{align}
x_{\text{L}} - x_{\text{R}} & = -\lambda_{1,0} + \frac{i(w_1 - w_2)R}{2}\ln\varepsilon + \frac{i(w_1+w_2)R}{4}\ln\varepsilon
\\ & = -\pi\left(p + \frac{p_{\text{s}}}{\ln\varepsilon}\right) +\frac{i(w_1+w_2)R}{12}\ln\varepsilon,
\end{align}
as claimed in the text. Matching the zero-modes near infinity produces the same condition after mapping it to the origin via an inversion, which flips chirality and hence replaces $p_{\text{s}}$ with $-p_{\text{s}}$.

Finally, we may compute the equal-time commutator of two exponential primaries from their mode expansions. From \eqref{compact exponential expansion}, the mode expansion of $\mathcal{O}_{n,w}(y_1,y_2)$ on the strip is
\begin{align}
\normal{e^{i(\frac{n}{R} + wR)X_{\text{L}}(y) + i(\frac{n}{R} - wR)X_{\text{R}}(\barred{y})}} & = A(n,w,y_1,y_2)e^{\frac{in}{R}x}e^{-\pi i wRp}e^{\big(\frac{in}{R}y_2 + wR y_1\big)x_{\text{s}}}e^{-\frac{\pi i w R}{\ln\varepsilon}p_{\text{s}}}
\\ \notag & \hspace{10pt} \times e^{i\sqrt{2}\sum_{\ell=1}^{\infty}\frac{\alpha_{-\ell}}{\ell}f_{-\ell}(n,w)}e^{-i\sqrt{2}\sum_{m=1}^{\infty}\frac{\alpha_m}{m}f_m(n,w)},
\end{align}
where $A(n,w,y_1,y_2)$ is a $c$-number multiplicative coefficient not important for this calculation, and $f_m(n_3,w_3) \equiv f_m\big(z(y),\barred{z}(\barred{y})\big)$ is the quantity defined in \eqref{oscillator coefficient}. Now consider swapping the order of the exponentials in the equal-time product $\mathcal{O}_{n_3,w_3}(y_1,y_2)\mathcal{O}_{n_4,w_4}(y_1,y_2')$. The non-oscillator exponentials result in
\begin{equation}
e^{\frac{in_3}{R}x}e^{-\pi i w_3 Rp}e^{\frac{in_4}{R}x}e^{-\pi i w_4 Rp} = e^{\pi i(n_3 w_4 - n_4 w_3)}e^{\frac{in_4}{R}x}e^{-\pi i w_4 Rp}e^{\frac{in_3}{R}x}e^{-\pi i w_3 Rp}
\end{equation}
and
\begin{multline}
e^{\big(\frac{in_3}{R}y_2 + w_3 R y_1\big)x_{\text{s}}}e^{-\frac{\pi i w_3 R}{\ln\varepsilon}p_{\text{s}}}e^{\big(\frac{in_4}{R}y_2' + w_4 R y_1\big)x_{\text{s}}}e^{-\frac{\pi i w_4 R}{\ln\varepsilon}p_{\text{s}}} 
\\ = e^{\frac{\pi i}{\ln\varepsilon}(n_3 w_4 y_2 - n_4 w_3 y_2')}e^{\big(\frac{in_4}{R}y_2' + w_4 R y_1\big)x_{\text{s}}}e^{-\frac{\pi i w_4 R}{\ln\varepsilon}p_{\text{s}}}e^{\big(\frac{in_3}{R}y_2 + w_3 R y_1\big)x_{\text{s}}}e^{-\frac{\pi i w_3 R}{\ln\varepsilon}p_{\text{s}}}.
\end{multline}
Swapping the order of the oscillators, meanwhile, gives
\begin{multline}
e^{i\sqrt{2}\sum_{\ell=1}^{\infty}\frac{\alpha_{-\ell}}{\ell}f_{-\ell}(n_3,w_3)}e^{-i\sqrt{2}\sum_{m=1}^{\infty}\frac{\alpha_m}{m}f_m(n_3,w_3)}e^{i\sqrt{2}\sum_{\ell'=1}^{\infty}\frac{\alpha_{-\ell'}}{\ell'}f_{-\ell'}(n_4,w_4)}e^{-i\sqrt{2}\sum_{m'=1}^{\infty}\frac{\alpha_{m'}}{m'}f_{m'}(n_4,w_4)}
\\ \hspace{15pt} \hspace{-1pt}=\hspace{-1pt}  e^{2\hspace{-1pt}F}\hspace{-1pt}e^{i\hspace{-1pt}\sqrt{2}\hspace{-1pt}\sum_{\ell=1}^{\infty}\hspace{-2pt}\frac{\alpha_{-\ell}}{\ell}\hspace{-2pt}f_{-\ell}\hspace{-1pt}(\hspace{-1pt}n_3,\hspace{-1pt}w_3\hspace{-1pt})}\hspace{-2pt}e^{\hspace{-1pt}-\hspace{-1pt}i\hspace{-1pt}\sqrt{2}\hspace{-1pt}\sum_{m=1}^{\infty}\hspace{-2pt}\frac{\alpha_m}{m}\hspace{-2pt}f_m\hspace{-1pt}(\hspace{-1pt}n_3,\hspace{-1pt}w_3\hspace{-1pt})}\hspace{-2pt}e^{i\hspace{-1pt}\sqrt{2}\hspace{-1pt}\sum_{\ell'=1}^{\infty}\hspace{-2pt}\frac{\alpha_{-\ell'}}{\ell'}\hspace{-2pt}f_{-\ell'}\hspace{-1pt}(\hspace{-1pt}n_4,\hspace{-1pt}w_4\hspace{-1pt})}\hspace{-2pt}e^{\hspace{-1pt}-\hspace{-1pt}i\hspace{-1pt}\sqrt{2}\hspace{-1pt}\sum_{m'=1}^{\infty}\hspace{-3pt}\frac{\alpha_{m'}}{m'}\hspace{-2pt}f_{m\hspace{-1pt}'}\hspace{-1pt}(\hspace{-1pt}n_4,\hspace{-1pt}w_4\hspace{-1pt})}\hspace{-1pt},
\end{multline}
where the commutator in the exponential is
\begin{align}
F & = \sum_{m=1}^{\infty}\frac{1}{m}\left[f_m(n_3,w_3;y_1,y_2)f_{-m}(n_4,w_4;y_1,y_2') - f_{-m}(n_3,w_3;y_1,y_2)f_m(n_4,w_4;y_1,y_2')\right]
\\ & = 2i\hspace{-3pt}\sum_{m=1}^{\infty}\hspace{-3pt}\frac{1}{m}\hspace{-3pt}\left[\hspace{-1pt}n_3 w_4\sin\hspace{-3pt}\left(\hspace{-3pt}\frac{\pi m\hspace{-1pt}(\hspace{-1pt}y_2\hspace{-3pt}-\hspace{-2pt}\ln\hspace{-1pt}\varepsilon)}{2\ln\varepsilon}\hspace{-3pt}\right)\hspace{-3pt}\cos\hspace{-3pt}\left(\hspace{-3pt}\frac{\pi m\hspace{-1pt}(\hspace{-1pt}y_2'\hspace{-3pt}-\hspace{-2pt}\ln\hspace{-1pt}\varepsilon)}{2\ln\varepsilon}\hspace{-3pt}\right) \hspace{-3pt}-\hspace{-3pt} n_4 w_3\hspace{-1pt}\cos\hspace{-3pt}\left(\hspace{-3pt}\frac{\pi m\hspace{-1pt}(\hspace{-1pt}y_2\hspace{-3pt}-\hspace{-2pt}\ln\hspace{-1pt}\varepsilon)}{2\ln\varepsilon}\hspace{-3pt}\right)\hspace{-3pt}\sin\hspace{-3pt}\left(\hspace{-3pt}\frac{\pi m\hspace{-1pt}(\hspace{-1pt}y_2'\hspace{-3pt}-\hspace{-2pt}\ln\hspace{-1pt}\varepsilon)}{2\ln\varepsilon}\hspace{-3pt}\right)\hspace{-3pt}\right]
\\ & = - \pi i\left[n_3 w_4 \left(\theta(y_2 - y_2') + \frac{y_2 - \ln\varepsilon}{2\ln\varepsilon}\right) - n_4 w_3 \left(\theta(y_2' - y_2) + \frac{y_2'-\ln\varepsilon}{2\ln\varepsilon}\right)\right].
\end{align}
Thus the oscillators contribute a factor
\begin{equation}
e^{2F} = e^{\pi i(n_3 w_4 - n_4 w_3)}e^{-\frac{\pi i}{\ln\varepsilon}(n_3 w_4 y_2 - n_4 w_3 y_2')},
\end{equation}
so that the total equal-time commutator is
\begin{align}
\mathcal{O}_{\hspace{-1pt}n_3,w_3}\hspace{-1pt}(\hspace{-1pt}y_1,\hspace{-1pt}y_2\hspace{-1pt})\mathcal{O}_{\hspace{-1pt}n_4,w_4}\hspace{-1pt}(\hspace{-1pt}y_1,\hspace{-1pt}y_2'\hspace{-1pt}) & = e^{\pi i(\hspace{-1pt}n_3 w_4 \hspace{-1pt}-\hspace{-1pt} n_4 w_3\hspace{-1pt})}\hspace{-1pt}e^{\hspace{-1pt}\frac{\pi i}{\ln\varepsilon}\hspace{-1pt}(\hspace{-1pt}n_3 w_4 y_2 \hspace{-1pt}-\hspace{-1pt} n_4 w_3 y_2'\hspace{-1pt})}\hspace{-1pt}e^{2\hspace{-1pt}F}\hspace{-2pt}\mathcal{O}_{\hspace{-1pt}n_4,w_4}\hspace{-1pt}(\hspace{-1pt}y_1,\hspace{-1pt}y_2'\hspace{-1pt})\mathcal{O}_{\hspace{-1pt}n_3,w_3}\hspace{-1pt}(\hspace{-1pt}y_1,\hspace{-1pt}y_2\hspace{-1pt})
\\  & = \mathcal{O}_{n_4,w_4}(y_1,y_2')\mathcal{O}_{n_3,w_3}(y_1,y_2).
\end{align}
Therefore the mode expansions of the exponential primaries in angular quantization automatically obey the correct operator algebra, as claimed.

For the bosonized $bc$ ghost system, the story is nearly identical. The bosonized exponential primary $\mathcal{O}_{n_{\text{gh}},\widetilde{n}_{\text{gh}}}$ has strip mode expansion
\begin{align}
\mathcal{O}_{n_{\text{gh}},\widetilde{n}_{\text{gh}}} & = B(\hspace{-1pt}n_{\text{gh}},\hspace{-1pt}\widetilde{n}_{\text{gh}},\hspace{-1pt}y_1,\hspace{-1pt}y_2\hspace{-1pt})e^{\hspace{-1pt}-\hspace{-1pt}\frac{i}{\sqrt{2}}(\hspace{-1pt}n_{\text{gh}}\hspace{-1pt}+\hspace{-1pt}\widetilde{n}_{\text{gh}}\hspace{-1pt})h}e^{\frac{\pi i}{\sqrt{2}}(\hspace{-1pt}n_{\text{gh}} \hspace{-1pt}-\hspace{-1pt} \widetilde{n}_{\text{gh}}\hspace{-1pt})p_h}e^{\hspace{-1pt}-\hspace{-1pt}\frac{1}{\sqrt{2}}(\hspace{-1pt}n_{\text{gh}}y - \widetilde{n}_{\text{gh}}\barred{y})h_{\text{s}}}e^{\frac{\pi i}{\sqrt{2}\ln\varepsilon}(\hspace{-1pt}n_{\text{gh}}-\widetilde{n}_{\text{gh}}\hspace{-1pt})p_{\text{s}}}
\\ & \hspace{10pt} \times e^{i\sqrt{2}\sum_{\ell=1}^{\infty}\frac{h_{-\ell}}{\ell}f_{-\ell}(n_{\text{gh}},\widetilde{n}_{\text{gh}})}e^{-i\sqrt{2}\sum_{m=1}^{\infty}\frac{h_m}{m}f_m(n_{\text{gh}},\widetilde{n}_{\text{gh}})},
\end{align}
where $f_m(n_{\text{gh}},\widetilde{n}_{\text{gh}})$ is the same as the expression \eqref{oscillator coefficient} with the replacements $\frac{n}{R} = -\frac{n_{\text{gh}}+\widetilde{n}_{\text{gh}}}{\sqrt{2}}$ and $w R = -\frac{n_{\text{gh}}-\widetilde{n}_{\text{gh}}}{\sqrt{2}}$. Therefore, the equal-time commutator is
\begin{align}
\mathcal{O}_{\hspace{-1pt}n_{\text{gh}},\widetilde{n}_{\text{gh}}}\hspace{-1pt}(\hspace{-1pt}y_1,\hspace{-1pt}y_2\hspace{-1pt})\mathcal{O}_{\hspace{-1pt}n'_{\text{gh}},\widetilde{n}'_{\text{gh}}}\hspace{-1pt}(\hspace{-1pt}y_1,\hspace{-1pt}y_2'\hspace{-1pt}) & = \mathcal{O}_{\hspace{-1pt}n'_{\text{gh}},\widetilde{n}'_{\text{gh}}}\hspace{-1pt}(\hspace{-1pt}y_1,\hspace{-1pt}y_2'\hspace{-1pt})\mathcal{O}_{\hspace{-1pt}n_{\text{gh}},\widetilde{n}_{\text{gh}}}\hspace{-1pt}(\hspace{-1pt}y_1,\hspace{-1pt}y_2\hspace{-1pt})
\\ & \hspace{-40pt}\times e^{\hspace{-1pt}-\pi i(\hspace{-1pt}n_{\text{gh}}\widetilde{n}'\hspace{-1pt}_{\text{gh}} \hspace{-2pt}-\hspace{-1pt} \widetilde{n}_{\text{gh}}n'_{\text{gh}} \hspace{-1pt})}\hspace{-1pt}e^{\hspace{-1pt}\frac{\pi i}{2\ln\varepsilon}\hspace{-1pt}[\hspace{-1pt}(n_{\text{gh}}\hspace{-1pt}+\hspace{-1pt}\widetilde{n}_{\text{gh}})(n_{\text{gh}}'\hspace{-2pt}-\hspace{-1pt}\widetilde{n}'_{\text{gh}}) y_2 \hspace{-1pt}-\hspace{-1pt} (n_{\text{gh}}\hspace{-1pt}-\hspace{-1pt}\widetilde{n}_{\text{gh}})(n_{\text{gh}}'\hspace{-2pt}+\hspace{-1pt}\widetilde{n}'_{\text{gh}}) y_2'\hspace{-1pt}]}\hspace{-1pt}e^{2\hspace{-1pt}F},
\end{align}
where the oscillator commutator in the exponential is
\begin{equation}
F \hspace{-2pt}=\hspace{-2pt} -\frac{\pi i}{2}\hspace{-3pt}\left[\hspace{-2pt}\left(\hspace{-1pt}n_{\text{gh}}\hspace{-3pt}+\hspace{-3pt}\widetilde{n}_{\text{gh}}\hspace{-1pt}\right)\hspace{-3pt}\left(\hspace{-1pt}n'_{\text{gh}} \hspace{-3pt}-\hspace{-3pt} \widetilde{n}'_{\text{gh}}\hspace{-1pt}\right)\hspace{-4pt}\left(\hspace{-2pt}\theta(\hspace{-1pt}y_2 \hspace{-3pt}-\hspace{-3pt} y_2'\hspace{-1pt}) \hspace{-2pt}+\hspace{-2pt} \frac{y_2 \hspace{-3pt}-\hspace{-3pt} \ln\hspace{-1pt}\varepsilon}{2\ln\varepsilon}\hspace{-3pt}\right) \hspace{-3pt}-\hspace{-3pt} \left(\hspace{-1pt}n_{\text{gh}}\hspace{-3pt}-\hspace{-3pt}\widetilde{n}_{\text{gh}}\hspace{-1pt}\right)\hspace{-3pt}\left(\hspace{-1pt}n'_{\text{gh}} \hspace{-3pt}+\hspace{-3pt} \widetilde{n}'_{\text{gh}}\hspace{-1pt}\right)\hspace{-4pt}\left(\hspace{-2pt}\theta(\hspace{-1pt}y_2' \hspace{-3pt}-\hspace{-3pt} y_2\hspace{-1pt}) \hspace{-2pt}+\hspace{-2pt} \frac{y_2'\hspace{-3pt}-\hspace{-3pt}\ln\hspace{-1pt}\varepsilon}{2\ln\varepsilon}\hspace{-2pt}\right)\hspace{-2pt}\right]\hspace{-1pt}.
\end{equation}
As before, the $(y_2-\ln\varepsilon)/2\ln\varepsilon$ and $(y_2'-\ln\varepsilon)/2\ln\varepsilon$ terms in $F$ exactly cancel the commutator contributions from the non-oscillator terms. Now, however, the $\theta(y_2-y_2')$ and $\theta(y_2'-y_2)$ terms in $F$ do not simply give unity because of the extra $1/2$ in front. Instead, we may write, for example,
\begin{multline}
-\frac{\pi i}{2}\left[\left(\hspace{-1pt}n_{\text{gh}}\hspace{-3pt}+\hspace{-3pt}\widetilde{n}_{\text{gh}}\hspace{-1pt}\right)\hspace{-3pt}\left(\hspace{-1pt}n'_{\text{gh}} \hspace{-3pt}-\hspace{-3pt} \widetilde{n}'_{\text{gh}}\hspace{-1pt}\right)\theta(y_2 - y_2') - \left(\hspace{-1pt}n_{\text{gh}}\hspace{-3pt}-\hspace{-3pt}\widetilde{n}_{\text{gh}}\hspace{-1pt}\right)\hspace{-3pt}\left(\hspace{-1pt}n'_{\text{gh}} \hspace{-3pt}+\hspace{-3pt} \widetilde{n}'_{\text{gh}}\hspace{-1pt}\right)\theta(y_2'-y_2)\right]
\\ = -\frac{\pi i}{2}\left(\hspace{-1pt}n_{\text{gh}}\hspace{-3pt}+\hspace{-3pt}\widetilde{n}_{\text{gh}}\hspace{-1pt}\right)\hspace{-3pt}\left(\hspace{-1pt}n'_{\text{gh}} \hspace{-3pt}-\hspace{-3pt} \widetilde{n}'_{\text{gh}}\hspace{-1pt}\right) + \pi i \left(n_{\text{gh}}n'_{\text{gh}} - \widetilde{n}_{\text{gh}}\widetilde{n}'_{\text{gh}}\right)\theta(y_2'-y_2),
\end{multline}
where the second term is always an integer times $\pi i$, and hence contributes unity, but the first term is not. The total equal-time commutator is hence
\begin{equation}\label{ghost exponential (anti)commutator}
\mathcal{O}_{\hspace{-1pt}n_{\text{gh}},\widetilde{n}_{\text{gh}}}\hspace{-1pt}(\hspace{-1pt}y_1,\hspace{-1pt}y_2\hspace{-1pt})\mathcal{O}_{\hspace{-1pt}n'_{\text{gh}},\widetilde{n}'_{\text{gh}}}\hspace{-1pt}(\hspace{-1pt}y_1,\hspace{-1pt}y_2'\hspace{-1pt}) = (-1)^{(n_{\text{gh}}+\widetilde{n}_{\text{gh}})(n'_{\text{gh}}+\widetilde{n}'_{\text{gh}})}\mathcal{O}_{\hspace{-1pt}n'_{\text{gh}},\widetilde{n}'_{\text{gh}}}\hspace{-1pt}(\hspace{-1pt}y_1,\hspace{-1pt}y_2'\hspace{-1pt})\mathcal{O}_{\hspace{-1pt}n_{\text{gh}},\widetilde{n}_{\text{gh}}}\hspace{-1pt}(\hspace{-1pt}y_1,\hspace{-1pt}y_2\hspace{-1pt}),
\end{equation}
where we freely replaced $(n_{\text{gh}}+\widetilde{n}_{\text{gh}})(n'_{\text{gh}}-\widetilde{n}'_{\text{gh}})$ with $(n_{\text{gh}}+\widetilde{n}_{\text{gh}})(n'_{\text{gh}}+\widetilde{n}'_{\text{gh}})$ since their difference is an even integer. The integer $n_{\text{gh}}+\widetilde{n}_{\text{gh}}$ is the total ghost number of $\mathcal{O}_{n_{\text{gh}},\widetilde{n}_{\text{gh}}}$, and the ghost number modulo $2$ is the ``fermion'' number for the $bc$ system. Therefore, the result \eqref{ghost exponential (anti)commutator} says that the mode expansions of two $bc$ ghost system primaries anticommute if and only if both their total ghost numbers are odd, otherwise they commute. Therefore, these mode expansions also obey the correct operator algebra, showing that angular quantization automatically takes into account the 2-cocycles in bosonization that must be added by hand in radial quantization.

\end{document}